\documentclass[11pt,a4paper]{article}
\pdfoutput=1
\usepackage{float}
\usepackage{jheppub}
\usepackage[T1]{fontenc}
\usepackage{lmodern,booktabs,mathtools,microtype,longtable,mathrsfs}
\usepackage{changepage}
\usepackage[section]{placeins}
\usepackage{tikz}
\usetikzlibrary{arrows.meta,decorations.markings,decorations.pathmorphing}
\allowdisplaybreaks[2]
\renewcommand\afterTocRuleSpace{\clearpage}

\newcommand{\dd}{\mathrm d}
\newcommand{\ii}{\mathrm i}
\newcommand{\Tr}{\operatorname{Tr}}
\newcommand{\TJJ}{\langle TJJ\rangle}
\newcommand{\TTT}{\langle TTT\rangle}
\newcommand{\nsJ}{n_{\mathrm s}^{TJJ}}
\newcommand{\nfJ}{n_{\mathrm f}^{TJJ}}
\newcommand{\nsT}{n_{\mathrm s}^{TTT}}
\newcommand{\nfT}{n_{\mathrm f}^{TTT}}

\newcommand{\intk}{\int\!\frac{\dd^3k}{(2\pi)^3}}

\numberwithin{equation}{section}
\hypersetup{pdftitle={Conformal collider data in QED3, scalar QED3, and QED3-GNY},
pdfauthor={Yongwei Guo, Zhijin Li and Ke Ren},
pdfsubject={Energy-flux projections and squashed-sphere free energy}}
\title{\boldmath $1/N$ corrections to $\langle TJJ\rangle$ and $\langle TTT\rangle$
in QED$_3$, QED$_3$--GNY and scalar QED$_3$}
\author[a,b]{Yongwei Guo}
\author[a,b]{\unskip, Zhijin Li}
\author[c]{and Ke Ren}
\affiliation[a]{School of Physics, Southeast University, 2 Sipailow Street, Nanjing 210096, China}
\affiliation[b]{Shing-Tung Yau Center, Southeast University, 2 Sipailow Street, Nanjing 210096, China}
\affiliation[c]{School of Physics and Astronomy, Sun Yat-Sen University, Zhuhai 519082, China}
\emailAdd{guoyw33@seu.edu.cn}
\emailAdd{zhijin\_li@seu.edu.cn}
\emailAdd{renk9@mail.sysu.edu.cn}
\abstract{Three-dimensional Abelian gauge theories have broad applications in deconfined quantum criticality, Dirac spin liquids and cuprate superconductivity. Correlation functions of conserved currents $J$ and the stress tensor $T$ carry fundamental information about these theories. We compute the first subleading large-flavor corrections to the parity-even three-point functions $\langle TJJ\rangle$ and $\langle TTT\rangle$ in QED$_3$, QED$_3$--Gross--Neveu--Yukawa, and critical scalar QED$_3$. For $\langle TJJ\rangle$, we study both global flavor currents and the topological $U(1)$ current, which measures the magnetic flux of the dynamical gauge field. We evaluate the flat-space energy flux in directions where the leading free-field profile vanishes. At first subleading order the flux in these directions is determined by four-particle production, yielding analytic corrections to the conformal collider parameters. We reproduce the corrections independently from squashed-sphere free-energy derivatives. Together with known two-point normalizations, the collider parameters fix the coefficients of both independent tensor structures in each correlator. Our results are useful for studying conductivity and shear response near quantum critical points and for conformal bootstrap studies of conserved-current and stress-tensor correlators in strongly coupled gauge theories.}
\begin{document}
\maketitle
\clearpage
\pagenumbering{arabic}
\raggedbottom
\section{Introduction and summary of results}\label{sec:intro}

Three-dimensional Abelian gauge theories provide fundamental
field-theory descriptions of strongly interacting quantum matter in
two spatial dimensions. Their applications include deconfined quantum
criticality, Dirac spin liquids, and proposed descriptions of
high-temperature superconductivity. Compared with the $O(N)$ vector
and Gross--Neveu--Yukawa (GNY) models studied in
ref.~\cite{GuoLiThreePoint}, dynamical gauge fields introduce
fractionalized excitations and monopole operators. These ingredients
allow gauge theories to describe phases and transitions beyond a
conventional theory of local order parameters.

We count $N_f$ two-component Dirac fermions and $N_b$ complex scalars.
At $N_f=2$, QED$_3$--GNY is proposed to be dual to critical noncompact
scalar QED$_3$ with $N_b=2$, the $NCCP^1$ description of the
N\'eel--valence-bond-solid (VBS) deconfined quantum critical point
(DQCP). The proposed emergent $SO(5)$ symmetry combines the three
N\'eel and two VBS components into one order parameter
\cite{SenthilDQCP,WangDQCP,BoyackQEDGN}.
Without the GNY scalar, $N_f=2$ QED$_3$ is instead proposed to
describe the easy-plane DQCP, with possible emergent $O(4)$ symmetry
\cite{WangDQCP}. These proposals belong to the three-dimensional
duality web connecting bosonic and fermionic gauge theories
\cite{SeibergDualityWeb,KarchTongDuality,SenthilDualitiesReview}.
They motivate studying universal observables beyond critical exponents.

At $N_f=4$, QED$_3$ is a candidate continuum description of Dirac spin
liquids in frustrated quantum magnets, with electrically neutral
Dirac spinons interacting through an emergent Abelian gauge field
\cite{HermeleSpinLiquid,SongSpinLiquids,HuDiracSpinLiquid}.
QED$_3$ also has a longstanding role in proposed descriptions of
cuprate superconductivity: algebraic spin liquids model the underdoped
regime \cite{RantnerWenCuprates,RantnerWenSpinCorrelations}, while
nodal quasiparticles coupled to vortex fluctuations appear in
phase-disordered $d$-wave superconductors
\cite{FranzTesanovicQED3,FranzTesanovicVafekQED3}.
The two four-component fermions often used in this literature
correspond to four two-component flavors. Applying continuum results
to a material requires matching the physical currents and accounting
for perturbations such as velocity anisotropy.

A distinctive observable is the topological $U(1)$ current
$j_{\mathrm{top}}$, whose charge measures the magnetic flux of the
dynamical gauge field in units of $2\pi$. This symmetry is present
when monopole deformations are absent or irrelevant at the fixed
point. In the $NCCP^1$ description, it rotates the VBS order parameter,
represented by a unit-flux monopole \cite{SenthilDQCP,WangDQCP}.
Dualities can exchange topological and flavor currents, making their
separate three-point functions physically important. Chowdhury et
al.~\cite{ChowdhuryTransport} obtained their leading large-flavor
collider data and explained their connection to particle--vortex
transport; here we determine the first interaction corrections.

The transport motivations for $\TJJ$ and $\TTT$ were discussed in
detail in ref.~\cite{GuoLiThreePoint}. Briefly, $\TJJ$ fixes the
stress-tensor contribution to the high-frequency electrical or spin
conductivity, while $\TTT$ enters the shear response and spectral sum
rules near quantum criticality
\cite{ChowdhuryTransport,KatzEtAl,ChowdhuryDavidPrakash}.
The topological-current correlator has the corresponding role in its
conductivity. Thermal one-point functions and additional dynamical
information are needed for a complete finite-temperature response.

These three-point coefficients also provide input for the conformal
bootstrap. Scalar and monopole correlators already constrain QED$_3$
and DQCP candidates
\cite{ChesterPufuQEDBootstrap,LiDQCPBootstrap,LiPolandGauge,AlbayrakQEDBootstrap,ChesterSuDQCP,LiShenDQCP}.
Current two-point normalizations determine Ward-identity data, but
systems with external conserved currents and stress tensors also
require independent three-point structures
\cite{DymarskyCurrentBootstrap,ReehorstMixed,HeNonAbelianCurrents,DymarskyStressBootstrap,ChangEtAl,ErramilliMitchellStress,DommesEtAl}.
Our previous large-$N$ predictions for the $O(N)$ collider parameters,
extrapolated to $O(2)$, agree well with the joint bootstrap region
under weaker spectral assumptions \cite{GuoLiThreePoint,DommesEtAl}.
This encourages extending the calculation to gauge theories and
provides a benchmark for future mixed-current bootstrap studies.

We consider QED$_3$, QED$_3$--GNY, and critical scalar QED$_3$ at zero
Chern--Simons level. The GNY field is one real flavor-singlet scalar.
The flavor current $J$ is associated with a Hermitian generator $Q$,
with $\Tr Q=0$ and $\Tr Q^2=1$; its eigenvalues are denoted by $q_i$.
The topological current is normalized in eq.~\eqref{eq:topcurrent}.
Throughout, we assume that the theories considered here flow to conformal infrared fixed points. Our results are controlled at large flavor number; applications at small flavor number require extrapolation.

We obtain analytic first subleading corrections to the conformal
collider parameters $\gamma$, $\gamma_{\mathrm{top}}$, and $t_4$ from
flat-space energy-flux projections. An independent calculation from
squashed-sphere free-energy derivatives reproduces all nine corrections.
As a further check in QED$_3$, we applied the Euclidean momentum-space
subgraph expansion used in our previous work~\cite{GuoLiThreePoint}.
The resulting $1/N_f$ coefficients agree exactly with those obtained
from the Lorentzian energy-flux calculation.
Combined with the known current and stress-tensor two-point normalizations
\cite{HuhStrackSachdevCurrents,HuhStrackCentralCharges,GiombiTarnopolskyKlebanov,GuoLi},
the collider parameters determine the coefficients of both independent parity-even tensor
structures in each correlator.
We summarize these results and their collider bounds below.

\subsection{Summary of results}\label{sec:summary-results}

We specify the separated-point parity-even coefficients by
\begin{align}
 \TJJ&=\nsJ\TJJ_{\mathrm s}+\nfJ\TJJ_{\mathrm f},\label{eq:Jstructures}\\
 \TTT&=\nsT\TTT_{\mathrm s}+\nfT\TTT_{\mathrm f}.\label{eq:Tstructures}
\end{align}
The reference correlators are those of one complex scalar and one
two-component Dirac fermion. Their two-point normalizations agree:
\begin{equation}
 C_J^{(0)}=\frac1{8\pi^2},\qquad C_T^{(0)}=\frac3{16\pi^2}.
 \label{eq:free-norms}
\end{equation}
The flavor invariant is already removed from $C_J$.
Consequently $\nsJ+\nfJ=C_J/C_J^{(0)}$, while
$\nsT+\nfT=C_T/C_T^{(0)}$. The collider parameters have the canonical
definitions
\begin{equation}
 \gamma=\frac{\nfJ-\nsJ}{12(\nsJ+\nfJ)},\qquad
 t_4=4\frac{\nsT-\nfT}{\nsT+\nfT}.
 \label{eq:collider}
\end{equation}
The scalar endpoint is $(-1/12,4)$ and the fermion endpoint is $(1/12,-4)$.

For the topological $U(1)$ current normalized in eq.~\eqref{eq:topcurrent},
the BF insertion exchanges the leading scalar and fermion endpoints.
The collider parameters for both currents and the stress tensor are
\begin{align}
 \gamma_{\mathrm{QED}_3}
 &=\frac1{12}+\frac1{N_f}\left(\frac13-\frac{152}{27\pi^2}\right)
 +O(N_f^{-2}),\label{eq:qed-gamma}\\
 \gamma_{\mathrm{top},\mathrm{QED}_3}
 &=-\frac1{12}+\frac1{N_f}\left(-\frac13+\frac{152}{27\pi^2}\right)
 +O(N_f^{-2}),\label{eq:qed-top-gamma}\\
 t_{4,\mathrm{QED}_3}
 &=-4+\frac1{N_f}\left(\frac{6792064}{45\pi^2}-15264\right)
 +O(N_f^{-2}),\label{eq:qed-t4}\\[1mm]
 \gamma_{\mathrm{sQED}_3}
 &=-\frac1{12}+\frac1{N_b}\left(-\frac14+\frac{52}{9\pi^2}\right)
 +O(N_b^{-2}),\label{eq:sqed-gamma}\\
 \gamma_{\mathrm{top},\mathrm{sQED}_3}
 &=\frac1{12}+\frac1{N_b}\left(\frac{143}{12}-\frac{1076}{9\pi^2}\right)
 +O(N_b^{-2}),\label{eq:sqed-top-gamma}\\
 t_{4,\mathrm{sQED}_3}
 &=4+\frac1{N_b}\left(\frac{48536}{3}-\frac{2398208}{15\pi^2}\right)
 +O(N_b^{-2}),\label{eq:sqed-t4}\\[1mm]
 \gamma_{\mathrm{QED}_3\text{-}\mathrm{GNY}}
 &=\frac1{12}+\frac1{N_f}\left(\frac14-\frac{46}{9\pi^2}\right)
 +O(N_f^{-2}),\label{eq:gny-gamma}\\
 \gamma_{\mathrm{top},\mathrm{QED}_3\text{-}\mathrm{GNY}}
 &=-\frac1{12}+\frac1{N_f}\left(-\frac{143}{12}+\frac{1094}{9\pi^2}\right)
 +O(N_f^{-2}),\label{eq:gny-top-gamma}\\
 t_{4,\mathrm{QED}_3\text{-}\mathrm{GNY}}
 &=-4+\frac1{N_f}\left(\frac{2399648}{15\pi^2}-\frac{48536}{3}\right)
 +O(N_f^{-2}).\label{eq:gny-t4}
\end{align}
Each ratio is expanded consistently in the inverse flavor number.
Independent flat-space energy-flux projections and
squashed-sphere source derivatives give all nine correction coefficients.
Their numerical values are listed in table~\ref{tab:collider}.
\begin{table}[ht]
\centering
\begin{tabular}{lrrr}
\toprule
Theory&$\gamma$&$\gamma_{\mathrm{top}}$&$t_4$\\
\midrule
QED$_3$&$-0.2370674042$&$0.2370674042$&$28.8880856558$\\
Critical scalar QED$_3$&$0.3354112833$&$-0.1968437332$&$-20.6182120349$\\
QED$_3$--GNY&$-0.2678638275$&$0.3994861005$&$30.3450456646$\\
\bottomrule
\end{tabular}
\caption{Coefficients multiplying $1/N_f$ or $1/N_b$ in
 eqs.~\eqref{eq:qed-gamma}--\eqref{eq:gny-t4}.
 The exact expressions follow from the flat-space phase-space integrals
 and agree with the independent sphere calculation.}
\label{tab:collider}
\end{table}

The topological-current calculation is given in
section~\ref{sec:topological}.
Its collider parameter and the known two-point normalization fix both
separated-point parity-even structures of
$\langle Tj_{\mathrm{top}}j_{\mathrm{top}}\rangle$ to relative order $N^{-1}$.
Their coefficients are listed in table~\ref{tab:topJstructures}.
Changing the current leaves $t_4$ unchanged.

The independent two-point results are written as
\begin{equation}
 \frac{C_J}{C_J^{(0)}}=1+\frac{c_J}{N}+O(N^{-2}),\qquad
 \frac{C_T}{C_T^{(0)}}=N+c_T+O(N^{-1}).
 \label{eq:centralcharges}
\end{equation}
Here $N$ is the corresponding $N_f$ or $N_b$.
For the topological current in eq.~\eqref{eq:topcurrent}, the normalization is
\begin{equation}
 \frac{C_J^{\mathrm{top}}}{C_J^{(0)}}
 =\frac{64}{\pi^2N}\left[1+\frac{c_J^{\mathrm{top}}}{N}+O(N^{-2})\right].
 \label{eq:topcentralcharges}
\end{equation}
The coefficients for both currents are given in table~\ref{tab:norms}
\cite{GiombiTarnopolskyKlebanov,GuoLi}.
The stress-tensor coefficient $c_T$ is unchanged by the choice of current.
Combining these independent normalizations with the collider ratios fixes
both coefficients in each parity-even correlator.
\begin{table}[ht]
\centering
\begin{tabular}{lccc}
\toprule
Theory & $c_J$ & $c_J^{\mathrm{top}}$ & $c_T$\\
\midrule
QED$_3$ & $-4 + \frac{368}{9 \pi^{2}}$ & $4-\frac{368}{9\pi^2}$ & $-4 + \frac{2096}{45 \pi^{2}}$\\
Scalar QED$_3$ & $-4 + \frac{208}{3 \pi^{2}}$ & $4-\frac{272}{3\pi^2}$ & $-4 + \frac{2476}{45 \pi^{2}}$\\
QED$_3$--GNY & $-4 + \frac{112}{3 \pi^{2}}$ & $4-\frac{176}{3\pi^2}$ & $-4 + \frac{2116}{45 \pi^{2}}$\\
\bottomrule
\end{tabular}
\caption{Two-point corrections for the flavor current, topological current,
 and stress tensor in eqs.~\eqref{eq:centralcharges} and~\eqref{eq:topcentralcharges}.}
\label{tab:norms}
\end{table}
\begin{table}[ht]
\centering
\begin{tabular}{lcc}
\toprule
Theory & Correction to $\nsJ$ & Correction to $\nfJ$\\
\midrule
QED$_3$ & $-2 + \frac{304}{9 \pi^{2}}$ & $-2 + \frac{64}{9 \pi^{2}}$\\
Scalar QED$_3$ & $- \frac{5}{2} + \frac{104}{3 \pi^{2}}$ & $- \frac{3}{2} + \frac{104}{3 \pi^{2}}$\\
QED$_3$--GNY & $- \frac{3}{2} + \frac{92}{3 \pi^{2}}$ & $- \frac{5}{2} + \frac{20}{3 \pi^{2}}$\\
\bottomrule
\end{tabular}
\caption{Coefficients multiplying $1/N$ in the flavor-current structures. The leading pair is $(0,1)$ for fermions and $(1,0)$ for scalars.}
\label{tab:Jstructures}
\end{table}

For the topological current, eq.~\eqref{eq:topstructureextraction}
gives the coefficients in table~\ref{tab:topJstructures}.
The displayed corrections are of relative order $N^{-1}$ after
factoring out the leading normalization $64/(\pi^2N)$.
\begin{table}[ht]
\centering
\begin{tabular}{lcc}
\toprule
Theory & Correction to $n_{\mathrm s}^{\mathrm{top}}$
 & Correction to $n_{\mathrm f}^{\mathrm{top}}$\\
\midrule
QED$_3$ & $6-\frac{224}{3\pi^2}$ & $-2+\frac{304}{9\pi^2}$\\
Scalar QED$_3$ & $-\frac{143}{2}+\frac{2152}{3\pi^2}$ & $\frac{151}{2}-\frac{808}{\pi^2}$\\
QED$_3$--GNY & $\frac{151}{2}-\frac{788}{\pi^2}$ & $-\frac{143}{2}+\frac{2188}{3\pi^2}$\\
\bottomrule
\end{tabular}
\caption{Coefficients multiplying $64/(\pi^2N^2)$ in the topological-current
 structures. The leading pair is $64(1,0)/(\pi^2N_f)$ for fermions and
 $64(0,1)/(\pi^2N_b)$ for scalars. Terms of order $N^{-3}$ are omitted.}
\label{tab:topJstructures}
\end{table}

\begin{table}[ht]
\centering
\begin{tabular}{lcc}
\toprule
Theory & Correction to $\nsT$ & Correction to $\nfT$\\
\midrule
QED$_3$ & $-1908 + \frac{849008}{45 \pi^{2}}$ & $1904 - \frac{282304}{15 \pi^{2}}$\\
Scalar QED$_3$ & $\frac{6055}{3} - \frac{896852}{45 \pi^{2}}$ & $- \frac{6067}{3} + \frac{299776}{15 \pi^{2}}$\\
QED$_3$--GNY & $- \frac{6067}{3} + \frac{299956}{15 \pi^{2}}$ & $\frac{6055}{3} - \frac{897752}{45 \pi^{2}}$\\
\bottomrule
\end{tabular}
\caption{Order-one corrections to the stress-tensor structures. The leading pair is $(0,N_f)$ for fermions and $(N_b,0)$ for scalars.}
\label{tab:Tstructures}
\end{table}
For bootstrap comparisons, the displayed coefficients must be converted
to the chosen unit-normalized operator conventions using $C_J$ and
$C_T$. The conformal collider parameters specify the ratios of the two
independent tensor-structure coefficients.
Thus central-charge bounds in conformal gauge-theory studies
\cite{LiPolandGauge,AlbayrakQEDBootstrap} and the present three-point
predictions provide complementary inputs.

Each row of tables~\ref{tab:Jstructures}, \ref{tab:topJstructures},
and~\ref{tab:Tstructures} sums to $c_J$, $c_J^{\mathrm{top}}$, and $c_T$,
respectively. These equalities check the normalization conventions. 

\subsection*{Collider bounds and the large-flavor expansion}

Positive energy flux requires $|\gamma|\leq1/12$ and $|t_4|\leq4$
\cite{HofmanMaldacena,HofmanEtAlProof,HartmanKunduTajdini}. For the current correlator, the bound on $\gamma$ also follows from the holographic causality analysis of Myers, Sachdev, and Singh~\cite{MyersSachdevSinghTransport}.
All first corrections displayed above move inward from their leading
endpoints.
This sign check applies at sufficiently large flavor number.
It does not bound the omitted terms at small flavor number.
For example, the leading QED$_3$ expansion gives $t_4\simeq10.4440$
at $N_f=2$. That value lies outside the collider interval.
It therefore diagnoses the failure of this truncation as a controlled
small-flavor prediction, rather than a violation by the exact CFT.

Section~\ref{sec:sources} introduces the background sources and explains
why the chosen energy-flux projections isolate the leading corrections.
Section~\ref{sec:current} evaluates the flavor and topological
current cuts, and section~\ref{sec:stress} treats the stress tensor.
Sections~\ref{sec:sphere} and~\ref{sec:sphere-evaluation} give the
independent sphere calculation, beginning with quadratic responses
that determine $C_J$ and $C_T$ and concluding with the inverse-photon
response for the topological current in
section~\ref{sec:topological-sphere}. We discuss the results in
section~\ref{sec:discussion}.
The appendices provide the phase-space conventions, exact radial integrals,
induced vertices, sphere regularization, and the metric response of
photon--auxiliary mixing.

\section{Gauge theories and energy-flux projections}\label{sec:sources}

We determine the three-point coefficients by measuring the angular
distribution of energy in a state created by a current or the stress
tensor. At leading order in the large-flavor expansion, the distribution
has the shape of one of the two free-field structures in
eqs.~\eqref{eq:Jstructures} and~\eqref{eq:Tstructures}.
The two shapes vanish in different directions. By placing the detector
at a zero of the leading distribution, we isolate the contribution of
the other structure. This is the flux we calculate at order $1/N$.
The two-point normalization fixes the sum of the two coefficients, so
the selected flux determines both when combined with that normalization.
We first obtain the production vertices by coupling the theories to
background sources. Sections~\ref{sec:current} and~\ref{sec:stress}
then apply the projection, and sections~\ref{sec:sphere}
and~\ref{sec:sphere-evaluation} extract the same data independently
from the free energy on a squashed sphere.

\subsection{Background sources and Feynman rules}

We use background sources to define the correlators and their contact
terms. These contact terms distinguish derivatives of the source
functional from correlators of separate operator insertions
\cite{OsbornPetkou,ErdmengerOsborn,BMSMomentum,BMSRenormalised}.

Let $B_\mu$ be the external flavor source and $a_\mu$ the dynamical photon.
The metric and flavor source enter the generating functional
\begin{equation}
 W[g,B]=-\log Z[g,B],\qquad
 Z[g,B]=\int\mathcal D\Phi\,\exp[-S[\Phi;g,B]],
 \label{eq:W}
\end{equation}
where $\Phi$ denotes the dynamical fields collectively.
The operator insertions are given by
\begin{equation}
 T^{\mu\nu}=-\frac2{\sqrt g}\frac{\delta S}{\delta g_{\mu\nu}},
 \qquad J^\mu=\frac1{\sqrt g}\frac{\delta S}{\delta B_\mu}.
 \label{eq:operators}
\end{equation}
For three source directions $s_i$, write $S_i=\partial_iS$ and
$S_{ij}=\partial_i\partial_jS$, with all derivatives evaluated at zero
sources. Three differentiations of eq.~\eqref{eq:W} then give
\begin{equation}
 W_{123}=\langle S_{123}\rangle
 -\langle S_{12}S_3+S_{13}S_2+S_{23}S_1\rangle_c
 +\langle S_1S_2S_3\rangle_c.
 \label{eq:thirdsource}
\end{equation}
The second derivatives generate contact vertices, with two source
insertions at the same point. Their correlators with the third insertion
can be nonzero when it is separated from that point. These semilocal
terms can contain nonanalytic two-point functions in momentum space.


We now apply these source variations to the matter actions. The same
actions determine the induced propagators that enter the production
amplitudes.
For fermions we use the symmetric Dirac action
\begin{equation}
 S_f=\int\dd^3x\sqrt g\left[
 \frac12\bar\psi_i\gamma^a e_a{}^\mu
 \overleftrightarrow{D}_\mu\psi_i+\sigma\bar\psi_i\psi_i\right],
 \quad D_\mu=\nabla_\mu-\ii a_\mu-\ii B_\mu Q.
 \label{eq:fermionaction}
\end{equation}
The scalar $\sigma$ is absent in QED$_3$.
It is the real Gross--Neveu auxiliary field in QED$_3$--GNY.
It couples to the flavor-singlet fermion bilinear
$\sum_i\bar\psi_i\psi_i$ in eq.~\eqref{eq:fermionaction}.
Related gauge--Yukawa critical points have been studied in dimensional
expansions \cite{IhrigQEDGN,ZerfQEDGNY}, while the singlet-coupled
three-dimensional theory is studied directly at large flavor number
in ref.~\cite{BoyackQEDGN}.
We use $N_f$ two-component Dirac fermions throughout.
The critical quadratic coupling of $\sigma$ is tuned before computing source
derivatives. Irrelevant local kinetic terms do not enter the leading
infrared propagators.

The critical scalar action is
\begin{equation}
 S_b=\int\dd^3x\sqrt g\left[
 g^{\mu\nu}(D_\mu\phi_i)^*D_\nu\phi_i
 +\frac R8\phi_i^*\phi_i+\ii\sigma\phi_i^*\phi_i\right].
 \label{eq:scalaraction}
\end{equation}
The factor $\ii$ specifies the auxiliary-field contour.
The curvature coupling supplies the conformal improvement.
The critical mass is tuned together with the auxiliary-field saddle.

The induced-propagator expansion is the large-flavor organization used
in the current and stress-tensor two-point calculations
\cite{HuhStrackSachdevCurrents,HuhStrackCentralCharges,GiombiTarnopolskyKlebanov,GuoLi}.
At leading order, the inverse propagators are generated by matter loops.
Each loop sums over the matter flavors, so the resulting propagators are
of order $1/N$. In flat space they are
\begin{align}
 D_{\mu\nu}(p)&=\frac{16}{N|p|}
 \left[\delta_{\mu\nu}-(1-\xi)\frac{p_\mu p_\nu}{p^2}\right],
 \label{eq:photonprop}\\
 D_{\sigma,f}(p)&=\frac8{N_f|p|},\qquad
 D_{\sigma,b}(p)=\frac{8|p|}{N_b}.
 \label{eq:sigmaprop}
\end{align}
In eq.~\eqref{eq:photonprop}, $N=N_f$ for fermions and $N=N_b$ for
scalars. The two scalar Yukawa vertices contribute $\ii^2=-1$.
Their exchange is therefore equivalent to a propagator $-8|p|/N_b$
when the vertices are written without $\ii$.

To obtain the source vertices, we vary the scalar and fermion kinetic
densities, which contain $\sqrt g\,g^{-1}$
and $\sqrt g\,g^{-1/2}$, respectively. In the symmetric vielbein
convention their linear variations in a symmetric metric direction $h$ are
\begin{equation}
 (\sqrt g\,g^{-1})_h=\tfrac12\Tr(h)I-h,\qquad
 (\sqrt g\,g^{-1/2})_h=\tfrac12\Tr(h)I-\tfrac12h.
 \label{eq:linear-source-vertices}
\end{equation}
These variations also act on the gauge vertices.
They generate the mixed stress-tensor--photon vertices. For example,
the mixed vertex cancels the longitudinal part of the fermion
emission diagrams in eq.~\eqref{eq:TemissionWard}.
The scalar action has an additional gauge seagull. For flavor-source
profiles $b,c$, its second derivative is
\begin{equation}
 S_{B_bB_c}=2\int\dd^3x\sqrt g\,
 g^{\mu\nu}b_\mu c_\nu\,\phi^*Q^2\phi.
 \label{eq:seagull}
\end{equation}
Replacing one flavor generator by the identity gives the flavor--photon
contact. The fermion kinetic action is linear in the gauge field and
has no such elementary seagull.

The curvature term in eq.~\eqref{eq:scalaraction} supplies the improved
stress-tensor vertex. Its contribution is essential for the trace
identity of the scalar induced vertex. This vertex comes from varying
the matter loop that generates the photon kernel. Varying the matter
propagators gives the triangle, and varying the current vertices gives
the mixed-vertex bubbles.
We display their sum and their separate transverse components in
section~\ref{sec:stress} and appendix~\ref{app:induced}.

\subsection{Energy-flux projections}\label{sec:energy-flux-projections}

Let $\boldsymbol n$ be a unit vector in the two-dimensional spatial plane.
At large radius $r$, the detector integrates the outgoing energy flow
in direction $\boldsymbol n$ over the retarded time $u=t-r$.
Its definition at future null infinity is \cite{HofmanMaldacena,BelitskyEventShapes}
\begin{equation}
 \mathcal E(\boldsymbol n)=
 \int_{-\infty}^{\infty}\!\dd u\,
 \lim_{r\to\infty}r\,n_iT^{0i}(r+u,r\boldsymbol n).
 \label{eq:detector}
\end{equation}
The limit is taken in Wightman matrix elements with smeared insertions.
We then take the momentum-space limit for a state of energy $E$ and
zero spatial momentum.
A spatial current polarization $\epsilon_i$ or symmetric traceless
stress-tensor polarization $e_{ij}$ creates the state. Positivity of this
observable is supported by the lightcone-bootstrap derivation of the
collider bounds \cite{LiMeltzerPoland,HofmanEtAlProof} and by the
causality derivation of averaged null energy \cite{HartmanKunduTajdini}.
The stress tensor in the detector and the two operators creating the bra
and ket make this measurement a projection of $\TJJ$ or $\TTT$.
Dividing by the norm of the state gives \cite{HofmanMaldacena}
\begin{align}
 \frac{\langle J(\epsilon)^\dagger\mathcal E(\boldsymbol n)
 J(\epsilon)\rangle}{\langle J(\epsilon)^\dagger J(\epsilon)\rangle}
 &=\frac E{2\pi}\left[1-24\gamma
 \left(\frac{|\epsilon\cdot\boldsymbol n|^2}{|\epsilon|^2}
 -\frac12\right)\right],\label{eq:currentflux}\\
 \frac{\langle T(e)^\dagger\mathcal E(\boldsymbol n)T(e)\rangle}
 {\langle T(e)^\dagger T(e)\rangle}
 &=\frac E{2\pi}\left[1+t_4
 \left(\frac{|e_{ij}n_in_j|^2}{e_{ij}^*e_{ij}}-\frac14\right)\right].
 \label{eq:stressflux}
\end{align}
Here $J(\epsilon)=\epsilon_iJ_i$ and $T(e)=e_{ij}T_{ij}$;
the insertions carry the fixed source momentum.
Their angular averages are $E/(2\pi)$, as required by energy conservation.

For example, in QED$_3$ the flavor current has the free fermion value
$\gamma=1/12$ at leading order. Choose the detector parallel to a real
polarization $\epsilon$. In units of the angular average $E/(2\pi)$,
eqs.~\eqref{eq:currentflux} and~\eqref{eq:collider} give
\begin{equation*}
 1-12\gamma
 =\frac{2\nsJ}{\nsJ+\nfJ}
 =-\frac{12\gamma_1}{N_f}+O(N_f^{-2}),\qquad
 \gamma=\frac1{12}+\frac{\gamma_1}{N_f}+O(N_f^{-2}).
\end{equation*}
The flux therefore vanishes at leading order and directly gives
$\gamma_1$ at the next order. For a leading scalar current profile,
the perpendicular detector similarly measures
$2\nfJ/(\nsJ+\nfJ)$. For $e=\operatorname{diag}(1,-1)$,
the fermion and scalar stress-tensor profiles are $1-\cos4\theta$ and
$1+\cos4\theta$, respectively, with $\theta$ measured from the first
spatial axis. Their zeros differ by $\pi/4$, so the same procedure
isolates the contribution of the other stress-tensor structure.
For the topological current, the coupling to the photon rotates the
current polarization by $\pi/2$ and exchanges the two leading profiles,
as explained in section~\ref{sec:topological}.

The special detector direction also simplifies the perturbative
calculation. At leading order, the source produces a back-to-back
matter--antimatter pair. The free flux is a sum of squared production
amplitudes with positive energy weights. At a zero of that flux, the
tree-level two-particle amplitude vanishes for the configuration in
which one member of the pair points toward the detector. Its interference
with the first virtual correction thus vanishes in the selected flux.
The square of the virtual correction can first contribute at relative
order $N^{-2}$.

The dressed photon and, where present, the auxiliary scalar have
continuous spectra from matter--antimatter pairs. The cut separates
the production amplitude from its complex conjugate. In the
perturbative prescription of appendix~\ref{app:lorentzian-cuts}, cutting
one of these dressed lines puts the two members of a second matter pair
on shell. The first new final state therefore contains four matter
particles.
For example, in QED$_3$ the flavor-current emission amplitude carries
a factor $q_i/N_f$ when the current creates a pair of flavor $i$.
Squaring this factor, summing $\sum_i q_i^2=1$, and summing over the
flavor of the second pair gives an order-$1/N_f$ contribution.
The leading flavor-current state norm is of order one. For the stress
tensor the flavor sum is of order $N^2$ and the leading norm is of
order $N$, giving the same relative $1/N$ correction. The topological
current has the same counting once the common external photon factor
cancels from the flux-to-norm ratio.

\begin{figure}[htbp]
\centering
\begin{tikzpicture}[font=\small,line width=.75pt,
 matter/.style={postaction={decorate},decoration={markings,
 mark=at position .55 with {\arrow{Stealth[length=1.7mm,width=1.5mm]}}}},
 source/.style={densely dotted},
 amplitude/.style={draw,rounded corners=7pt,fill=black!6,
 minimum width=1.4cm,minimum height=2.65cm,inner sep=0pt}]
\node[amplitude] (L) at (0,0) {$\mathcal M_{\mathcal O}$};
\node[amplitude] (R) at (6,0) {$\mathcal M_{\mathcal O}^{*}$};
\draw[source] (-1.75,0)--(L.west);
\draw[source] (R.east)--(7.75,0);
\node[left] at (-1.75,0) {$\mathcal O(q)$};
\node[right] at (7.75,0) {$\mathcal O^{\dagger}(q)$};
\draw[matter] (.7,1.05)--(5.3,1.05);
\draw[matter] (5.3,.35)--(.7,.35);
\draw[matter] (.7,-.35)--(5.3,-.35);
\draw[matter] (5.3,-1.05)--(.7,-1.05);
\foreach \y/\lab in {1.05/1,.35/2,-.35/3,-1.05/4}
 {\node[above,inner sep=3pt] at (2.05,\y) {$k_{\lab}$};}
\draw[dashed,line width=.65pt] (3,-1.6)--(3,1.65);
\node[above] at (3,1.7) {$\mathcal E(\boldsymbol n)$};
\node[below,align=center] at (3,-1.65)
 {four on-shell matter lines};
\node[below] at (0,-1.65) {production};
\node[below] at (6,-1.65) {conjugate};
\end{tikzpicture}
\caption{Inclusive cut through four fundamental matter lines for the selected energy flux, with
$\mathcal O=J(\epsilon)$, $T(e)$, or $j_{\mathrm{top}}(\epsilon)$.
The shaded blocks denote the complete production amplitude and its
complex conjugate. The detector insertion is summed over the four
on-shell matter lines, with the energy and directional weight of each
line. All emission, contact, and induced-vertex contributions are summed
before squaring the amplitude. At relative order $N^{-1}$, this resolved
cut determines the collider correction at the selected zero of the
leading two-particle flux.}
\label{fig:four-particle-cut}
\end{figure}
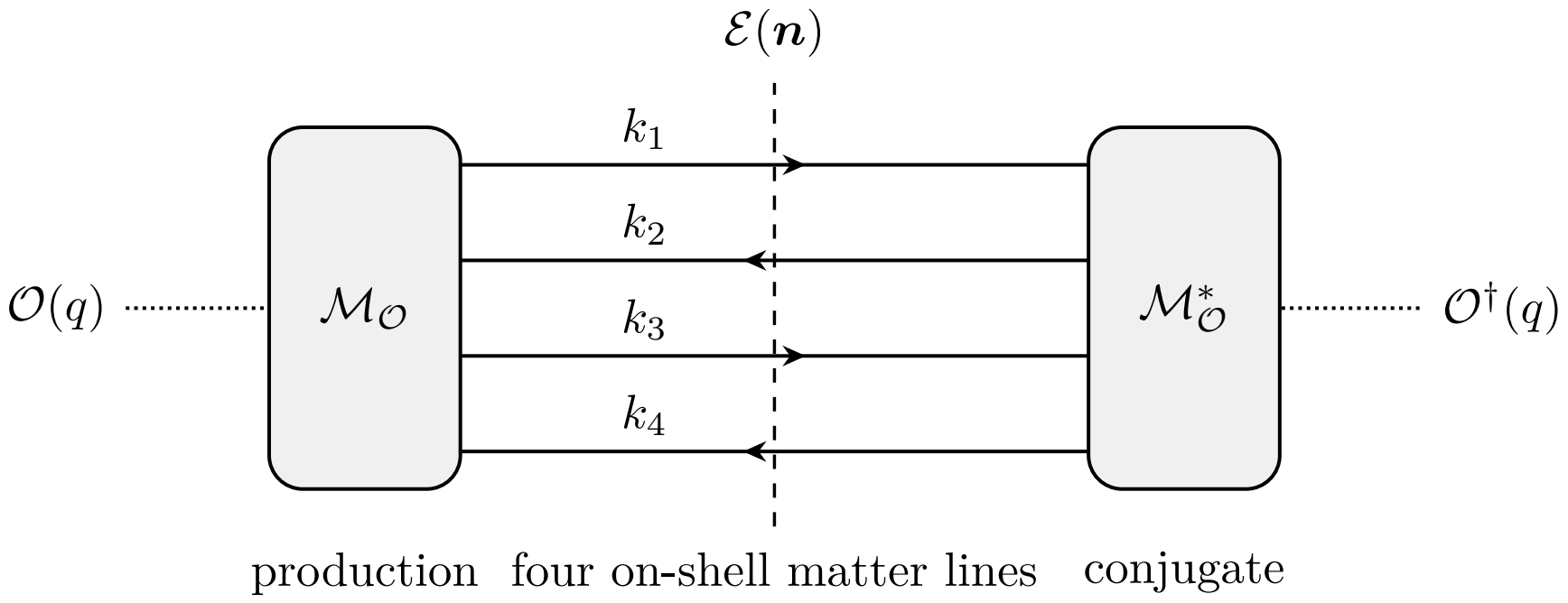

Figure~\ref{fig:four-particle-cut} shows how the flux is obtained from
this final state. We add the allowed emission, contact and induced-vertex
diagrams at the amplitude level. An induced vertex joining two dressed
lines can contribute at the same order as ordinary emission because
its internal matter loop supplies a factor of $N$.
We then integrate the squared amplitude
over the four-particle phase space, weighting each particle by the
energy it carries into the detector direction. Since the selected
leading flux is zero, only the leading state norm is needed for this
first correction.

\section{\texorpdfstring{$\langle TJJ\rangle$}{TJJ} correlators with flavor or topological currents}\label{sec:current}

We determine $\TJJ$ for the flavor current and
$\langle Tj_{\mathrm{top}}j_{\mathrm{top}}\rangle$ for the conserved topological current. Both have the conformal structures of
eq.~\eqref{eq:Jstructures}, but their source couplings and flavor sums
are different. We begin with the flavor-current cuts and then turn to
the topological-current calculation in section~\ref{sec:topological}.

\subsection{Conformal structures and the flavor-current energy-flux projection}\label{sec:flavor-projection}

The leading scalar and fermion structures and their transport
interpretation were studied in ref.~\cite{ChowdhuryTransport}.
Here we extract their interaction-induced admixture. This uses the same
finite-dimensional conformal matching principle as the Euclidean
subgraph approach \cite{LiSubgraphs,GuoLiThreePoint}, but a different
projection: an inclusive Lorentzian cut replaces the hard and soft
momentum expansion. We retain the distinction because the organization
of intermediate contributions and source contacts differs.

For a mixed bootstrap with a physical global current as an external
operator, these coefficients determine stress-tensor exchange in the
$J\times J$ OPE. Mixed scalar--current and current--stress-tensor
studies already constrain this information
\cite{ReehorstMixed,DommesEtAl}. The Ward identity relates one
combination to $C_J$, but the independent ratio $\gamma$ must also be
specified. The corrections below therefore supply additional data for
such systems, beyond a comparison of current two-point coefficients.

The ratio in eq.~\eqref{eq:collider} can be measured in a direction
where the free-theory flux vanishes. For fermions, the current polarization
is parallel to the detector. For scalars, it is perpendicular to the
detector. The opposite free structure gives twice the angular average in
each selected direction. The leading pair-production amplitude vanishes
in the detected configuration, eliminating its interference with the
first virtual correction. At relative order $N^{-1}$, the selected flux
is therefore determined by inclusive cuts through four fundamental
matter lines, as explained in section~\ref{sec:energy-flux-projections}.

Set the source energy to one and write its momentum as $q=(1,0,0)$.
The four outgoing massless momenta are $k_1,\ldots,k_4$, with
$q=\sum_a k_a$. The pair momenta and masses are
\begin{equation}
 P=k_1+k_2,\quad Q=k_3+k_4,\qquad
 x=\sqrt{P^2},\quad y=\sqrt{Q^2},\qquad x,y\geq0,\quad x+y\leq1.
 \label{eq:pairkinematics}
\end{equation}
We use Minkowski signature $(+,-,-)$ for these invariants.
Gamma-matrix expressions below use the equivalent complex Euclidean
momenta $\widehat k=(k^0,-\ii k^1,-\ii k^2)$.
Thus $(\widehat k)^2=k^2$ and all hatted contractions use $\delta_{\mu\nu}$.
Appendix~\ref{app:lorentzian-cuts} specifies the Feynman boundary values
represented in this tensor basis.

After the overall spatial rotation is integrated, the phase-space measure is
\begin{equation}
 \dd\Phi_4=\frac{\dd x\,\dd y}{64\pi^2}
 \frac{\dd\alpha}{2\pi}\frac{\dd\beta}{2\pi}.
 \label{eq:phasespace}
\end{equation}
The angles $\alpha,\beta$ describe the two pair decays.
This measure follows by factorizing three two-body phase spaces,
$\dd\Phi_2=\dd\theta/(8\pi M)$, and the two invariant-mass measures.
The free current norm in this convention is $1/8$.

\subsection{Flavor-current calculation}

\subsubsection*{QED$_3$}

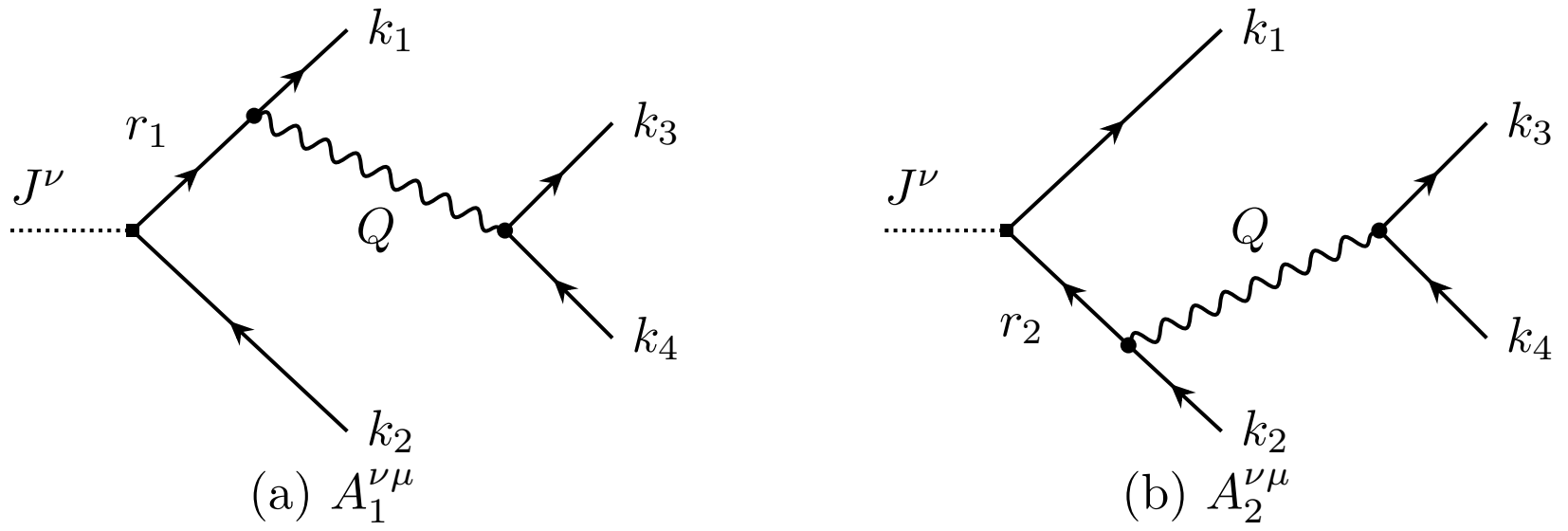
\begin{figure}[H]
\centering
\begin{tikzpicture}[font=\small,
 matter/.style={line width=.75pt,postaction={decorate},
 decoration={markings,mark=at position .55 with {\arrow{Stealth[length=1.7mm,width=1.5mm]}}}},
 photon/.style={line width=.75pt,decorate,
 decoration={snake,amplitude=.6mm,segment length=2.3mm}},
 source/.style={densely dotted,line width=.75pt}]
\begin{scope}[xshift=0cm]
\coordinate (v) at (0,0);
\coordinate (w) at (2.6,0);
\coordinate (k1) at (1.5,1.4);\coordinate (k2) at (1.5,-1.4);
\coordinate (k3) at (3.35,.75);\coordinate (k4) at (3.35,-.75);
\draw[source] (-.85,0)--(v);
\node[above] at (-.65,.05) {$J^\nu$};
\coordinate (e) at (.85,.8);
\draw[matter] (v)--(e);\draw[matter] (e)--(k1);
\draw[matter] (k2)--(v);
\draw[photon] (e)--(w);
\node[above left] at (.4,.45) {$r_1$};
\fill (e) circle (1.6pt);
\draw[matter] (w)--(k3);\draw[matter] (k4)--(w);
\fill (v) +(-.045,-.045) rectangle +(.045,.045);\fill (w) circle (1.6pt);
\node[right] at (k1) {$k_1$};\node[right] at (k2) {$k_2$};
\node[right] at (k3) {$k_3$};\node[right] at (k4) {$k_4$};
\node at (1.7,0) {$Q$};
\node at (1.4,-1.85) {(a) $A_1^{\nu\mu}$};
\end{scope}
\begin{scope}[xshift=6.1cm]
\coordinate (v) at (0,0);
\coordinate (w) at (2.6,0);
\coordinate (k1) at (1.5,1.4);\coordinate (k2) at (1.5,-1.4);
\coordinate (k3) at (3.35,.75);\coordinate (k4) at (3.35,-.75);
\draw[source] (-.85,0)--(v);
\node[above] at (-.65,.05) {$J^\nu$};
\coordinate (e) at (.85,-.8);
\draw[matter] (v)--(k1);
\draw[matter] (e)--(v);\draw[matter] (k2)--(e);
\draw[photon] (e)--(w);
\node[below left] at (.4,-.45) {$r_2$};
\fill (e) circle (1.6pt);
\draw[matter] (w)--(k3);\draw[matter] (k4)--(w);
\fill (v) +(-.045,-.045) rectangle +(.045,.045);\fill (w) circle (1.6pt);
\node[right] at (k1) {$k_1$};\node[right] at (k2) {$k_2$};
\node[right] at (k3) {$k_3$};\node[right] at (k4) {$k_4$};
\node at (1.7,0) {$Q$};
\node at (1.4,-1.85) {(b) $A_2^{\nu\mu}$};
\end{scope}
\end{tikzpicture}
\caption{Photon-emission diagrams for the fermion flavor-current
production amplitude in eq.~\eqref{eq:Jfermionamp}.
The current creates a fermion--antifermion pair of flavor $i$.
The fermion emits the photon in panel (a), and the antifermion emits it
in panel (b). The photon creates the second pair of flavor $j$.
The panels correspond to $A_1^{\nu\mu}$ and $A_2^{\nu\mu}$ in
 eq.~\eqref{eq:Jfermiondiagrams}.
The dotted line denotes the current insertion with incoming momentum $q$.
The wavy line denotes the leading large-$N_f$ dressed photon propagator
with momentum $Q=k_3+k_4$.
All external matter momenta are outgoing; the solid-line arrows indicate
fermion flow. The internal momenta are $r_1=q-k_2$ and $r_2=q-k_1$.}
\label{fig:flavor-fermion-emission}
\end{figure}

Write $r_1=q-k_2$ and $r_2=q-k_1$.
Figure~\ref{fig:flavor-fermion-emission} shows photon emission from
the fermion and antifermion lines.

The corresponding photon-emission matrices are
\begin{align}
 A_1^{\nu\mu}&=\gamma^\mu
 \frac{\gamma^a\widehat r_{1a}}{r_1^2}\gamma^\nu,\nonumber\\
 A_2^{\nu\mu}&=-\gamma^\nu
 \frac{\gamma^a\widehat r_{2a}}{r_2^2}\gamma^\mu.
 \label{eq:Jfermiondiagrams}
\end{align}
Their matrix elements multiply the current of the second pair,
$J_{34}^\mu=\bar u_3\gamma^\mu v_4$.
The resulting reduced amplitude is
\begin{equation}
 M_f^\nu=\frac1y\bar u_1(A_1^{\nu\mu}+A_2^{\nu\mu})v_2J_{34,\mu}.
 \label{eq:Jfermionamp}
\end{equation}
The physical amplitude includes the overall factor $16q_i/N_f$.
The leading flavor sum uses $\sum_iq_i^2=1$ and contributes $N_f$.
The interference obtained by moving the flavor source to the other pair
is proportional to $\sum_{i\ne j}q_iq_j=-1$ and enters one order later.

On-shell spinors obey $\bar u_1\gamma^a\widehat k_{1a}=0$ and
$\gamma^a\widehat k_{2a}v_2=0$. Substituting these equations into
eq.~\eqref{eq:Jfermiondiagrams} gives
\begin{align}
 \bar u_1\widehat Q_\mu(A_1^{\nu\mu}+A_2^{\nu\mu})v_2
 &=\bar u_1(\gamma^\nu-\gamma^\nu)v_2=0,\nonumber\\
 \bar u_1\widehat q_\nu(A_1^{\nu\mu}+A_2^{\nu\mu})v_2
 &=\bar u_1(\gamma^\mu-\gamma^\mu)v_2=0.
 \label{eq:Jfermionwards}
\end{align}
The first contraction removes longitudinal photon exchange.
The second establishes conservation of the flavor-source current.

For particle $a$ at spatial angle $\theta_a$, the detector projection is
$\cos\theta_a M_f^1+\sin\theta_a M_f^2$.
Its integrated energy weight is
\begin{equation}
 I_{J,f}=\int_0^1\dd x\int_0^{1-x}\dd y\,
 \left\langle\sum_{a=1}^4 k_a^0
 |\cos\theta_a M_f^1+\sin\theta_a M_f^2|^2\right\rangle_{\alpha,\beta}
 =\frac{19}{9}-\frac{\pi^2}{8}.
 \label{eq:IJfermion}
\end{equation}
Appendix~\ref{app:integrals} gives its radial integrand and reduction.
The phase-space factor, free norm, and opposite-structure flux give
\begin{align}
 \frac{2\nsJ}{\nsJ+\nfJ}
 &=\frac{256}{N_f}\frac{8}{64\pi^2}I_{J,f}+O(N_f^{-2}),\nonumber\\
 \gamma_{\mathrm{QED}_3}
 &=\frac1{12}-\frac{8I_{J,f}}{3\pi^2N_f}+O(N_f^{-2})\nonumber\\
 &=\frac1{12}+\frac1{N_f}\left(\frac13-\frac{152}{27\pi^2}\right)
 +O(N_f^{-2}).
 \label{eq:Jqedextraction}
\end{align}
The last line is eq.~\eqref{eq:qed-gamma}.

\subsubsection*{Critical scalar QED$_3$}

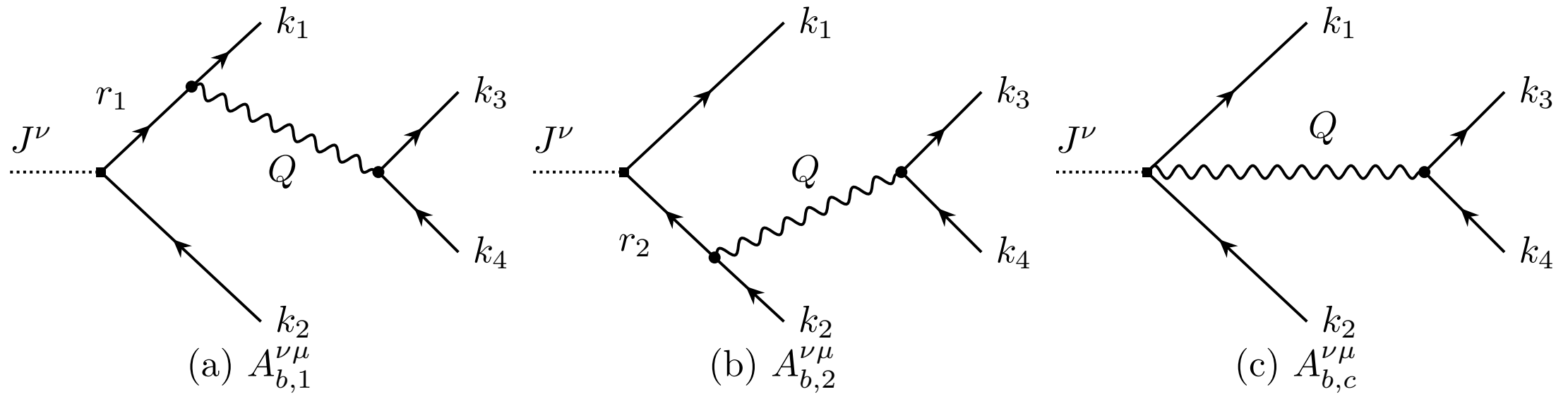
\begin{figure}[H]
\centering
\begin{tikzpicture}[font=\small,
 matter/.style={line width=.75pt,postaction={decorate},
 decoration={markings,mark=at position .55 with {\arrow{Stealth[length=1.7mm,width=1.5mm]}}}},
 photon/.style={line width=.75pt,decorate,
 decoration={snake,amplitude=.6mm,segment length=2.3mm}},
 source/.style={densely dotted,line width=.75pt}]
\begin{scope}[xshift=0cm]
\coordinate (v) at (0,0);
\coordinate (w) at (2.6,0);
\coordinate (k1) at (1.5,1.4);\coordinate (k2) at (1.5,-1.4);
\coordinate (k3) at (3.35,.75);\coordinate (k4) at (3.35,-.75);
\draw[source] (-.85,0)--(v);
\node[above] at (-.65,.05) {$J^\nu$};
\coordinate (e) at (.85,.8);
\draw[matter] (v)--(e);\draw[matter] (e)--(k1);
\draw[matter] (k2)--(v);
\draw[photon] (e)--(w);
\node[above left] at (.4,.45) {$r_1$};
\fill (e) circle (1.6pt);
\draw[matter] (w)--(k3);\draw[matter] (k4)--(w);
\fill (v) +(-.045,-.045) rectangle +(.045,.045);\fill (w) circle (1.6pt);
\node[right] at (k1) {$k_1$};\node[right] at (k2) {$k_2$};
\node[right] at (k3) {$k_3$};\node[right] at (k4) {$k_4$};
\node at (1.7,0) {$Q$};
\node at (1.4,-1.85) {(a) $A_{b,1}^{\nu\mu}$};
\end{scope}
\begin{scope}[xshift=4.9cm]
\coordinate (v) at (0,0);
\coordinate (w) at (2.6,0);
\coordinate (k1) at (1.5,1.4);\coordinate (k2) at (1.5,-1.4);
\coordinate (k3) at (3.35,.75);\coordinate (k4) at (3.35,-.75);
\draw[source] (-.85,0)--(v);
\node[above] at (-.65,.05) {$J^\nu$};
\coordinate (e) at (.85,-.8);
\draw[matter] (v)--(k1);
\draw[matter] (e)--(v);\draw[matter] (k2)--(e);
\draw[photon] (e)--(w);
\node[below left] at (.4,-.45) {$r_2$};
\fill (e) circle (1.6pt);
\draw[matter] (w)--(k3);\draw[matter] (k4)--(w);
\fill (v) +(-.045,-.045) rectangle +(.045,.045);\fill (w) circle (1.6pt);
\node[right] at (k1) {$k_1$};\node[right] at (k2) {$k_2$};
\node[right] at (k3) {$k_3$};\node[right] at (k4) {$k_4$};
\node at (1.7,0) {$Q$};
\node at (1.4,-1.85) {(b) $A_{b,2}^{\nu\mu}$};
\end{scope}
\begin{scope}[xshift=9.8cm]
\coordinate (v) at (0,0);
\coordinate (w) at (2.6,0);
\coordinate (k1) at (1.5,1.4);\coordinate (k2) at (1.5,-1.4);
\coordinate (k3) at (3.35,.75);\coordinate (k4) at (3.35,-.75);
\draw[source] (-.85,0)--(v);
\node[above] at (-.65,.05) {$J^\nu$};
\draw[matter] (v)--(k1);\draw[matter] (k2)--(v);
\draw[photon] (v)--(w);
\draw[matter] (w)--(k3);\draw[matter] (k4)--(w);
\fill (v) +(-.045,-.045) rectangle +(.045,.045);\fill (w) circle (1.6pt);
\node[right] at (k1) {$k_1$};\node[right] at (k2) {$k_2$};
\node[right] at (k3) {$k_3$};\node[right] at (k4) {$k_4$};
\node[above] at (1.65,.12) {$Q$};
\node at (1.4,-1.85) {(c) $A_{b,c}^{\nu\mu}$};
\end{scope}
\end{tikzpicture}
\caption{Photon-emission and seagull diagrams for the scalar flavor
current. The panels correspond to $A_{b,1}^{\nu\mu}$,
$A_{b,2}^{\nu\mu}$, and $A_{b,c}^{\nu\mu}$ in
 eq.~\eqref{eq:Jscalardiagrams}.
Panels (a) and (b) show emission from the scalar and antiscalar lines.
Panel (c) contains the flavor--photon seagull vertex.
The dotted line denotes the current insertion with incoming momentum $q$.
The wavy line denotes the leading large-$N_b$ dressed photon propagator
with momentum $Q=k_3+k_4$.
The first pair has flavor $i$, and the second has flavor $j$.
All external matter momenta are outgoing; the solid-line arrows indicate
charge flow. The internal momenta are $r_1=q-k_2$ and $r_2=q-k_1$.}
\label{fig:flavor-scalar-emission}
\end{figure}

Figure~\ref{fig:flavor-scalar-emission} shows the two scalar emission
diagrams and the flavor--photon seagull diagram.

The scalar emission tensor has three contributions:
\begin{align}
 A_{b,1}^{\nu\mu}&=\frac{(r_1-k_2)^\nu(k_1+r_1)^\mu}{r_1^2},\nonumber\\
 A_{b,2}^{\nu\mu}&=-\frac{(k_1-r_2)^\nu(k_2+r_2)^\mu}{r_2^2},\nonumber\\
 A_{b,c}^{\nu\mu}&=-2\delta^{\nu\mu}.
 \label{eq:Jscalardiagrams}
\end{align}
The last term is the flavor--photon seagull.
It is required by both current Ward identities:
\begin{align}
 Q_\mu(A_{b,1}+A_{b,2}+A_{b,c})^{\nu\mu}
 &=(r_1-k_2)^\nu-(k_1-r_2)^\nu-2Q^\nu=0,\nonumber\\
 q_\nu(A_{b,1}+A_{b,2}+A_{b,c})^{\nu\mu}
 &=(k_1+r_1)^\mu+(k_2+r_2)^\mu-2q^\mu=0.
 \label{eq:Jscalarwards}
\end{align}
These cancellations take place before any angular integration.
They also establish invariance under longitudinal changes of the photon
propagator.

The reduced photon amplitude is
$M_b^\nu=(A_{b,1}+A_{b,2}+A_{b,c})^{\nu\mu}(k_3-k_4)_\mu/y$.
The orthogonal detector projection gives
\begin{equation}
 I_{J,b}=\int_0^1\dd x\int_0^{1-x}\dd y\,
 \left\langle\sum_{a=1}^4 k_a^0
 |-\sin\theta_a M_b^1+\cos\theta_a M_b^2|^2\right\rangle_{\alpha,\beta}
 =\frac{22}{9}-\frac{\pi^2}{8}.
 \label{eq:IJscalar}
\end{equation}
Its contribution to the coefficient of $1/N_b$ in $\gamma$ is
$8I_{J,b}/(3\pi^2)=-1/3+176/(27\pi^2)$.

With the same overall photon normalization, its auxiliary counterpart is
\begin{equation}
 M_{\sigma,b}^\nu=\frac y2\left[
 \frac{(r_1-k_2)^\nu}{r_1^2}
 +\frac{(k_1-r_2)^\nu}{r_2^2}\right].
 \label{eq:Jscalarauxamp}
\end{equation}
The two emissions are shown in figure~\ref{fig:flavor-auxiliary-production}.

\begin{figure}[htbp]
\centering
\begin{tikzpicture}[font=\small,line width=.75pt,
 matter/.style={postaction={decorate},decoration={markings,
  mark=at position .55 with {\arrow{Stealth[length=1.7mm,width=1.5mm]}}}},
 auxiliary/.style={densely dashed},
 source/.style={densely dotted}]
\begin{scope}
 \coordinate (s) at (0,0);
 \coordinate (e) at (1,.9);
 \coordinate (v) at (2.5,0);
 \coordinate (k1) at (4,1.65);
 \coordinate (k2) at (4,-1.65);
 \coordinate (k3) at (4,.55);
 \coordinate (k4) at (4,-.55);
 \draw[source] (-1,0)--(s);
 \node[above] at (-.55,.08) {$J^\nu(q)$};
 \draw[matter] (s)--(e);
 \draw[matter] (e)--(k1);
 \draw[matter] (k2)--(s);
 \draw[auxiliary] (e)--node[above right,pos=.48,inner sep=2pt] {$\sigma$} (v);
 \draw[matter] (v)--(k3);
 \draw[matter] (k4)--(v);
 \fill (s) +(-.045,-.045) rectangle +(.045,.045);
 \fill (e) circle (1.7pt);
 \fill (v) circle (1.7pt);
 \node[right,inner sep=3pt] at (k1) {$k_1$};
 \node[right,inner sep=3pt] at (k2) {$k_2$};
 \node[right,inner sep=3pt] at (k3) {$k_3$};
 \node[right,inner sep=3pt] at (k4) {$k_4$};
 \node at (1.75,-2.12) {$(a)$ Emission from the particle};
\end{scope}
\begin{scope}[xshift=6.4cm]
 \coordinate (s) at (0,0);
 \coordinate (e) at (1,-.9);
 \coordinate (v) at (2.5,0);
 \coordinate (k1) at (4,1.65);
 \coordinate (k2) at (4,-1.65);
 \coordinate (k3) at (4,.55);
 \coordinate (k4) at (4,-.55);
 \draw[source] (-1,0)--(s);
 \node[above] at (-.55,.08) {$J^\nu(q)$};
 \draw[matter] (s)--(k1);
 \draw[matter] (k2)--(e);
 \draw[matter] (e)--(s);
 \draw[auxiliary] (e)--node[below right,pos=.48,inner sep=2pt] {$\sigma$} (v);
 \draw[matter] (v)--(k3);
 \draw[matter] (k4)--(v);
 \fill (s) +(-.045,-.045) rectangle +(.045,.045);
 \fill (e) circle (1.7pt);
 \fill (v) circle (1.7pt);
 \node[right,inner sep=3pt] at (k1) {$k_1$};
 \node[right,inner sep=3pt] at (k2) {$k_2$};
 \node[right,inner sep=3pt] at (k3) {$k_3$};
 \node[right,inner sep=3pt] at (k4) {$k_4$};
 \node at (1.75,-2.12) {$(b)$ Emission from the antiparticle};
\end{scope}
\end{tikzpicture}
\caption{Auxiliary-scalar emission in flavor-current production.
The dotted source ends at the current insertion (square); solid arrows
show matter charge flow, and the dashed line is the dressed $\sigma$
propagator. Both emissions are summed before squaring, with the vertices
and relative signs in eqs.~\eqref{eq:Jscalarauxamp} and~\eqref{eq:JGNamp}
for critical scalar QED$_3$ and QED$_3$--GNY, respectively.
All matter momenta are outgoing.}
\label{fig:flavor-auxiliary-production}
\end{figure}

The source contraction vanishes since the two fractions contract to
$1$ and $-1$. The sign of the complete auxiliary amplitude follows
from the contour convention in eq.~\eqref{eq:scalaraction}; its common
phase cancels from the squared flux.
The auxiliary-scalar exchange gives an additional coefficient
$1/12-20/(27\pi^2)$.
The photon current of the second pair changes sign under particle--antiparticle
exchange, while the scalar bilinear does not.
Their interference is therefore odd under this exchange.
The phase-space measure and detector weight are even, so the integrated
interference vanishes. Substitution into eq.~\eqref{eq:collider} gives
\begin{align}
 \gamma_{\mathrm{sQED}_3}
 =-\frac1{12}+\frac1{N_b}\left(-\frac14+\frac{52}{9\pi^2}\right)
 +O(N_b^{-2}).
 \label{eq:Jscalarextraction}
\end{align}

\subsubsection*{QED$_3$--GNY}

The auxiliary-emission diagrams in
figure~\ref{fig:flavor-auxiliary-production} also apply to the GNY scalar.
The Gross--Neveu exchange adds the reduced amplitude
\begin{equation}
 M_\sigma^\nu=\frac1{2y}
 \bar u_1\left[
 \frac{\gamma^a\widehat r_{1a}}{r_1^2}\gamma^\nu
 -\gamma^\nu\frac{\gamma^a\widehat r_{2a}}{r_2^2}\right]v_2
 (\bar u_3v_4).
 \label{eq:JGNamp}
\end{equation}
The factor $1/2$ follows from the ratio of the auxiliary and photon
propagators in eqs.~\eqref{eq:photonprop} and~\eqref{eq:sigmaprop}.
The full current amplitude is $M_f+M_\sigma$ in one spinor convention.
Its photon--scalar interference cancels under charge conjugation of one
pair. The remaining scalar square supplies
$-1/12+14/(27\pi^2)$ to the coefficient of $1/N_f$.
Consequently
\begin{align}
 \gamma_{\mathrm{QED}_3\text{-}\mathrm{GNY}}
 =\frac1{12}+\frac1{N_f}\left(\frac14-\frac{46}{9\pi^2}\right)
 +O(N_f^{-2}).
 \label{eq:Jgnyextraction}
\end{align}
The combined-amplitude computation checks the interference tensor for
all current polarizations. It also checks Coulomb and covariant exchange
propagators before phase-space integration.

\subsection{Topological-current calculation and results}
\label{sec:topological}
\label{sec:topflat}

The topological current provides an additional global-current observable in
each of the three theories. Its leading collider data were computed in
ref.~\cite{ChowdhuryTransport}; here we determine their first interaction
corrections.
The topological current measures magnetic flux.
The corresponding source insertion and coupling are
\begin{equation}
 j_{\mathrm{top}}^\mu=\frac{\ii}{4\pi}
 \epsilon^{\mu\nu\rho}F_{\nu\rho},\qquad
 S_{\mathrm{BF}}=\frac{\ii}{2\pi}\int B_{\mathrm{top}}\wedge\dd a,
 \label{eq:topcurrent}
\end{equation}
where $F=\dd a$, and $\epsilon^{\mu\nu\rho}$ is the metric volume tensor.
The factor $\ii$ implements the Euclidean source convention.

The BF source couples directly to the photon, so its response probes
the topological-current correlator. The flavor source instead couples
to the matter fields through the traceless generator $Q$.
The topological current can become an ordinary matter current in a
particle--vortex dual description \cite{SeibergDualityWeb,WangDQCP}.
We first extract its collider parameter from flat-space energy flux.
The independent inverse-photon-kernel calculation on the squashed
sphere is given in section~\ref{sec:topological-sphere}.

Consider the BF insertion in eq.~\eqref{eq:topcurrent}. At $q=(1,0,0)$, its spatial curl rotates the current
polarization by $\pi/2$. This rotation exchanges the leading scalar and
fermion collider endpoints. The common external photon factor cancels
between the energy flux and the state norm.
We write $s=+1$ for fermions and $s=-1$ for scalars, with $N=N_f$
or $N_b$, respectively.

After factoring out the common external photon factor, the detector projections are therefore
the same projections used for the flavor current in
section~\ref{sec:flavor-projection}:
\begin{equation}
 \mathcal P_{a,f}M=\cos\theta_a M^1+\sin\theta_a M^2,\qquad
 \mathcal P_{a,b}M=-\sin\theta_a M^1+\cos\theta_a M^2.
 \label{eq:topprojections}
\end{equation}
Here $\theta_a$ is the spatial angle of particle $a$.
These projections select a direction where the leading pair-production
amplitude vanishes, eliminating its interference with the first virtual
correction. At relative order $N^{-1}$, the selected flux is therefore
determined by inclusive cuts through four fundamental matter lines in
the resolved description of section~\ref{sec:energy-flux-projections}.

The reduced matter vertex couples equally to every flavor.
Its four-particle amplitude includes coherent attachments to both pairs.
These attachments separate into two charge-conjugation sectors.
The sectors have equal integrated squares, while their interference is
odd under charge conjugation of one pair and integrates to zero.
For one sector with reduced amplitude $\mathcal M_r$, where $r=f,b$,
write its detector integral as
\begin{equation}
 I_{\mathrm{top}}=\int_0^1\dd x\int_0^{1-x}\dd y\,
 \left\langle\sum_{a=1}^4 k_a^0
 |\mathcal P_{a,r}\mathcal M_r|^2\right\rangle_{\alpha,\beta}.
 \label{eq:topcutintegral}
\end{equation}
The variables and angular average are those of
eqs.~\eqref{eq:pairkinematics} and~\eqref{eq:phasespace}.
After the common external photon factor is removed, the production
amplitude has an additional factor $16/N$. There are $N(N-1)/2$ unordered
pairs of distinct flavors. The leading two-particle normalization after
removing the common external factors is $N/8$. This reduced normalization
differs from $C_J^{\mathrm{top}}$, which is of order $1/N$.
Identical-flavor states contribute one order later. The selected flux, divided by its angular average, is
therefore
\begin{align}
 1+12s\gamma_{\mathrm{top}}
 &=\frac{[N(N-1)/2](16/N)^2\,2}{N/8}
   \frac{I_{\mathrm{top}}}{64\pi^2}+O(N^{-2})\nonumber\\
 &=\frac{32I_{\mathrm{top}}}{\pi^2N}+O(N^{-2}),\nonumber\\
 \gamma_{\mathrm{top}}
 &=-\frac{s}{12}+\frac{8sI_{\mathrm{top}}}{3\pi^2N}+O(N^{-2}).
 \label{eq:topflatextraction}
\end{align}
The factor two counts the charge-conjugation sectors.

\subsubsection*{QED$_3$}

In QED$_3$, one sector is the photon-emission amplitude
$M_f$ of eq.~\eqref{eq:Jfermionamp}.
There is no auxiliary exchange, and charge conjugation forbids an
induced three-photon vertex. Equation~\eqref{eq:IJfermion} then gives
\begin{equation}
 I_{\mathrm{top},\mathrm{QED}_3}=I_{J,f}
 =\frac{19}{9}-\frac{\pi^2}{8}.
 \label{eq:topqedcut}
\end{equation}
Its substitution into eq.~\eqref{eq:topflatextraction} determines the
QED$_3$ result.

\subsubsection*{Critical scalar QED$_3$ and QED$_3$--GNY}

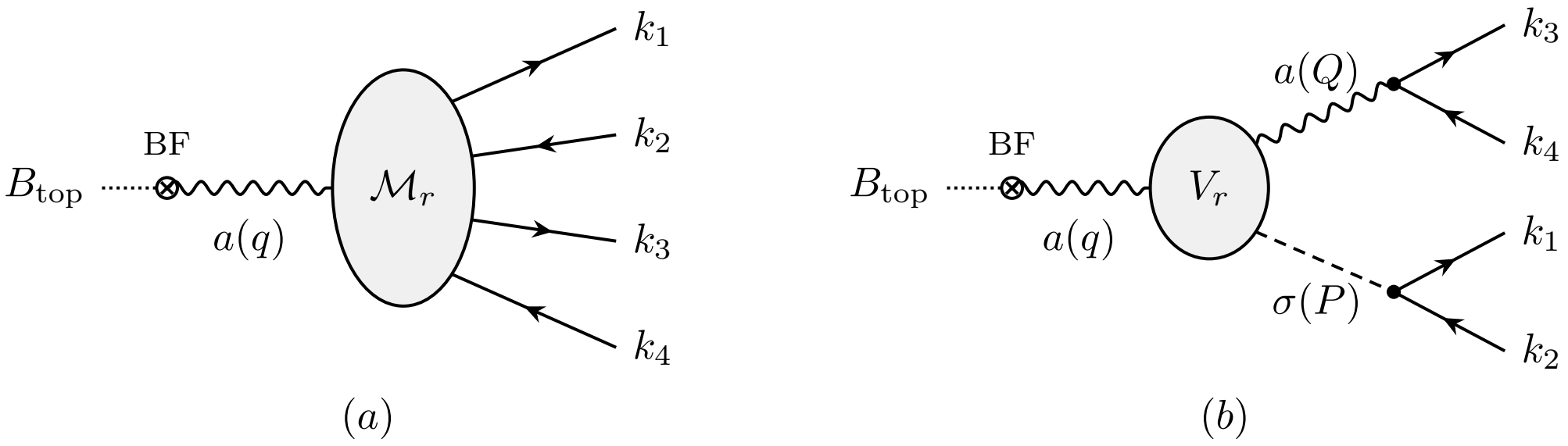
\begin{figure}[htbp]
\centering
\begin{tikzpicture}[font=\small,line width=.75pt,
 topmatter/.style={postaction={decorate},decoration={markings,
   mark=at position .56 with {\arrow{Stealth[length=1.7mm,width=1.5mm]}}}},
 topphoton/.style={decorate,decoration={snake,amplitude=.55mm,segment length=2.2mm}},
 topsigma/.style={dash pattern=on 3pt off 2pt},
 topsource/.style={densely dotted},
 topblob/.style={inner sep=0pt},
 topvertex/.style={inner sep=0pt}]
\begin{scope}
 \coordinate (bf) at (-.45,0);
 \node[left] at (-1.0,0) {$B_{\mathrm{top}}$};
 \draw[topsource] (-1.0,0)--(bf);
 \draw[fill=white] (bf) circle (2.5pt);
 \draw (-.50,-.05)--(-.40,.05) (-.50,.05)--(-.40,-.05);
 \node[above] at (-.45,.14) {\scriptsize BF};
 \draw[topphoton] (bf)--(.98,0);
 \node[below] at (.25,-.12) {$a(q)$};
 \draw[fill=black!6] (1.55,0) ellipse (.6cm and 1cm);
 \node[topblob] at (1.55,0) {$\mathcal M_r$};
 \draw[topmatter] (1.96,.73)--(3.35,1.35);
 \draw[topmatter] (3.35,.45)--(2.13,.27);
 \draw[topmatter] (2.13,-.27)--(3.35,-.45);
 \draw[topmatter] (3.35,-1.35)--(1.96,-.73);
 \node[right] at (3.35,1.35) {$k_1$};
 \node[right] at (3.35,.45) {$k_2$};
 \node[right] at (3.35,-.45) {$k_3$};
 \node[right] at (3.35,-1.35) {$k_4$};
 \node at (1.25,-1.95) {$(a)$};
\end{scope}
\begin{scope}[xshift=7.15cm]
 \coordinate (bf) at (-.45,0);
 \node[left] at (-1.0,0) {$B_{\mathrm{top}}$};
 \draw[topsource] (-1.0,0)--(bf);
 \draw[fill=white] (bf) circle (2.5pt);
 \draw (-.50,-.05)--(-.40,.05) (-.50,.05)--(-.40,-.05);
 \node[above] at (-.45,.14) {\scriptsize BF};
 \draw[topphoton] (bf)--(.72,0);
 \node[below] at (.12,-.12) {$a(q)$};
 \draw[fill=black!6] (1.22,0) ellipse (.5cm and .6cm);
 \node[topvertex] at (1.22,0) {$V_r$};
 \draw[topphoton] (1.61,.37)--(2.78,.88);
 \node[above] at (2.13,.66) {$a(Q)$};
 \draw[topsigma] (1.61,-.37)--(2.78,-.88);
 \node[below] at (2.13,-.66) {$\sigma(P)$};
 \fill (2.78,.88) circle (1.7pt);
 \fill (2.78,-.88) circle (1.7pt);
 \draw[topmatter] (2.78,.88)--(3.72,1.38);
 \draw[topmatter] (3.72,.38)--(2.78,.88);
 \draw[topmatter] (2.78,-.88)--(3.72,-.38);
 \draw[topmatter] (3.72,-1.38)--(2.78,-.88);
 \node[right] at (3.72,1.38) {$k_3$};
 \node[right] at (3.72,.38) {$k_4$};
 \node[right] at (3.72,-.38) {$k_1$};
 \node[right] at (3.72,-1.38) {$k_2$};
 \node at (1.35,-1.95) {$(b)$};
\end{scope}
\end{tikzpicture}
\caption{Topological-current production through the BF source.
(a) A BF insertion (crossed circle) couples the source to a dressed
photon; the shaded block is the coherent amplitude in
 eq.~\eqref{eq:topcoherentamp}. The other charge-conjugation sector
follows by pair interchange.
(b) The induced photon--auxiliary contribution
$M_{\mathrm{ind},r}(Q;P)$ in critical scalar QED$_3$ and QED$_3$--GNY.
The vertex $V_r(Q,P)$ includes both loop orientations and the scalar
seagull where applicable. Wavy and dashed lines are dressed photons
and auxiliary scalars, with $Q=k_3+k_4$ and $P=k_1+k_2$.
Solid arrows indicate charge flow; all matter momenta are outgoing.}
\label{fig:topological-production}
\end{figure}

The auxiliary field adds a coherent emission amplitude.
It also permits an induced vertex with one source current, one photon,
and one auxiliary field. In QED$_3$--GNY, the auxiliary scalar couples
to the flavor-singlet fermion bilinear, so this vertex survives the
flavor sum.
For both theories, one charge-conjugation sector is
\begin{equation}
 \mathcal M_r^\nu=M_r^\nu(12;34)
 -M_{\sigma,r}^\nu(34;12)+M_{\mathrm{ind},r}^\nu(Q;P).
 \label{eq:topcoherentamp}
\end{equation}
The notation $(12;34)$ places the source on the first pair.
The auxiliary amplitudes are given in eqs.~\eqref{eq:Jscalarauxamp}
and~\eqref{eq:JGNamp}, with $M_{\sigma,f}=M_\sigma$.
Their relative minus sign in eq.~\eqref{eq:topcoherentamp} follows
from the auxiliary propagator and vertices in the common photon
normalization. The second sector follows by exchanging the two pairs.
Figure~\ref{fig:topological-production} shows the BF coupling and the
induced photon--auxiliary production channel.

To evaluate the induced term, let $V_r^{\nu\mu}(P,Q)$ denote the
reduced vertex with outgoing photon momentum $P$ and auxiliary momentum
$Q=q-P$. All loop contractions below use the complex Euclidean
momenta of section~\ref{sec:flavor-projection}, and
$\int_k=\int\dd^3k/(2\pi)^3$.
For the scalar vertex, the two triangle orientations and the seagull give
\begin{equation}
 V_b^{\nu\mu}(P,Q)=2\int_k
 \frac{(2k-\widehat q)^\nu(2k-\widehat P)^\mu}
 {k^2(k-\widehat P)^2(k-\widehat q)^2}
 -\frac{\delta^{\nu\mu}}{4y}.
 \label{eq:topscalartriangle}
\end{equation}
The seagull is required for transversality.

For the fermion vertex, the relative sign is fixed by the same Fourier
convention as the open emission amplitudes.
With $S_F(x-z)=\int_k e^{-\ii k\cdot(x-z)}S_F(k)$,
the outgoing factors are $e^{\ii P\cdot z}$ and $e^{\ii Q\cdot y}$,
and the source factor is $e^{-\ii q\cdot x}$.
For loop momenta $k,\ell,m$, their combined phase is
\begin{align}
(-k+m-q)\cdot x+(k-\ell+P)\cdot z
 +(\ell-m+Q)\cdot y.
\end{align}
The position integrals impose $\ell=k+P$ and $m=k+q$.
Thus the closed-loop sign and the two orientations give
\begin{align}
 V_f^{\nu\mu}(P,Q)
 &=-2\int_k\frac{\operatorname{tr}[\gamma^\nu
 (\gamma^a k_a)\gamma^\mu
 (\gamma^b(k+\widehat P)_b)(\gamma^c(k+\widehat q)_c)]}
 {k^2(k+\widehat P)^2(k+\widehat q)^2}\nonumber\\
 &=2\int_k\frac{\operatorname{tr}[\gamma^\nu
 (\gamma^a k_a)\gamma^\mu
 (\gamma^b(k-\widehat P)_b)(\gamma^c(k-\widehat q)_c)]}
 {k^2(k-\widehat P)^2(k-\widehat q)^2}.
 \label{eq:topfermiontriangle}
\end{align}
The second line follows from $k\mapsto-k$.
Each of the three fermion numerators supplies one minus sign.

Use the pair frame of eq.~\eqref{eq:fourmomenta}, in which
$\widehat P=(E_P,-\ii K,0)$ and
$E_P=(1+x^2-y^2)/2$. Evaluating the loop integrals gives
\begin{align}
 V_b^{\nu\mu}(P,Q)
 &=\frac{1+x-y}{4(1+x+y)}
 \begin{pmatrix}
 0&0&0\\
 -\ii K/(xy)&-E_P/(xy)&0\\
 0&0&-1/y
 \end{pmatrix},\nonumber\\
 V_f^{\nu\mu}(P,Q)
 &=\frac{1+x-y}{4(1+x+y)}
 \begin{pmatrix}
 0&0&0\\
 0&0&-\ii\\
 -K/x&\ii E_P/x&0
 \end{pmatrix}.
 \label{eq:toptrianglevalues}
\end{align}
Both matrices obey $\widehat q_\nu V_r^{\nu\mu}=0$ and
$V_r^{\nu\mu}\widehat P_\mu=0$.
These identities remove longitudinal photon exchange.

Attaching the photon and auxiliary propagators gives
\begin{align}
 M_{\mathrm{ind},b}^\nu(Q;P)
 &=-\frac{8x}{y}V_b^{\nu\mu}(Q,P)J_{34,\mu},\nonumber\\
 M_{\mathrm{ind},f}^\nu(Q;P)
 &=-\frac{8}{xy}V_f^{\nu\mu}(Q,P)J_{34,\mu}
 (\bar u_1v_2).
 \label{eq:topinducedamp}
\end{align}
Here $J_{34}^\mu=(\widehat k_3-\widehat k_4)^\mu$ for scalars
and $J_{34}^\mu=\bar u_3\gamma^\mu v_4$ for fermions.
All terms are continued to the same Euclidean tensor basis before
forming their interference. The matrices $V_r(Q,P)$ follow from
eq.~\eqref{eq:toptrianglevalues} by $x\leftrightarrow y$ and
$K\mapsto-K$.

\subsubsection*{Angular and radial integrals}

Substituting eq.~\eqref{eq:topcoherentamp} into
eq.~\eqref{eq:topcutintegral} separates the photon and auxiliary squares
from the induced square and interference terms.
Let $\Delta I_r$ be the integral of the latter terms, so that its
integrand is
\begin{align}
 &|\mathcal P_{a,r}\mathcal M_r|^2
 -|\mathcal P_{a,r}M_r(12;34)|^2
 -|\mathcal P_{a,r}M_{\sigma,r}(34;12)|^2\nonumber\\
 &=|\mathcal P_{a,r}M_{\mathrm{ind},r}|^2
 -2\operatorname{Re}\big[
 \mathcal P_{a,r}M_r(\mathcal P_{a,r}M_{\sigma,r})^*\big]\nonumber\\
 &\quad+2\operatorname{Re}\big[
 \mathcal P_{a,r}M_r(\mathcal P_{a,r}M_{\mathrm{ind},r})^*\big]\nonumber\\
 &\quad-2\operatorname{Re}\big[\mathcal P_{a,r}M_{\sigma,r}
  (\mathcal P_{a,r}M_{\mathrm{ind},r})^*\big].
 \label{eq:topinterferences}
\end{align}
The pair arguments on the right-hand side are those on the first line.
The separate squares already computed in section~\ref{sec:current}
then give
\begin{align}
 I_{\mathrm{top},\mathrm{sQED}_3}
 &=\frac{13}{6}-\frac{3\pi^2}{32}+\Delta I_b,\nonumber\\
 I_{\mathrm{top},\mathrm{QED}_3\text{-}\mathrm{GNY}}
 &=\frac{23}{12}-\frac{3\pi^2}{32}+\Delta I_f.
 \label{eq:topcutbaselines}
\end{align}

The angular averages use the same rational trigonometric integrals as
appendix~\ref{app:integrals}.
For example, $\langle(a-b\cos\alpha)^{-1}\rangle
=(a^2-b^2)^{-1/2}$ for $a>|b|$; differentiation in $a$ gives higher
denominator powers. After these averages, set $u=x+y$ and $v=x-y$.
The measure is $\dd x\,\dd y=\dd u\,\dd v/2$, with
$0\leq u\leq1$ and $-u\leq v\leq u$.
Integration over $v$ yields
\begin{equation}
 \Delta I_r=\int_0^1\dd u\left[
 R_r(u)+C_r(u)\log\frac{1-u}{1+u}\right],
 \label{eq:topradialintegral}
\end{equation}
where the rational functions are
\begin{align}
 R_b(u)&=u^2-\frac{22u}{3}+\frac{65}{3}
 -\frac{28}{u+1}-\frac{68}{3(u+1)^2}+\frac{32}{(u+1)^3},\nonumber\\
 C_b(u)&=\frac{u^3}{2}-14+\frac{48}{u+1}
 -\frac{48}{(u+1)^2}+\frac{3}{2u},\nonumber\\
 R_f(u)&=-u^2-\frac{10u}{3}+\frac{65}{3}
 -\frac{32}{u+1}-\frac{56}{3(u+1)^2}+\frac{32}{(u+1)^3},\nonumber\\
 C_f(u)&=-\frac{u^3}{2}+3u-16+\frac{48}{u+1}
 -\frac{48}{(u+1)^2}+\frac{3}{2u}.
 \label{eq:topradialfunctions}
\end{align}
The apparent pole at $u=0$ multiplies a logarithm that vanishes linearly.
The combined integrals are finite at both endpoints.

The logarithmic moments needed in eq.~\eqref{eq:topradialintegral} are
\begin{equation}
 \int_0^1\dd u\,
 \begin{pmatrix}u^3\\u\\1\\1/u\\1/(1+u)\\1/(1+u)^2\end{pmatrix}
 \log\frac{1-u}{1+u}
 =\begin{pmatrix}-2/3\\-1\\-2\log2\\-\pi^2/4\\-\pi^2/12\\-1/2\end{pmatrix}.
 \label{eq:toplogmoments}
\end{equation}
These moments follow by integration by parts or by integrating the
convergent series
$\log[(1-u)/(1+u)]=-2\sum_{n=0}^{\infty}u^{2n+1}/(2n+1)$.
Substituting the moments and the elementary rational integrals gives
\begin{align}
 \Delta I_b
 &=(19-28\log2)
 +\left(\frac{71}{3}+28\log2-\frac{35\pi^2}{8}\right)
 =\frac{128}{3}-\frac{35\pi^2}{8},\nonumber\\
 \Delta I_f
 &=\left(\frac{67}{3}-32\log2\right)
 +\left(\frac{64}{3}+32\log2-\frac{35\pi^2}{8}\right)
 =\frac{131}{3}-\frac{35\pi^2}{8}.
 \label{eq:topmixcutvalues}
\end{align}
The first parentheses contain the rational integrals.
The second parentheses contain the logarithmic integrals, whose
$\log2$ terms cancel those in the first parentheses.
Equation~\eqref{eq:topcutbaselines} consequently becomes
\begin{align}
 I_{\mathrm{top},\mathrm{sQED}_3}
 &=\frac{13}{6}-\frac{3\pi^2}{32}
 +\frac{128}{3}-\frac{35\pi^2}{8}
 =\frac{269}{6}-\frac{143\pi^2}{32},\nonumber\\
 I_{\mathrm{top},\mathrm{QED}_3\text{-}\mathrm{GNY}}
 &=\frac{23}{12}-\frac{3\pi^2}{32}
 +\frac{131}{3}-\frac{35\pi^2}{8}
 =\frac{547}{12}-\frac{143\pi^2}{32}.
 \label{eq:topfinalcuts}
\end{align}
Together with eq.~\eqref{eq:topqedcut}, these integrals determine all
three coefficients through eq.~\eqref{eq:topflatextraction}.
This is an independent energy-flux projection of
$\langle Tj_{\mathrm{top}}j_{\mathrm{top}}\rangle$.
It gives the same collider parameters as the sphere derivatives.

Substituting the cut integrals into eq.~\eqref{eq:topflatextraction} gives the
following results, independently reproduced by the sphere extraction
in eq.~\eqref{eq:topgammaextraction}:
\begin{align}
 \gamma_{\mathrm{top},\mathrm{QED}_3}
 &=-\frac1{12}+\frac1{N_f}\left(-\frac13+\frac{152}{27\pi^2}\right)
 +O(N_f^{-2}),\nonumber\\
 \gamma_{\mathrm{top},\mathrm{sQED}_3}
 &=\frac1{12}+\frac1{N_b}\left(\frac{143}{12}-\frac{1076}{9\pi^2}\right)
 +O(N_b^{-2}),\nonumber\\
 \gamma_{\mathrm{top},\mathrm{QED}_3\text{-}\mathrm{GNY}}
 &=-\frac1{12}+\frac1{N_f}\left(-\frac{143}{12}+\frac{1094}{9\pi^2}\right)
 +O(N_f^{-2}).
 \label{eq:topresults}
\end{align}
All three collider corrections point inward from their leading endpoints.
In QED$_3$, $\gamma_{\mathrm{top}}=-\gamma+O(N_f^{-2})$ at this order.
The corresponding sign reversal of the flavor result does not hold
for the subleading terms in scalar QED$_3$ or QED$_3$--GNY:
coherent emission and the induced photon--auxiliary vertex contribute
in the singlet gauge channel. As for the flavor results, these
expansions apply at large flavor number.

The known two-point normalization $C_J^{\mathrm{top}}$
\cite{GiombiTarnopolskyKlebanov,GuoLi} and $\gamma_{\mathrm{top}}$
determine the full separated-point parity-even three-point function.
Using the reference structures of eq.~\eqref{eq:Jstructures}, write
\begin{equation}
 \langle Tj_{\mathrm{top}}j_{\mathrm{top}}\rangle
 =n_{\mathrm s}^{\mathrm{top}}\TJJ_{\mathrm s}
 +n_{\mathrm f}^{\mathrm{top}}\TJJ_{\mathrm f}.
 \label{eq:topstructures}
\end{equation}
The sum of its coefficients is $C_J^{\mathrm{top}}/C_J^{(0)}$.
Solving this constraint together with eq.~\eqref{eq:collider} gives
\begin{equation}
 n_{\mathrm s}^{\mathrm{top}}
 =\frac{C_J^{\mathrm{top}}}{2C_J^{(0)}}(1-12\gamma_{\mathrm{top}}),\qquad
 n_{\mathrm f}^{\mathrm{top}}
 =\frac{C_J^{\mathrm{top}}}{2C_J^{(0)}}(1+12\gamma_{\mathrm{top}}).
 \label{eq:topstructureextraction}
\end{equation}
The resulting coefficients to relative order $N^{-1}$ are summarized in
table~\ref{tab:topJstructures} in section~\ref{sec:summary-results}.

\section{The stress-tensor three-point function}\label{sec:stress}

\subsection{Conformal structures and stress-tensor production amplitudes}

The parity-even stress-tensor structures are fixed kinematically by conformal
symmetry \cite{OsbornPetkou,ErdmengerOsborn}. Their independent coefficient
also enters the stress-tensor contribution to shear spectral sum rules
\cite{ChowdhuryDavidPrakash}; determining it requires more information
than the two-point normalization. The induced gauge vertices below
supply that information at the same order as matter emission.

In transport, $t_4$ enters the stress-tensor contribution to the high-frequency
shear response extracted from the retarded $T_{xy}$ two-point function 
\cite{ChowdhuryDavidPrakash,LucasSierensWitczak}. In a bootstrap system
with external stress tensors, it instead parametrizes the independent
$\TTT$ coupling entering stress-tensor exchange. Recent mixed stress-tensor and current studies demonstrate that
this conformal collider parameter can be constrained directly
\cite{ChangEtAl,DommesEtAl}.

\begin{figure}[htbp]
\centering
\begin{tikzpicture}[font=\small,line width=.75pt,
 matter/.style={postaction={decorate},decoration={markings,
 mark=at position .55 with {\arrow{Stealth[length=1.7mm,width=1.5mm]}}}},
 photon/.style={decorate,decoration={snake,amplitude=.65mm,segment length=2.3mm}},
 source/.style={densely dotted},
 induced/.style={draw,circle,fill=black!6,minimum size=8mm,inner sep=0pt}]
\begin{scope}
\coordinate (v) at (0,0);\coordinate (e) at (.65,.62);
\coordinate (w) at (2.45,0);
\draw[source] (-.8,0)--(v);\node[above] at (-.55,.06) {$T(e)$};
\draw[matter] (v)--(e);\draw[matter] (e)--(1.12,1.18);
\draw[matter] (1.12,-1.18)--(v);
\draw[photon] (e)--(w);\node[above] at (1.65,.45) {$Q$};
\draw[matter] (w)--(3.18,.7);\draw[matter] (3.18,-.7)--(w);
\foreach \v in {e,w} \fill (\v) circle (1.6pt);
\fill (v) +(-.045,-.045) rectangle +(.045,.045);
\node[right] at (1.12,1.18) {$k_1$};\node[right] at (1.12,-1.18) {$k_2$};
\node[right] at (3.18,.7) {$k_3$};\node[right] at (3.18,-.7) {$k_4$};
\node at (1.35,-1.75) {(a) particle emission};
\end{scope}
\begin{scope}[xshift=6.2cm]
\coordinate (v) at (0,0);\coordinate (e) at (.65,-.62);
\coordinate (w) at (2.45,0);
\draw[source] (-.8,0)--(v);\node[above] at (-.55,.06) {$T(e)$};
\draw[matter] (v)--(1.12,1.18);
\draw[matter] (1.12,-1.18)--(e);\draw[matter] (e)--(v);
\draw[photon] (e)--(w);\node[below] at (1.65,-.45) {$Q$};
\draw[matter] (w)--(3.18,.7);\draw[matter] (3.18,-.7)--(w);
\foreach \v in {e,w} \fill (\v) circle (1.6pt);
\fill (v) +(-.045,-.045) rectangle +(.045,.045);
\node[right] at (1.12,1.18) {$k_1$};\node[right] at (1.12,-1.18) {$k_2$};
\node[right] at (3.18,.7) {$k_3$};\node[right] at (3.18,-.7) {$k_4$};
\node at (1.35,-1.75) {(b) antiparticle emission};
\end{scope}
\begin{scope}[yshift=-3.9cm]
\coordinate (v) at (0,0);\coordinate (w) at (2.45,0);
\draw[source] (-.8,0)--(v);\node[above] at (-.55,.06) {$T(e)$};
\draw[matter] (v)--(1.12,1.18);\draw[matter] (1.12,-1.18)--(v);
\draw[photon] (v)--(w);\node[above] at (1.65,.12) {$Q$};
\draw[matter] (w)--(3.18,.7);\draw[matter] (3.18,-.7)--(w);
\fill (v) +(-.07,-.07) rectangle +(.07,.07);
\fill (w) circle (1.6pt);
\node[right] at (1.12,1.18) {$k_1$};\node[right] at (1.12,-1.18) {$k_2$};
\node[right] at (3.18,.7) {$k_3$};\node[right] at (3.18,-.7) {$k_4$};
\node at (1.35,-1.75) {(c) mixed metric--photon vertex};
\end{scope}
\begin{scope}[xshift=6.2cm,yshift=-3.9cm]
\node[induced] (G) at (0,0) {$\Gamma_e$};
\coordinate (u) at (1.5,.7);\coordinate (v) at (1.5,-.7);
\draw[source] (-1.2,0)--(G.west);\node[above] at (-.95,.06) {$T(e)$};
\draw[photon] (G)--(u);\draw[photon] (G)--(v);
\node[above] at (.85,.48) {$P$};\node[below] at (.85,-.48) {$Q$};
\draw[matter] (u)--(2.65,1.25);\draw[matter] (2.65,.25)--(u);
\draw[matter] (v)--(2.65,-.25);\draw[matter] (2.65,-1.25)--(v);
\fill (u) circle (1.6pt);\fill (v) circle (1.6pt);
\node[right] at (2.65,1.25) {$k_1$};\node[right] at (2.65,.25) {$k_2$};
\node[right] at (2.65,-.25) {$k_3$};\node[right] at (2.65,-1.25) {$k_4$};
\node at (1.35,-1.75) {(d) induced two-photon vertex};
\end{scope}
\end{tikzpicture}
\caption{Four-particle production by a stress-tensor source.
Panels (a)--(c) are the ordinary and mixed-vertex terms in
 eq.~\eqref{eq:Temission}; their pair-interchanged counterparts must also
be included. Panel (d) shows the full induced vertex attached to both
matter pairs, the last term of eq.~\eqref{eq:fullstressamp}.
The dotted line is the metric source, wavy lines are dressed photons,
and solid arrows indicate matter charge flow, with all $k_a$ outgoing.
Here $P=k_1+k_2$ and $Q=k_3+k_4$.
All four panels contribute at the same large-flavor order.
The shaded vertex includes the triangle and contact bubbles in
figure~\ref{fig:induced}. The same topologies apply to scalar matter
with the appropriate improved vertices.}
\label{fig:emission}
\end{figure}
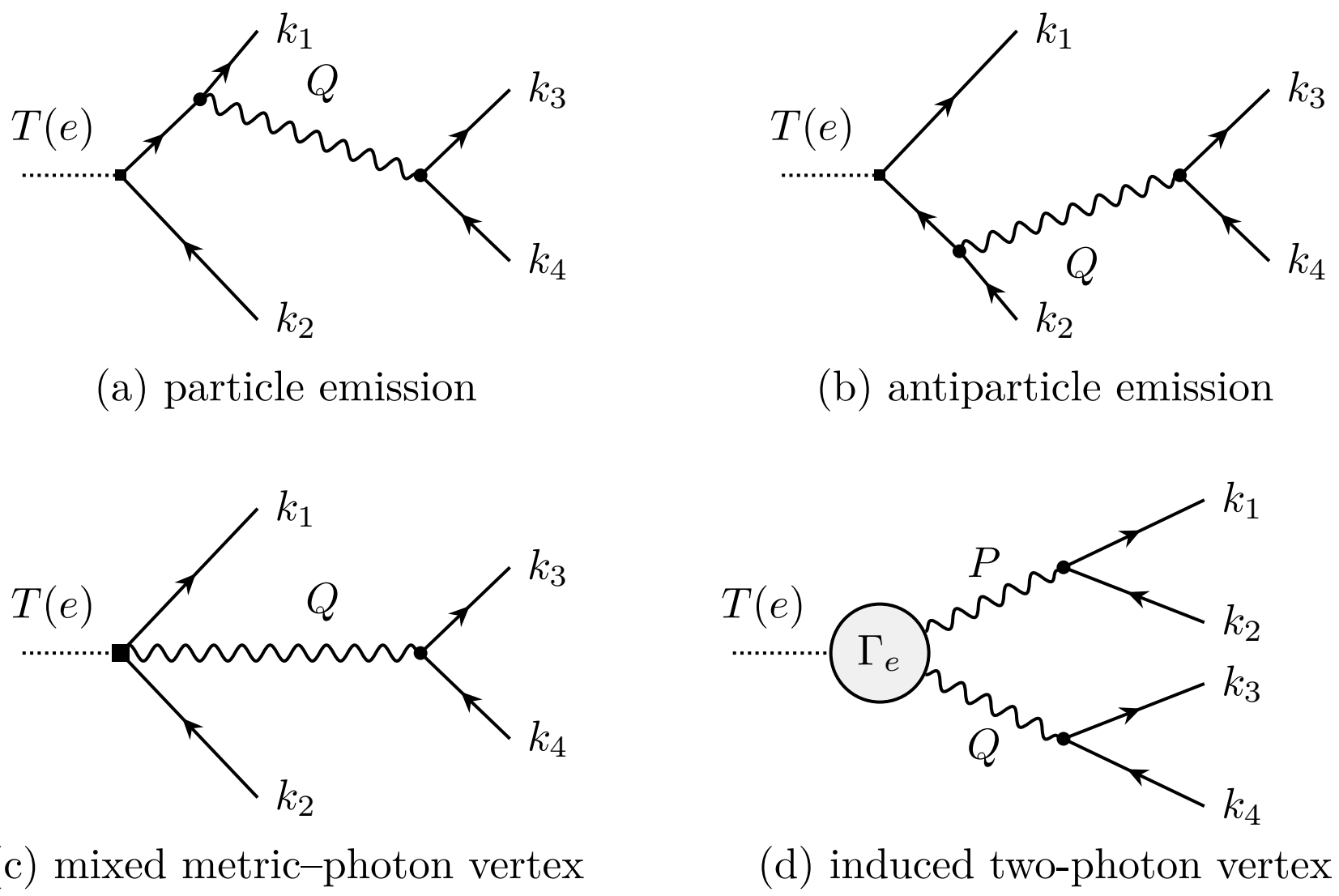

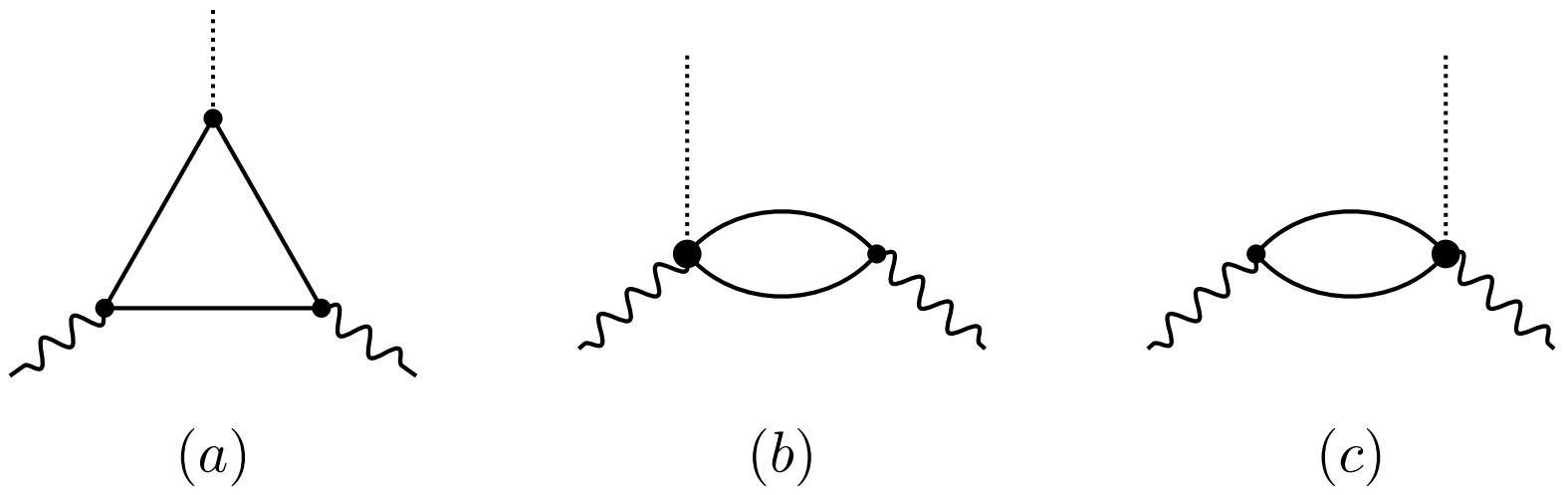
\begin{figure}[ht]
\centering
\begin{tikzpicture}[scale=.9,
 photon/.style={thick,decorate,decoration={snake,amplitude=.8mm,segment length=2.5mm}},
 source/.style={densely dotted,line width=.75pt}]
\begin{scope}[xshift=0cm]
\coordinate (t) at (0,1);\coordinate (l) at (-.8,-.4);\coordinate (r) at (.8,-.4);
\draw[thick] (t)--(l)--(r)--cycle;
\draw[source] (0,1.8)--(t);
\draw[photon] (l)--(-1.5,-.9);\draw[photon] (r)--(1.5,-.9);
\fill (t) circle (2pt);\fill (l) circle (2pt);\fill (r) circle (2pt);
\node at (0,-1.5) {$(a)$};
\end{scope}
\begin{scope}[xshift=4.2cm]
\draw[thick] (-.7,0) to[bend left=50] (.7,0);
\draw[thick] (-.7,0) to[bend right=50] (.7,0);
\draw[source] (-.7,0)--(-.7,1.5);
\draw[photon] (-.7,0)--(-1.5,-.7);\draw[photon] (.7,0)--(1.5,-.7);
\fill (-.7,0) circle (3pt);\fill (.7,0) circle (2pt);
\node at (0,-1.5) {$(b)$};
\end{scope}
\begin{scope}[xshift=8.4cm]
\draw[thick] (-.7,0) to[bend left=50] (.7,0);
\draw[thick] (-.7,0) to[bend right=50] (.7,0);
\draw[source] (.7,0)--(.7,1.5);
\draw[photon] (-.7,0)--(-1.5,-.7);\draw[photon] (.7,0)--(1.5,-.7);
\fill (-.7,0) circle (2pt);\fill (.7,0) circle (3pt);
\node at (0,-1.5) {$(c)$};
\end{scope}
\end{tikzpicture}
\caption{The induced stress-tensor--photon--photon vertex.
Panel $(a)$ includes both matter-loop orientations.
Panels $(b)$ and $(c)$ are the mixed-vertex contact bubbles.
Their explicit expressions are the three terms in
eqs.~\eqref{eq:inducedfermion} and~\eqref{eq:inducedscalar}.
With photon momentum $P$ on the left and $Q$ on the right,
the panels give $\mathcal T_{a,e}$, $\mathcal C_{a,e}^{Q}$,
and $\mathcal C_{a,e}^{P}$ in eq.~\eqref{eq:inducedseparation}.
Their evaluated projections are listed in appendix~\ref{app:induced},
eqs.~\eqref{eq:contactprojectiontable}--\eqref{eq:triangle-b-zw}.}
\label{fig:induced}
\end{figure}

The free scalar and fermion stress-tensor correlators have the same
two-point normalization in eq.~\eqref{eq:free-norms}.
Their energy-flux profiles have different zeros.
Choosing a zero of the leading profile isolates the opposite structure
at first subleading order, as in the current calculation.
For a spatial polarization $e$, the resulting four-particle amplitude
contains ordinary emissions, mixed metric--photon vertices, and a
stress-tensor insertion on the induced photon kernel.
All three types contribute at the same flavor order.
Figures~\ref{fig:emission} and~\ref{fig:induced} show the corresponding
complete photon-mediated production topologies and the decomposition
of the induced vertex.

For a traceless fermion stress-tensor polarization $e$, the ordinary vertex is
\begin{equation}
 V_e(r,s)=\frac12\gamma^a e_{ab}(r+s)^b.
 \label{eq:fermionTvertex}
\end{equation}
The arguments in this subsection are complex Euclidean momenta.
We suppress hats inside $V_e$ and the gamma-matrix products.
The three emission and mixed-vertex contributions on the first pair are
\begin{align}
 \mathcal A_{e,1}^\mu&=\gamma^\mu
 \frac{\gamma^a r_{1a}}{r_1^2}V_e(r_1,-k_2),\nonumber\\
 \mathcal A_{e,2}^\mu&=-V_e(k_1,-r_2)
 \frac{\gamma^a r_{2a}}{r_2^2}\gamma^\mu,\nonumber\\
 \mathcal A_{e,c}^\mu&=-e_{a\mu}\gamma^a.
 \label{eq:Temission}
\end{align}
Their photon contraction reduces to
\begin{equation}
 \bar u_1 Q_\mu(\mathcal A_{e,1}^\mu+
 \mathcal A_{e,2}^\mu+\mathcal A_{e,c}^\mu)v_2
 =\bar u_1[\gamma^a e_{ab}Q^b-\gamma^a e_{ab}Q^b]v_2=0.
 \label{eq:TemissionWard}
\end{equation}
The contact term in Fig.~\ref{fig:emission}(c) cancels the longitudinal part of the two emission diagrams, so the sum of Figs.~\ref{fig:emission}(a--c) is gauge invariant.

\subsection{\texorpdfstring{QED$_3$: the induced vertex and its contact bubbles}{QED3: the induced vertex and its contact bubbles}}

For one Dirac flavor, the current bubble is
\begin{equation}
 \Pi^{\mu\nu}(p)=\frac{|p|}{16}
 \left(\delta^{\mu\nu}-\frac{p^\mu p^\nu}{p^2}\right).
 \label{eq:bubble}
\end{equation}
Its metric derivative is the sum of a triangle and two contact bubbles:
\begin{align}
 \Gamma_e^{\mu\nu}(P,Q)
 ={}&-2\intk
 \frac{\operatorname{tr}[V_e(k,k-q)(\gamma^a k_a)\gamma^\mu
 (\gamma^b(k-P)_b)\gamma^\nu(\gamma^c(k-q)_c)]}
 {k^2(k-P)^2(k-q)^2}\nonumber\\
 &+e_{\mu\rho}\Pi^{\rho\nu}(Q)
 +\Pi^{\mu\rho}(P)e_{\rho\nu}.
 \label{eq:inducedfermion}
\end{align}
The factor two in the first line sums the two loop orientations.
The two terms in the second line arise from the mixed vertices.
Tensor reduction gives $P_\mu\Gamma_e^{\mu\nu}=0$ and
$\Gamma_e^{\mu\nu}Q_\nu=0$.
The reduced transverse components are given in
appendix~\ref{app:induced}.

Let $\mathcal A_e=\mathcal A_{e,1}+\mathcal A_{e,2}+\mathcal A_{e,c}$.
The complete reduced four-fermion amplitude is
\begin{equation}
 \mathcal M_e=
 \frac{(\bar u_1\mathcal A_e^\mu v_2)J_{34,\mu}}{y}
 +\frac{J_{12,\mu}(\bar u_3\mathcal A_e^\mu v_4)}{x}
 +\frac{16}{xy}J_{12,\mu}\Gamma_e^{\mu\nu}J_{34,\nu}.
 \label{eq:fullstressamp}
\end{equation}
The physical amplitude is $16\mathcal M_e/N_f$.
The last term has two photon propagators and a matter loop of order $N_f$.
It therefore has the same overall order as the emission terms.
Its sign is fixed by the metric-source Ward identity.

For the two spatial polarizations $e_+=\operatorname{diag}(0,1,-1)$ and
$(e_\times)_{12}=(e_\times)_{21}=1$, define the angular weight
\begin{equation}
 \mathcal S_f(x,y)=
 \left\langle\sum_a k_a^0
 |\cos(2\theta_a)\mathcal M_++\sin(2\theta_a)\mathcal M_\times|^2
 \right\rangle_{\alpha,\beta}.
 \label{eq:stressweight}
\end{equation}
The norm of the state created by the free stress tensor is $1/32$
per Dirac flavor.
The unordered distinct-flavor sum gives
$N_f(N_f-1)(16/N_f)^2/2=128+O(N_f^{-1})$.
Together with eq.~\eqref{eq:phasespace}, these factors give
\begin{align}
 \nsT&=\frac{32}{\pi^2}\int_0^1\dd x\int_0^{1-x}\dd y\,
 \mathcal S_f(x,y)+O(N_f^{-1})\nonumber\\
 &=\frac{16}{\pi^2}\left(\frac{53063}{45}-\frac{477\pi^2}{4}\right)
 +O(N_f^{-1})\nonumber\\
 &=-1908+\frac{849008}{45\pi^2}+O(N_f^{-1}).
 \label{eq:fermionstresscut}
\end{align}
The first line fixes the phase-space normalization.
The second inserts the analytic angular and radial integrations.
Substitution into eq.~\eqref{eq:collider} gives
$t_{4,\mathrm{QED}_3}=-4+8\nsT/N_f+O(N_f^{-2})$ and
eq.~\eqref{eq:qed-t4}.

\subsection{\texorpdfstring{Critical scalar QED$_3$}{Critical scalar QED3}}

The scalar amplitudes use the same complex Euclidean momenta and
contractions as the fermion amplitudes.
For two outgoing scalar momenta $a,b$, the improved stress-tensor vertex is
\begin{equation}
 V_{b,e}(a,b)=-2a^\mu e_{\mu\nu}b^\nu
 +\Tr(e)\,a\cdot b
 +\frac14[(a+b)^\mu e_{\mu\nu}(a+b)^\nu-\Tr(e)(a+b)^2].
 \label{eq:scalarTvertex}
\end{equation}
The last term follows from the curvature coupling in
eq.~\eqref{eq:scalaraction}.
The emitted photon couples through
\begin{align}
 \mathcal A_{b,e}^\mu={}&
 \frac{(k_1+r_1)^\mu V_{b,e}(r_1,k_2)}{r_1^2}
 -\frac{(k_2+r_2)^\mu V_{b,e}(k_1,r_2)}{r_2^2}\nonumber\\
 &-2e^\mu{}_{\nu}(k_1-k_2)^\nu+\Tr(e)(k_1-k_2)^\mu.
 \label{eq:scalarTemission}
\end{align}
Its last line is the mixed stress-tensor--photon vertex.
The induced scalar loop also contains mixed-vertex bubbles.
Writing $k'=k-q$, its full expression is
\begin{align}
 \Gamma_{b,e}^{\mu\nu}={}&2\intk
 \frac{[2k\cdot e\cdot k'-\Tr(e)k\cdot k'
 +(q\cdot e\cdot q-\Tr(e)q^2)/4]
 (2k-P)^\mu(2k-P-q)^\nu}
 {k^2(k-P)^2(k-q)^2}\nonumber\\
 &+(2e-\Tr(e)I)^{\mu}{}_{\rho}\Pi^{\rho\nu}(Q)
 +\Pi^{\mu\rho}(P)(2e-\Tr(e)I)_{\rho}{}^\nu.
 \label{eq:inducedscalar}
\end{align}
The scalar bubble has the same coefficient as eq.~\eqref{eq:bubble}.
Both photon contractions of eq.~\eqref{eq:inducedscalar} vanish.
Its trace polarization $e=I$ also vanishes after the triangle and
contact terms are combined.
The separate two-photon seagull bubble and the three-source tadpole
vanish as shown in appendix~\ref{app:induced}.

The complete amplitude has the form of eq.~\eqref{eq:fullstressamp},
with spinor currents replaced by scalar pair currents and with
eqs.~\eqref{eq:scalarTemission} and~\eqref{eq:inducedscalar}.
The detector polarization is rotated by $\pi/4$ relative to the fermionic
zero. Its integrated photon contribution is
\begin{equation}
 (\nfT)_A=\frac{16}{\pi^2}
 \left(\frac{53033}{45}-\frac{477\pi^2}{4}\right)
 =-1908+\frac{848528}{45\pi^2}.
 \label{eq:scalarstresscut}
\end{equation}
The auxiliary-scalar production topologies are shown in
figure~\ref{fig:stress-auxiliary-production}.
The auxiliary-scalar contribution is
$(\nfT)_\sigma=10160/(9\pi^2)-343/3$.
Its interference with photon exchange is odd under particle--antiparticle
exchange and integrates to zero. Adding both contributions gives
\begin{align}
 t_{4,\mathrm{sQED}_3}
 &=4-\frac8{N_b}\left[-1908+\frac{848528}{45\pi^2}
 +\frac{10160}{9\pi^2}-\frac{343}{3}\right]+O(N_b^{-2})\nonumber\\
 &=4+\frac1{N_b}\left(\frac{48536}{3}-\frac{2398208}{15\pi^2}\right)
 +O(N_b^{-2}).
 \label{eq:scalarstressresult}
\end{align}

\subsection{\texorpdfstring{QED$_3$--GNY}{QED3--GNY}}

\begin{figure}[htbp]
\centering
\begin{tikzpicture}[font=\small,line width=.75pt,
 matter/.style={postaction={decorate},decoration={markings,
 mark=at position .55 with {\arrow{Stealth[length=1.7mm,width=1.5mm]}}}},
 auxiliary/.style={dashed},source/.style={densely dotted},
 induced/.style={draw,circle,fill=black!6,minimum size=10mm,inner sep=0pt}]
\begin{scope}
\coordinate (s) at (0,0);\coordinate (e) at (.7,.64);
\coordinate (v) at (2.45,0);
\draw[source] (-.8,0)--(s);\node[above] at (-.55,.06) {$T(e)$};
\draw[matter] (s)--(e);\draw[matter] (e)--(1.2,1.2);
\draw[matter] (1.2,-1.2)--(s);
\draw[auxiliary] (e)--(v);\node[above] at (1.7,.45) {$\sigma(Q)$};
\draw[matter] (v)--(3.2,.7);\draw[matter] (3.2,-.7)--(v);
\foreach \w in {e,v} \fill (\w) circle (1.6pt);
\fill (s) +(-.045,-.045) rectangle +(.045,.045);
\node[right] at (1.2,1.2) {$k_1$};\node[right] at (1.2,-1.2) {$k_2$};
\node[right] at (3.2,.7) {$k_3$};\node[right] at (3.2,-.7) {$k_4$};
\node at (1.4,-1.7) {(a) auxiliary emission};
\end{scope}
\begin{scope}[xshift=6.2cm]
\node[induced] (G) at (0,0) {$\Gamma_e^{\sigma\sigma}$};
\coordinate (u) at (1.55,.7);\coordinate (v) at (1.55,-.7);
\draw[source] (-1.3,0)--(G.west);\node[above] at (-1.0,.06) {$T(e)$};
\draw[auxiliary] (G)--(u);\draw[auxiliary] (G)--(v);
\node[above] at (.95,.52) {$\sigma(P)$};\node[below] at (.95,-.52) {$\sigma(Q)$};
\draw[matter] (u)--(2.7,1.25);\draw[matter] (2.7,.25)--(u);
\draw[matter] (v)--(2.7,-.25);\draw[matter] (2.7,-1.25)--(v);
\fill (u) circle (1.6pt);\fill (v) circle (1.6pt);
\node[right] at (2.7,1.25) {$k_1$};\node[right] at (2.7,.25) {$k_2$};
\node[right] at (2.7,-.25) {$k_3$};\node[right] at (2.7,-1.25) {$k_4$};
\node at (1.4,-1.7) {(b) induced $T\sigma\sigma$ vertex};
\end{scope}
\end{tikzpicture}
\caption{Auxiliary-scalar channels in stress-tensor production for
QED$_3$--GNY and critical scalar QED$_3$.
Dashed lines are dressed auxiliary propagators.
Panel (a) shows one emission; attachments to both lines of each pair
are included in the complete amplitude. Panel (b) contains the induced
$T\sigma\sigma$ vertex attached to the two pairs.
As in figure~\ref{fig:emission}, all four matter momenta are outgoing,
with $P=k_1+k_2$ and $Q=k_3+k_4$.
Both topologies are needed at the first subleading order.}
\label{fig:stress-auxiliary-production}
\end{figure}
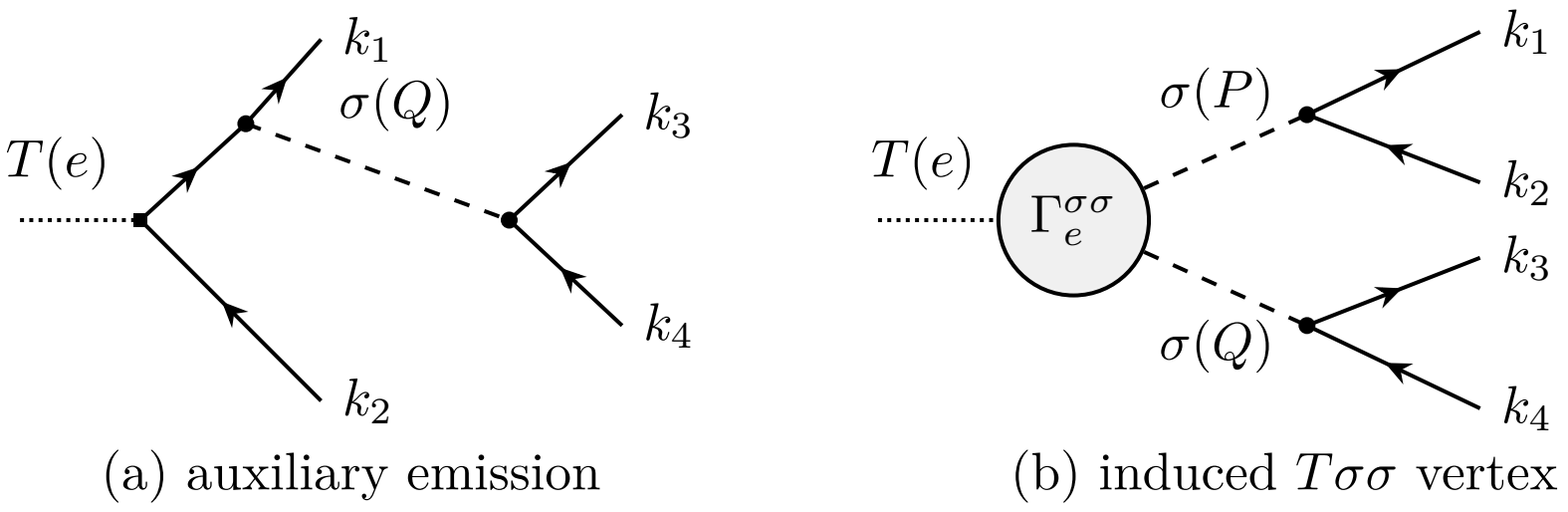

The GNY amplitude adds auxiliary-scalar emission from both fermion pairs.
It also adds the induced $T\sigma\sigma$ triangle.
The two production topologies are shown in
figure~\ref{fig:stress-auxiliary-production}.

The same external spinors are used for the photon and scalar amplitudes.
Their relative normalization is then fixed by eq.~\eqref{eq:sigmaprop}.
The source contractions cancel between auxiliary emission and the induced
triangle. The photon amplitude is odd under charge conjugation within
either pair, while the auxiliary amplitude is even. The detector weight
and phase-space measure are invariant under this transformation, so the
integrated interference vanishes.
The auxiliary square gives
$(\nsT)_\sigma=10172/(9\pi^2)-343/3$.
Hence
\begin{align}
 t_{4,\mathrm{QED}_3\text{-}\mathrm{GNY}}
 &=-4+\frac8{N_f}\left[-1908+\frac{849008}{45\pi^2}
 +\frac{10172}{9\pi^2}-\frac{343}{3}\right]+O(N_f^{-2})\nonumber\\
 &=-4+\frac1{N_f}\left(\frac{2399648}{15\pi^2}-\frac{48536}{3}\right)
 +O(N_f^{-2}).
 \label{eq:GNstressresult}
\end{align}

\section{Squashed-sphere responses and conformal correlators}\label{sec:sphere}

\subsection{Background sources and two-point functions}
\label{sec:sphere-free-two-point}

The relation between near-round-sphere free-energy responses and
stress-tensor data was developed in
refs.~\cite{BobevBuenoVreys,BuenoEtAl,BuenoEtAl2020}.
We use the source conventions of the background-field construction
\cite{ParmentierBobev,GuoLiThreePoint} and calibrate both projections
with free matter. In the gauge theories, the additional ingredients are
the integration over gauge-field configurations modulo gauge transformations
and, for critical scalars, the response of the saddle to both sources.

The independent calculation uses the unit-radius Hopf metric and a flavor
source along its fiber:
\begin{align}
 \dd s^2&=\frac14[\dd\theta^2+\sin^2\theta\,\dd\varphi^2
 +(1+\varepsilon)(\dd\psi+\cos\theta\,\dd\varphi)^2],\nonumber\\
 B&=\frac\lambda2(\dd\psi+\cos\theta\,\dd\varphi).
 \label{eq:spheresources}
\end{align}
The ranges are $0\leq\theta\leq\pi$, $0\leq\varphi<2\pi$, and
$0\leq\psi<4\pi$, with the usual sphere identifications.
The one-form $B$ is held fixed when the metric is varied.
The parity-even sphere free energy is
$F(\varepsilon,\lambda)=\operatorname{Re}W[g(\varepsilon),B(\lambda)]$.
The quadratic derivatives give the simplest examples of extracting
correlators from this free energy. For either source parameter
$x=\lambda$ or $x=\varepsilon$, differentiation of the path integral gives
\begin{equation}
 \partial_x^2F=\operatorname{Re}\left[
 \langle\partial_x^2S\rangle
 -\langle(\partial_xS)^2\rangle_{\mathrm{conn}}\right].
 \label{eq:spherequadraticvariation}
\end{equation}
The first action derivative inserts the current or stress tensor,
integrated against the corresponding source profile.
The second action derivative supplies its contact term.
Thus the quadratic free-energy response measures an integrated two-point
function with the contact prescription fixed by the action.

Conformal invariance fixes each two-point function up to its normalization.
For the profiles in eq.~\eqref{eq:spheresources}, the integrated responses are
\begin{equation}
 F_{\lambda\lambda}(0)=2\pi^4C_J,\qquad
 F_{\varepsilon\varepsilon}(0)=-\frac{\pi^4C_T}{3}.
 \label{eq:spheretwopoint}
\end{equation}
The current factor is calibrated below with the unit-charge scalar.
The metric factor follows from the integrated stress-tensor two-point
function on the round sphere
\cite{BobevBuenoVreys,BuenoEtAl,BuenoEtAl2020,ParmentierBobev}:
\begin{equation}
 C_J=\frac{F_{\lambda\lambda}(0)}{2\pi^4}
 =\frac{[\lambda^2]F(0,\lambda)}{\pi^4},\qquad
 C_T=-\frac{3F_{\varepsilon\varepsilon}(0)}{\pi^4}
 =-\frac{6[\varepsilon^2]F(\varepsilon,0)}{\pi^4}.
 \label{eq:spherecentralextraction}
\end{equation}
Here the brackets extract a Taylor coefficient, so each second derivative
contributes a factor of $2!$. The flavor source includes the generator
$Q$, with $\Tr Q^2=1$ as in section~\ref{sec:intro}.

\subsubsection*{Free-scalar example}

Consider first one unit-charge complex scalar with conformal curvature
coupling. Set the auxiliary mass to zero in
eq.~\eqref{eq:chargedscalarspectrum}.
The modes have $n=1,2,\ldots$ and
$m=-(n-1)/2,\ldots,(n-1)/2$ in unit steps.
Each fixed $m$ has degeneracy $n$.
The Gaussian integral then gives
\begin{equation}
 F_{\mathrm c}(\varepsilon,\lambda)-F_{\mathrm c}(0,0)
 =\sum_{n=1}^{\infty}\sum_{m=-(n-1)/2}^{(n-1)/2}
 n\log\left[
 \frac{n^2-\frac14-\frac{4m^2\varepsilon}{1+\varepsilon}
 -\frac\varepsilon4+\frac{4m\lambda+\lambda^2}{1+\varepsilon}}
 {n^2-\frac14}\right].
 \label{eq:scalarfreeenergydifference}
\end{equation}
The spectral sums in this example are evaluated by analytic
continuation, as in eq.~\eqref{eq:zeta-determinant}.
The magnetic sums are given in eq.~\eqref{eq:saddlemagneticmoments}.
Their use reduces both quadratic responses to sums over $n$.

The required sums follow from the cotangent expansion in
eq.~\eqref{eq:saddlegeneratingsum}. For a noninteger spectral parameter
$a$, that expansion gives
\begin{align}
 \sum_{n=1}^{\infty}\frac{n^2}{n^2-a^2}
 &=\zeta(0)+a^2\left[\frac1{2a^2}
 -\frac\pi{2a}\cot(\pi a)\right]
 =-\frac{\pi a}{2}\cot(\pi a),\nonumber\\
 \sum_{n=1}^{\infty}\frac{n^2}{(n^2-a^2)^2}
 &=\frac1{2a}\frac{\dd}{\dd a}
 \left[-\frac{\pi a}{2}\cot(\pi a)\right]
 =\frac{\pi^2}{4}\csc^2(\pi a)-\frac\pi{4a}\cot(\pi a).
 \label{eq:scalarquadraticsums}
\end{align}
The second sum is convergent. At the conformal value $a=1/2$,
these identities and $\zeta(-2)=0$ give
\begin{equation}
 \sum_{n=1}^{\infty}n^2=0,\qquad
 \sum_{n=1}^{\infty}\frac{n^2}{n^2-1/4}=0,\qquad
 \sum_{n=1}^{\infty}\frac{n^2}{(n^2-1/4)^2}=\frac{\pi^2}{4}.
 \label{eq:scalarquadraticsumvalues}
\end{equation}
These sums suffice for both correlators.

For the current, expand eq.~\eqref{eq:scalarfreeenergydifference}
to second order in $\lambda$ at $\varepsilon=0$:
\begin{align}
 [\lambda^2]F_{\mathrm c}(0,\lambda)
 &=\sum_{n,m}n\left[
 \frac1{n^2-1/4}-\frac{8m^2}{(n^2-1/4)^2}\right]\nonumber\\
 &=\sum_{n=1}^{\infty}\left[
 \frac{n^2}{n^2-1/4}
 -\frac{2n^2(n^2-1)}{3(n^2-1/4)^2}\right]
 =\frac{\pi^2}{8}.
 \label{eq:scalarCJexample}
\end{align}
The first term in the first line comes from the quadratic source vertex.
The second term comes from two linear current vertices.
The magnetic sum gives the second line.
Substituting eq.~\eqref{eq:scalarquadraticsumvalues} evaluates that line.
Consequently $F_{\lambda\lambda}(0)=2!\,\pi^2/8=\pi^2/4$.
Together with $C_J^{(0)}=1/(8\pi^2)$, this fixes the geometric factor
$F_{\lambda\lambda}(0)/C_J^{(0)}=2\pi^4$ in
eq.~\eqref{eq:spheretwopoint}.

For the stress tensor, the scalar eigenvalue at $\lambda=0$ expands as
\begin{equation}
 \Lambda_{nm}(\varepsilon)
 =n^2-\frac14-\left(4m^2+\frac14\right)\varepsilon
 +4m^2\varepsilon^2+O(\varepsilon^3).
 \label{eq:scalarquadraticmetricspectrum}
\end{equation}
Inserting this expansion into the same logarithm gives
\begin{align}
 [\varepsilon^2]F_{\mathrm c}(\varepsilon,0)
 &=\sum_{n,m}n\left[
 \frac{4m^2}{n^2-1/4}
 -\frac{(4m^2+1/4)^2}{2(n^2-1/4)^2}\right]\nonumber\\
 &=\sum_{n=1}^{\infty}\left[
 \frac{n^2(n^2-1)}{3(n^2-1/4)}
 -\frac{n^2(48n^4-120n^2+87)}{480(n^2-1/4)^2}\right]\nonumber\\
 &=\sum_{n=1}^{\infty}\left[
 \frac{7n^2}{30}-\frac{n^2}{20(n^2-1/4)}
 -\frac{n^2}{8(n^2-1/4)^2}\right]
 =-\frac{\pi^2}{32}.
 \label{eq:scalarCTexample}
\end{align}
The second line uses the second and fourth magnetic moments.
Partial fractions give the third line, which is evaluated with
eq.~\eqref{eq:scalarquadraticsumvalues}.
At these quadratic orders the scalar sums contain only even powers
in their large-$n$ expansions. There is therefore no $1/n$ term producing an additional finite
contribution through a spectral pole.

The two responses therefore give
\begin{equation}
 C_J=\frac{\pi^2/8}{\pi^4}=\frac1{8\pi^2},\qquad
 C_T=-\frac6{\pi^4}\left(-\frac{\pi^2}{32}\right)
 =\frac3{16\pi^2}.
 \label{eq:scalarcentralexample}
\end{equation}
This reproduces the normalizations in eq.~\eqref{eq:free-norms}.
At the critical scalar saddle, the mass shift starts at quadratic
source order. Its contribution to the leading determinant at that
order multiplies $\sum_nn^2/(n^2-1/4)=0$.
Thus the same leading responses apply at the critical saddle.
The subleading fluctuation determinants still depend on that saddle,
as described below.

\subsection{Three-point functions and conformal collider parameters}

One further metric derivative probes the three-point functions.
For the Hopf squashing linear in the covariant metric, the full
metric variation, including its semilocal terms, gives
\cite[section~2.2.3]{ParmentierBobev}
\begin{equation}
 F_{\varepsilon\varepsilon\varepsilon}(0)=\frac{\pi^4C_Tt_4}{630}.
 \label{eq:spherestress}
\end{equation}
For the mixed response, we hold the background one-form fixed and
use a conformal scheme preserving background gauge invariance and
diffeomorphisms. Match the two parity-even separated-point structures
to their free scalar and fermion representatives with the same
$C_J$. The difference of the source derivatives then obeys homogeneous
Ward identities. At a $T$--$J$ coincidence only a delta function
times a current two-point function can contribute by dimension and
spin. Its coefficient vanishes by the homogeneous current Ward
identity. At a $J$--$J$ coincidence, the allowed operators have
dimension at most one and have zero two-point function with $T$.
There is no parity-even dimension-three local counterterm quadratic
in the background one-form, and the gauge Chern--Simons term is metric
independent. The mixed derivative is therefore calibrated on the
same two free structures in each theory, including for the topological
current. Using the free responses below gives
\begin{equation}
 t_4=-210\frac{F_{\varepsilon\varepsilon\varepsilon}(0)}
 {F_{\varepsilon\varepsilon}(0)},\qquad
 \gamma=\frac58\frac{F_{\varepsilon\lambda\lambda}(0)}
 {F_{\lambda\lambda}(0)}.
 \label{eq:sphereratios}
\end{equation}
These are derived expressions for the canonical ratios in
eq.~\eqref{eq:collider}.

The conformal collider parameters extracted here also enter
spinning bootstrap systems \cite{ReehorstMixed,DommesEtAl}.
The sphere determines these parameters from integrated three-point
functions with a specified source-contact prescription. It therefore gives an independent
large-flavor determination of potential bootstrap benchmark data.

For one complex scalar and one Dirac fermion, respectively, the free
Taylor coefficients are
\begin{equation}
\begin{array}{c|llll}
 & [\varepsilon^2]F&[\varepsilon^3]F&[\lambda^2]F&[\varepsilon\lambda^2]F\\\hline
 \text{scalar}&-\pi^2/32~~&\pi^2/5040&\pi^2/8~~~~&-\pi^2/60\\
 \text{fermion}&-\pi^2/32&-\pi^2/5040~~&\pi^2/8&\pi^2/60
\end{array}
 \label{eq:freeTaylor}
\end{equation}
The current entries refer to a unit-charge flavor and scale with
$\Tr Q^2$. That invariant cancels from the mixed ratio.
The derivative factorials cancel in the current ratio.
The stress-tensor ratio has an additional factor $3!/2!=3$.

\section{Large-flavor evaluation on the squashed sphere}
\label{sec:sphere-evaluation}

We now evaluate the responses introduced in section~\ref{sec:sphere}
at first subleading order in the large-flavor expansion.
We first construct the matter kernels and the physical gauge determinant,
then evaluate their source-dependent harmonic sums. These determine the
flavor-current and stress-tensor data; the inverse-photon response gives
an independent determination of the topological-current data.

\subsection{Matter kernels and gauge fixing}

At first subleading order, the sphere free energy is obtained from
log determinants of the fluctuation kernels. Charge conjugation makes the photon--scalar
quadratic kernel vanish at zero flavor source. The same block structure
persists for the source response needed here after summing over flavors and using $\Tr Q=0$. Thus
\begin{align}
 F_{\mathrm{QED}_3}&=N_fF_{\mathrm D}+F_{A,f}+O(N_f^{-1}),\nonumber\\
 F_{\mathrm{sQED}_3}&=N_bF_{\mathrm c}+F_{A,b}+F_{\sigma,b}+O(N_b^{-1}),\nonumber\\
 F_{\mathrm{QED}_3\text{-}\mathrm{GNY}}
 &=N_fF_{\mathrm D}+F_{A,f}+F_{\sigma,f}+O(N_f^{-1}).
 \label{eq:spheredeterminants}
\end{align}
At nonzero flavor source, the leading terms in these expressions denote
the sums of free energies of matter fields with the individual charges $q_i$.
Their quadratic flavor responses are weighted by $\sum_iq_i^2=1$,
while their zero-source metric responses carry $N_f$ or $N_b$.
Integrating out the matter fields gives the matter effective action.
The photon kernel is its second functional derivative with respect to
$a_\mu$. The auxiliary-field kernel is its second functional derivative
with respect to the auxiliary field. These derivatives include the
seagulls generated by the matter action.

For scalar harmonics, use $n=1,2,\ldots$ and
$m=-(n-1)/2,\ldots,(n-1)/2$, with degeneracy $n$ at fixed $m$.
At vanishing flavor source the scalar operator has eigenvalues
\begin{equation}
 \Lambda_{nm}=n^2-\frac14-\frac{4m^2\varepsilon}{1+\varepsilon}
 -\frac\varepsilon4+\sigma_*(\varepsilon).
 \label{eq:scalarSpectrum}
\end{equation}
Here $\sigma_*$ is the real saddle value of the mass term $\ii\sigma$
in eq.~\eqref{eq:scalaraction}.
The critical saddle satisfies the analytically continued gap equation
$\Tr(-\nabla_g^2+R/8+\sigma_*)^{-1}=0$.
Solving it through cubic order gives
\begin{equation}
 \sigma_*(\varepsilon)=\frac{\varepsilon^2}{10}
 +\frac{202-21\pi^2}{84}\varepsilon^3+O(\varepsilon^4).
 \label{eq:saddle}
\end{equation}
This saddle is substituted into both scalar fluctuation kernels before
their metric derivatives are taken.

For fermions, the corresponding construction starts from the spinor
operator. Let $J_a$ be the spin-$j$ generators, with $n=2j+1$.
Writing $b=\sqrt{1+\varepsilon}$, the Hermitian Dirac operator is
\begin{equation}
 D_b=2\sigma_1J_1+2\sigma_2J_2+\frac{2\sigma_3J_3}{b}
 +\frac b2+\frac1b.
 \label{eq:DiracSphere}
\end{equation}
At zero gauge and flavor sources, our convention is
$D_b=\ii\gamma^a e_a{}^\mu\nabla_\mu$.
The fermion operator in eq.~\eqref{eq:fermionaction} therefore obeys
\[
 \gamma^a e_a{}^\mu\nabla_\mu+\sigma
 =-\ii(D_b+\ii\sigma).
\]
Thus the real Yukawa field enters as $\ii\sigma$ in the determinant
written in terms of $D_b$.
The overall factor $-\ii$ does not affect its parity-even free energy.
The Pauli and angular-momentum algebras give
\begin{equation}
 (D_b-b/2)^2=n^2-\frac{\varepsilon}{1+\varepsilon}(2J_3+\sigma_3)^2.
 \label{eq:DiracSquare}
\end{equation}
This identity reduces the spinor propagator to scalar resolvents.
The resulting matter loops give the photon and Gross--Neveu kernels.

The photon integral includes a gauge-fixing Jacobian and the
Faddeev--Popov determinant. After the nonconstant longitudinal modes
cancel, its free energy is
\begin{equation}
 F_A=\frac12\log\det K_T-\frac12\log\det{}'(-\nabla^2)
 +\frac12\log V.
 \label{eq:quotientdet}
\end{equation}
Here $K_T$ acts on transverse one-forms, the prime omits the constant
scalar mode, and $V=2\pi^2b$.
The scalar Laplacian and volume terms contribute
\begin{equation}
 [\varepsilon^2]\left[-\tfrac12\log\det{}'(-\nabla^2)+\tfrac12\log V\right]
 =\frac1{24},\qquad
 [\varepsilon^3](\cdots)=\frac{2\pi^2}{189}-\frac1{36}.
 \label{eq:ghostresponse}
\end{equation}
They are part of the metric response and cannot be omitted.
They have no flavor-source dependence and therefore do not contribute
to the mixed current response.

\subsection{Flavor-source response and spectral continuation}

The kernels are projected on complete harmonic representations.
Their source derivatives then reduce to finite matrix traces at each
representation. For example, the coefficient linear in squashing and
quadratic in flavor source contains
\begin{equation}
 \frac12\Tr[K_{00}^{-1}K_{12}
 -K_{00}^{-1}K_{10}K_{00}^{-1}K_{02}].
 \label{eq:mixedtrace}
\end{equation}
The subscripts label Taylor coefficients in $(\varepsilon,\lambda)$.
Their harmonic sum is evaluated by subtracting its large-representation
asymptotic expansion and analytically continuing the subtracted terms.
Source differentiation includes finite contributions from poles of the
analytically continued spectral sums. 

The photon kernel $K$ acts on gauge-field configurations modulo gauge
transformations. These equivalence classes are represented by transverse
one-forms. For a gauge slice with basis $U$, the matrix of the quadratic form is
$U^\dagger KU$. The matrix of the induced inner product is $U^\dagger P_TU$,
where $P_T$ projects onto transverse one-forms. The generalized
eigenvalue problem for these two matrices gives the transverse spectrum
of $K$. A common spectral prescription therefore identifies the
determinants in the different slices.

The explicit gauge comparisons use complete $n=1,2,3$ representations.
They contain 42 one-form components and 13 nonconstant gauge directions.
Landau and polar Coulomb slices give the same transverse spectrum.
Covariant parameters $\xi=1/2,1,3$ give the same flavor-source traces.
The full continued sums use this identification of the spectra.

For the flavor eigenvalues $q_i$ introduced in section~\ref{sec:intro},
the normalization of the flavor current gives
$\sum_iq_i=0$ and $\sum_iq_i^2=1$.
In the magnetic convention with linear source term $4mq_i\lambda$,
the scalar spectrum extends eq.~\eqref{eq:scalarSpectrum} to
\begin{align}
 \Lambda_{nm,i}(\varepsilon,\lambda)
 ={}&n^2-\frac14-\frac{4m^2\varepsilon}{1+\varepsilon}
 -\frac\varepsilon4
 +\frac{4mq_i\lambda+q_i^2\lambda^2}{1+\varepsilon}
 +\sigma_*(\varepsilon,\lambda).
 \label{eq:chargedscalarspectrum}
\end{align}
The metric and flavor sources change the critical saddle that enters
the matter spectrum. Its expansion at the required order is
\begin{equation}
 \sigma_*(\varepsilon,\lambda)
 =\sigma_*(\varepsilon,0)+\frac{\lambda^2}{N_b}
 \left[-\frac23+\left(\pi^2-\frac{418}{45}\right)\varepsilon\right]
 +O(\lambda^4,\varepsilon^2\lambda^2).
 \label{eq:mixedsaddle}
\end{equation}
The factor $1/N_b$ follows from $\sum_iq_i^2=1$ in the gap equation.
The coefficients follow by expanding that equation before continuing
its harmonic sums.
The gap-equation derivation is given in appendix~\ref{app:sphere-saddle}.

For fermions, the flavor source changes the Dirac operator by
\begin{equation}
 D_{b,i}(\lambda)=D_b+\frac{q_i\lambda}{b}\sigma_3.
 \label{eq:chargedDirac}
\end{equation}
The operator $2J_3+\sigma_3$ commutes with this Dirac operator.
The Pauli algebra gives the exact product
\begin{align}
 D_{b,i}(\lambda)\bigl[D_{b,i}(\lambda)-b\bigr]
 &=\bigl[D_{b,i}(\lambda)-b/2\bigr]^2-b^2/4\nonumber\\
 &=n^2-\frac{b^2}{4}
 -\frac{\varepsilon(2J_3+\sigma_3)^2}{1+\varepsilon}
 +\frac{2q_i\lambda(2J_3+\sigma_3)+q_i^2\lambda^2}{1+\varepsilon}.
 \label{eq:chargedDiracproduct}
\end{align}
The source-independent part follows from eq.~\eqref{eq:DiracSquare}.
The linear source term uses
$\{D_b-b/2,\sigma_3\}=2(2J_3+\sigma_3)/b$.
Since the denominator commutes with the numerator, the propagator is
\begin{align}
 D_{b,i}(\lambda)^{-1}
 ={}&[D_{b,i}(\lambda)-b]\bigg[
 n^2-\frac{b^2}{4}-\frac{\varepsilon(2J_3+\sigma_3)^2}{1+\varepsilon}
 \nonumber\\
 &\hspace{22mm}+\frac{2q_i\lambda(2J_3+\sigma_3)+q_i^2\lambda^2}
 {1+\varepsilon}\bigg]^{-1}.
 \label{eq:chargedDiracresolvent}
\end{align}
This resolvent supplies all six orders
$\varepsilon^r\lambda^s$ with $0\leq r\leq1$ and $0\leq s\leq2$.
Those orders enter the two matter propagators in each photon kernel,
including the products of their terms linear in $\lambda$.

These source-dependent propagators determine the kernel derivatives
and hence the continued harmonic sums. For the normalized flavor
generator, write the source-dependent
fluctuation contribution as
\begin{equation}
 F_1(\varepsilon,\lambda)-F_1(\varepsilon,0)
 =\frac{\lambda^2}{N}\bigl(f_{02}+\varepsilon f_{12}\bigr)
 +O(\lambda^4,\varepsilon^2\lambda^2).
 \label{eq:mixedfreecoefficients}
\end{equation}
Here $N$ is the appropriate flavor number, and the common factor $1/N$
is removed from the kernel Taylor coefficients below.
On the transverse modes of representation $n$, the round photon kernel
has eigenvalues $(n\pm1)/16$.
Factor out $n/16$ and use the dimensionless matrix $16K_{T,n}/n$.
For $(r,s)=(0,2)$ or $(1,2)$, the required mode traces are
\begin{align}
 a_{n,rs}&=\frac12[\varepsilon^r\lambda^s]
 \Tr_n\log(16K_{T,n}/n),\nonumber\\
 b_{n,rs}&=\frac14[\varepsilon^r\lambda^s]
 \Tr_n\bigl[\log(16K_{T,n}/n)\bigr]^2.
 \label{eq:mixedmodetraces}
\end{align}
The trace includes the left-representation degeneracy $n$.
The coefficient $a_{n,12}$ is the matrix trace in
 eq.~\eqref{eq:mixedtrace}, after this degeneracy is included.
The trace $b_{n,rs}$ supplies a finite contribution through a pole of
its analytically continued sum. To exhibit this contribution, let
$s_\zeta$ be the spectral regulator.
Expanding the mode zeta function at fixed source gives
\begin{align}
 [\varepsilon^r\lambda^s]\frac12\Tr_n K_{T,n}^{-s_\zeta}
 =\left(\frac n{16}\right)^{-s_\zeta}
 \left[-s_\zeta a_{n,rs}+s_\zeta^2b_{n,rs}
 +O(s_\zeta^3)\right].
 \label{eq:mixedzetaexpansion}
\end{align}
The continued free energy is minus the regulator derivative at zero.
Write the large-$n$ expansions as
$a_{n,rs}\sim\alpha_{0,rs}+\sum_{k\geq1}\alpha_{k,rs}n^{-k}$ and
$b_{n,rs}\sim\beta_{0,rs}+\sum_{k\geq1}\beta_{k,rs}n^{-k}$.
The $\beta_{1,rs}/n$ term sums to
$\beta_{1,rs}\zeta(1+s_\zeta)$.
This function has a pole at $s_\zeta=0$ with residue $\beta_{1,rs}$.
Multiplying by $s_\zeta^2$ and taking minus the regulator derivative
at zero gives the finite free-energy contribution $-\beta_{1,rs}$.
The corresponding $1/n$ coefficient in $a_{n,rs}$ is consistent with
zero at the precision of the recorded fits.
The finite sum with a truncated asymptotic tail is therefore
\begin{align}
 f_{rs}\simeq{}&\sum_{n=1}^{L}(a_{n,rs}-\alpha_{0,rs})
 +\alpha_{0,rs}\zeta(0)
 +\sum_{k=2}^{K}\alpha_{k,rs}\zeta(k,L+1)-\beta_{1,rs}.
 \label{eq:mixedcontinued}
\end{align}
Here $L$ is the summation cutoff and $K$ is the highest fitted inverse
power. The approximation comes from truncating the tail.
The finite contribution $-\beta_{1,rs}$ is included explicitly.

The pole residues that determine these finite contributions are
\begin{equation}
\begin{array}{l@{\qquad}rr}
 &\beta_{1,02}&\beta_{1,12}\\\hline
 \text{fermion-induced photon}&2.6666666667&-0.7349157290\\
 \text{scalar-induced photon}&0.0000000000&-0.0355555556
\end{array}
\end{equation}
These entries are numerical asymptotic coefficients.
The largest residual $1/n$ coefficient in the unweighted traces
$a_{n,rs}$ is below $1.5\times10^{-25}$ for these fits.
The resulting mixed coefficients are
\begin{table}[ht]
\centering
\begin{tabular}{lrr}
\toprule
Fluctuation sector&$f_{02}$&$f_{12}$\\
\midrule
Fermion-induced photon&$0.176308910566432$&$-0.444444444444444$\\
Scalar-induced photon&$4.176308910566432$&$0.088888888888889$\\
Critical scalar auxiliary&$-4/9$&$\pi^2/60-4/45$\\
Gross--Neveu auxiliary&$-4/9$&$2/45-\pi^2/60$\\
\bottomrule
\end{tabular}
\caption{Mixed flavor-source Taylor coefficients with the factor $1/N$
in eq.~\eqref{eq:mixedfreecoefficients} removed.
The photon entries are numerical continued sums.
The scalar entries include the source-dependent critical saddle.}
\label{tab:mixedsphere}
\end{table}

We checked numerical convergence by increasing the number of modes and varying the extrapolation of the spectral sums. These changes affect the photon contributions to \(\gamma\) by amounts well below the quoted precision.

\subsection{Two-point and three-point results}
\label{sec:sphere-interacting-two-point}

Write the interaction contribution at zero flavor source as
$F_1(\varepsilon)-F_1(0)=f_2\varepsilon^2+f_3\varepsilon^3+\cdots$.
Table~\ref{tab:spherecoeff} gives the independent determinant results.

\begin{table}[ht]
\centering
\begin{tabular}{lrr}
\toprule
Fluctuation sector&$f_2$&$f_3$\\
\midrule
Fermion-induced photon&$-0.221855005419386$&$0.012733955190174$\\
Scalar-induced photon&$-0.555188338752719$&$-0.006384748840967$\\
Critical scalar auxiliary&$0.069444444444444$&$-0.000625091326237$\\
Gross--Neveu auxiliary&$-0.013888888888889$&$0.000625091326237$\\
\bottomrule
\end{tabular}
\caption{Taylor coefficients of the fluctuation free energies.
The scalar entries include the critical saddle.}
\label{tab:spherecoeff}
\end{table}

The quadratic coefficients now determine the interacting two-point
functions by the same steps as in the scalar example.
The leading matter determinants give $\pi^2/8$ in the flavor response
and $-N\pi^2/32$ in the metric response.
The different flavor factors follow from $\Tr Q^2=1$ for the current
and from the sum over all $N$ matter fields for the stress tensor.
Adding the fluctuation determinants in eq.~\eqref{eq:spheredeterminants}
therefore gives
\begin{align}
 [\lambda^2]F(0,\lambda)
 &=\frac{\pi^2}{8}+\frac{f_{02}}{N}+O(N^{-2}),\nonumber\\
 [\varepsilon^2]F(\varepsilon,0)
 &=-\frac{N\pi^2}{32}+f_2+O(N^{-1}).
 \label{eq:sphereinteractingquadratic}
\end{align}
Here $f_{02}$ and $f_2$ are the sums of the applicable rows of
tables~\ref{tab:mixedsphere} and~\ref{tab:spherecoeff}.
Substitution into eq.~\eqref{eq:spherecentralextraction} gives
\begin{align}
 \frac{C_J}{C_J^{(0)}}
 &=\frac{8}{\pi^2}[\lambda^2]F(0,\lambda)
 =1+\frac{8f_{02}}{\pi^2N}+O(N^{-2}),\nonumber\\
 \frac{C_T}{C_T^{(0)}}
 &=-\frac{32}{\pi^2}[\varepsilon^2]F(\varepsilon,0)
 =N-\frac{32f_2}{\pi^2}+O(N^{-1}).
 \label{eq:spherecentralcorrections}
\end{align}
Comparison with the expansion in eq.~\eqref{eq:centralcharges}
then yields $c_J=8f_{02}/\pi^2$ and $c_T=-32f_2/\pi^2$.

The determinants needed for these coefficients stop at quadratic order.
For example, $\Tr Q=0$ removes the term linear in $\lambda$ from
each flavor-summed fluctuation kernel.
Writing the quadratic coefficient with its explicit flavor factor gives
$K_n(0,\lambda)=K_{00}+\lambda^2K_{02}/N+O(\lambda^4)$,
as in eq.~\eqref{eq:mixedfreecoefficients}.
The finite-representation current trace is consequently
\begin{equation}
 \frac N2[\lambda^2]\Tr_n\log K_n(0,\lambda)
 =\frac12\Tr_n(K_{00}^{-1}K_{02}).
 \label{eq:spherequadraticcurrenttrace}
\end{equation}
The metric trace also contains the product of two first derivatives:
\begin{equation}
 \frac12[\varepsilon^2]\Tr_n\log K_n(\varepsilon,0)
 =\frac12\Tr_n\left[
 K_{00}^{-1}K_{20}
 -\frac12K_{00}^{-1}K_{10}K_{00}^{-1}K_{10}\right].
 \label{eq:spherequadraticstresstrace}
\end{equation}
Both expressions follow by expanding the matrix logarithm through
second order. Their continued harmonic sums include the finite contributions from
spectral poles discussed above.
For the photon metric response, the scalar Laplacian and volume terms
in eq.~\eqref{eq:ghostresponse} must also be added.
They contribute $1/24$ to $f_2$ and have no flavor-source response.

For QED$_3$, inserting the photon entries gives the explicit example
\begin{align}
 c_{J,\mathrm{QED}_3}
 &\simeq\frac8{\pi^2}(0.176308910566432)
 \simeq0.1429106200,\nonumber\\
 c_{T,\mathrm{QED}_3}
 &\simeq-\frac{32}{\pi^2}(-0.221855005419386)
 \simeq0.7193155759.
 \label{eq:sphereQEDcentralexample}
\end{align}
Critical scalar QED adds its auxiliary-scalar determinant to the
scalar-induced photon determinant. QED$_3$--GNY instead adds the
Gross--Neveu determinant to the fermion-induced photon determinant.
Table~\ref{tab:spherecentralcharges} gives the resulting sector sums
and two-point corrections.

\begin{table}[ht]
\centering
\setlength{\tabcolsep}{4pt}
\begin{tabular}{lrrrr}
\toprule
Theory&$f_{02}$&$f_2$&$c_J$&$c_T$\\
\midrule
QED$_3$&$0.1763089106$&$-0.2218550054$&$0.1429106200$&$0.7193155759$\\
Critical scalar QED$_3$&$3.7318644661$&$-0.4857438943$&$3.0249353992$&$1.5749166822$\\
QED$_3$--GNY&$-0.2681355339$&$-0.2357438943$&$-0.2173424774$&$0.7643472130$\\
\bottomrule
\end{tabular}
\caption{Two-point data extracted from the quadratic sphere responses.
The current correction multiplies $1/N$, whereas the stress-tensor
correction is of order one in eq.~\eqref{eq:centralcharges}.}
\label{tab:spherecentralcharges}
\end{table}

These numerical sphere determinations agree with the independent
analytic two-point results in table~\ref{tab:norms}.
The comparison tests the quadratic source response, including the
critical saddle and integration over transverse gauge fields.
It does not use the cubic or mixed three-point coefficients.
Those coefficients enter only when the collider ratios are extracted
below.

Having checked the quadratic responses, we turn to the mixed and cubic
coefficients that determine the conformal collider parameters.
The Taylor coefficients give the flavor-current ratios directly.
Using the fermion and scalar signs in eq.~\eqref{eq:freeTaylor}, respectively,
\begin{align}
 \gamma_{f,b}
 &=\frac58\frac{\pm\pi^2/30+2f_{12}/N}
 {\pi^2/4+2f_{02}/N}+O(N^{-2})\nonumber\\
 &=\pm\frac1{12}+\frac5{\pi^2N}
 \left(f_{12}\mp\frac{2f_{02}}{15}\right)+O(N^{-2}).
 \label{eq:mixedratioexpanded}
\end{align}
Adding the appropriate rows of table~\ref{tab:mixedsphere} reproduces
the three current coefficients in table~\ref{tab:collider}.
This comparison uses the source spectra and continued determinants,
independently of the flat-space cut integrals.

Using eq.~\eqref{eq:freeTaylor}, the fermion and scalar ratios expand to
\begin{align}
 t_{4,f}&=-4+\frac{20160f_3-128f_2}{\pi^2N_f}+O(N_f^{-2}),\nonumber\\
 t_{4,b}&=4+\frac{20160f_3+128f_2}{\pi^2N_b}+O(N_b^{-2}).
 \label{eq:sphereexpand}
\end{align}
The determinant sectors are added according to
eq.~\eqref{eq:spheredeterminants} before applying these formulas.
The current ratio is expanded with the same flavor conventions.
The six flavor-current and stress-tensor sphere results agree with
the analytic flat-space values in table~\ref{tab:collider}.

\subsection{Topological-current response from the inverse photon kernel}
\label{sec:topological-sphere}

We now give the independent sphere determination of the topological-current data presented in section~\ref{sec:topological}. It uses the same metric
profile and collider relation as the flavor-current calculation, with
the BF source replacing the matter flavor source.

\subsubsection{BF source and the inverse photon kernel}

The BF source of eq.~\eqref{eq:topcurrent} couples linearly to the
photon and does not enter the matter covariant derivative.
On the sphere of eq.~\eqref{eq:spheresources}, let $\omega=(\dd\psi+\cos\theta\,\dd\varphi)/2$ and use
$B_{\mathrm{top}}=\lambda\omega$.
Choose the orientation for which $\star\dd\omega=2\omega$ at the round
metric. Then $\int\omega\wedge\dd\omega=4\pi^2$.
The photon component $a=t\omega$ consequently has coupling
\begin{equation}
 S_{\mathrm{BF}}=2\pi\ii\lambda t.
 \label{eq:topBFmode}
\end{equation}
This integral is metric independent, including away from the round sphere.

Let $\Gamma(\varepsilon,t)$ be the one-particle-irreducible photon
effective action on this mode, with the mean auxiliary field eliminated.
The residual sphere isometries make $\omega$ an isolated invariant mode.
Its quadratic coefficient has the large-flavor expansion
\begin{equation}
 [t^2]\Gamma(\varepsilon,t)
 =N\left(\frac{\pi^2}{8}+s\frac{\pi^2}{60}\varepsilon\right)
 +u_0+\varepsilon u_1+O(\varepsilon^2,N^{-1}).
 \label{eq:topGamma}
\end{equation}
Here $s=+1$ for fermions and $s=-1$ for scalars, and $N$ is the
corresponding $N_f$ or $N_b$.
The leading coefficients follow from eq.~\eqref{eq:freeTaylor}
with unit charge for every flavor.
The order-one coefficients $u_0,u_1$ include all fluctuation sectors.

The connected photon variance is the inverse of the Hessian of $\Gamma$.
Differentiating the free energy twice with respect to the BF source
amplitude $\lambda$ therefore gives
\begin{equation}
 F^{\mathrm{top}}_{\lambda\lambda}(\varepsilon,0)
 =4\pi^2\langle t^2\rangle_{\varepsilon,c}
 =\frac{4\pi^2}{\Gamma_{tt}(\varepsilon,0)}
 =\frac{2\pi^2}{[t^2]\Gamma(\varepsilon,t)}.
 \label{eq:topBFresponse}
\end{equation}
This response supplies the denominator of the collider ratio and
reproduces the normalization in eq.~\eqref{eq:topcentralcharges}.

One metric derivative measures the topological three-point function.
Because the BF coefficient is metric independent, the derivative acts
only on the inverse photon Hessian:
\begin{align}
 \gamma_{\mathrm{top}}
 &=\frac58\frac{F^{\mathrm{top}}_{\varepsilon\lambda\lambda}(0,0)}
 {F^{\mathrm{top}}_{\lambda\lambda}(0,0)}
 =-\frac58\frac{[\varepsilon t^2]\Gamma}{[t^2]\Gamma}\nonumber\\
 &=-\frac58\frac{sN\pi^2/60+u_1+O(N^{-1})}
 {N\pi^2/8+u_0+O(N^{-1})}\nonumber\\
 &=-\frac{s}{12}-\frac5{\pi^2N}
 \left(u_1-s\frac{2u_0}{15}\right)+O(N^{-2}).
 \label{eq:topgammaextraction}
\end{align}
This exchanges the scalar and fermion collider endpoints at leading order.
At the next order it requires the metric response of the full photon
effective action, not just the flavor-current response.

\subsubsection{Photon--auxiliary mixing}

The additional contribution arises because every matter field couples
with the same gauge charge. The mixed photon--auxiliary Hessian is
therefore linear in $t$ and survives the flavor sum.
The same term vanishes for the source coupled to the traceless flavor generator of section~\ref{sec:sphere}.
For critical scalar QED$_3$ and QED$_3$--GNY, with the latter defined in
eq.~\eqref{eq:fermionaction}, write the fluctuation Hessian as
\begin{equation}
 H(\varepsilon,t)=
 \begin{pmatrix}
 K(\varepsilon)+t^2K_{(2)}(\varepsilon)&\ii tR(\varepsilon)\\
 \ii tR(\varepsilon)^\dagger&B(\varepsilon)+t^2B_{(2)}(\varepsilon)
 \end{pmatrix}+O(t^3).
 \label{eq:topmixedHessian}
\end{equation}
Here $K$ is the photon kernel restricted to transverse one-forms,
which represent gauge-field configurations modulo gauge transformations.
The kernel $B$ is the auxiliary-field kernel.
The factors $\ii$ follow from the scalar integration contour or from
the Yukawa insertion in the convention for the Dirac operator explained
after eq.~\eqref{eq:DiracSphere}.
Expanding $\tfrac12\Tr\log H$ gives the mixed contribution
\begin{equation}
 u_{\mathrm{mix}}(\varepsilon)
 =\frac12\Tr\left[B(\varepsilon)^{-1}R(\varepsilon)^\dagger
 K(\varepsilon)^{-1}R(\varepsilon)\right].
 \label{eq:topmixtrace}
\end{equation}
Its positive sign comes from the two factors of $\ii$ and the negative
quadratic term in the matrix logarithm.

The round-sphere trace can be summed analytically.
For scalar harmonics with $n=1,2,\ldots$, let $P_{n,+}$ project
$\omega Y$ onto the transverse curl branch with eigenvalue $n+1$.
The normalized kernels on this branch are
\begin{equation}
\begin{array}{c|cc}
 &\text{critical scalar}&\text{Gross--Neveu}\\\hline
 B_0(n)&1/(8n)&n/8\\
 R_0Y&-P_{n,+}(\omega Y)/[2n(n+2)]&-P_{n,+}(\omega Y)/[2(n+2)]
\end{array},\qquad K_0\big|_{P_{n,+}}=\frac{n+1}{16}.
 \label{eq:toproundmixedkernels}
\end{equation}
For each of the $n$ left magnetic states,
$\sum_m\|P_{n,+}(\omega Y_{nm})\|^2=(n+2)/3$.
This identity follows by summing over three orthogonal unit one-forms
and taking the trace of the spin-one projector.
Inserting the scalar kernels into eq.~\eqref{eq:topmixtrace} gives
\begin{align}
 u_{\mathrm{mix}}(0)
 &=\frac12\sum_{n=1}^{\infty}
 n(8n)\frac{16}{n+1}\frac1{4n^2(n+2)^2}\frac{n+2}{3}\nonumber\\
 &=\frac{16}{3}\sum_{n=1}^{\infty}\frac1{(n+1)(n+2)}
 =\frac{16}{3}\sum_{n=1}^{\infty}
 \left(\frac1{n+1}-\frac1{n+2}\right)=\frac83.
 \label{eq:topmixroundsum}
\end{align}
The fermion kernels give the same summand: the change in $B_0^{-1}$
cancels the two additional factors of $n$ in $R_0$.

The first metric derivatives require the perturbed kernels.
Appendix~\ref{app:topological} gives their construction and the derivative
of eq.~\eqref{eq:topmixtrace}.
The converged sums are
\begin{equation}
 [\varepsilon]u_{\mathrm{mix},b}\simeq-0.6290769358751701,\qquad
 [\varepsilon]u_{\mathrm{mix},f}\simeq 0.0957436025418368.
 \label{eq:topmixmetricresults}
\end{equation}
These are numerical metric-response calculations.
They are not inferred from the round value $8/3$.

The diagonal terms are the same source responses tabulated in
table~\ref{tab:mixedsphere}, now for a mean photon rather than a
source coupled to the traceless flavor generator.
In the following sums, the labels $A$ and $\sigma$ select the photon
and auxiliary rows, and $b,f$ select scalar or fermion matter.
Combining these rows with eqs.~\eqref{eq:topmixroundsum} and
\eqref{eq:topmixmetricresults} gives
\begin{align}
 u_{0,\mathrm{QED}_3}&=f_{02,A,f},&
 u_{1,\mathrm{QED}_3}&=f_{12,A,f},\nonumber\\
 u_{0,\mathrm{sQED}_3}&=f_{02,A,b}+f_{02,\sigma,b}+\frac83,&
 u_{1,\mathrm{sQED}_3}&=f_{12,A,b}+f_{12,\sigma,b}
 +[\varepsilon]u_{\mathrm{mix},b},\nonumber\\
 u_{0,\mathrm{QED}_3\text{-}\mathrm{GNY}}
 &=f_{02,A,f}+f_{02,\sigma,f}+\frac83,&
 u_{1,\mathrm{QED}_3\text{-}\mathrm{GNY}}
 &=f_{12,A,f}+f_{12,\sigma,f}+[\varepsilon]u_{\mathrm{mix},f}.
 \label{eq:topsectorsums}
\end{align}
There is no auxiliary field in QED$_3$, so its mixed term is absent.

Substituting these sector sums into eq.~\eqref{eq:topgammaextraction}
reproduces the three topological collider parameters in
eq.~\eqref{eq:topresults}. In combination with the six flavor-current
and stress-tensor comparisons above, this gives independent sphere
checks of all nine corrections in table~\ref{tab:collider}.

\section{Discussion}\label{sec:discussion}

We have determined the first subleading large-flavor corrections to the parity-even correlators $\TJJ$, $\langle Tj_{\mathrm{top}}j_{\mathrm{top}}\rangle$, and $\TTT$ in QED$_3$, QED$_3$--GNY, and critical scalar QED$_3$. Flat-space energy-flux projections give analytic expressions for the conformal collider parameters $\gamma$, $\gamma_{\mathrm{top}}$, and $t_4$, which agree with an independent calculation from squashed-sphere free-energy derivatives. Combined with the known two-point normalizations, these results fix the coefficients of both independent parity-even tensor structures in each correlator. The resulting data provide input for quantum-critical transport and conformal bootstrap studies involving conserved currents and the stress tensor.

For transport, $\TJJ$ fixes the stress-tensor contribution to the
high-frequency conductivity, while $\TTT$ enters the shear response
and spectral sum rules
\cite{ChowdhuryTransport,KatzEtAl,ChowdhuryDavidPrakash}.
The correlator $\langle Tj_{\mathrm{top}}j_{\mathrm{top}}\rangle$
similarly fixes the stress-tensor contribution to the conductivity of the
topological current, whose charge measures the magnetic flux of the
dynamical gauge field. Under particle--vortex duality this current can represent
an ordinary global current in the dual description
\cite{SeibergDualityWeb}. The new coefficients thus provide universal
input for both flavor-current and topological-current response.
Thermal one-point functions and low-dimension scalar contributions
are also needed to determine the high-frequency expansion
\cite{KatzEtAl,LucasGazitPodolskyWitczak,LucasSierensWitczak}; the dc
conductivity requires further dynamical information
\cite{DamleSachdevTransport}.

The DQCP and spin-liquid applications give concrete targets for these
observables. The proposed $SO(5)$ DQCP relates $N_f=2$ QED$_3$--GNY
to $N_b=2$ noncompact scalar QED$_3$, whereas $N_f=2$ QED$_3$ is
associated with the easy-plane DQCP \cite{WangDQCP,BoyackQEDGN}.
The topological current probes the VBS phase-rotation symmetry in the
$NCCP^1$ description, and flavor and topological currents can be
exchanged by duality. Comparing their three-point functions requires
the correct current map and normalization. At $N_f=4$, QED$_3$ is a
candidate description of Dirac spin liquids
\cite{HermeleSpinLiquid,SongSpinLiquids,HuDiracSpinLiquid} and appears
in proposed theories of underdoped and phase-disordered cuprates
\cite{RantnerWenCuprates,FranzTesanovicVafekQED3}.
Connecting our continuum data to a material additionally requires
identifying its physical current and relevant perturbations.

For the conformal bootstrap, the independent conformal collider parameters supply
information beyond bounds on $C_J$, $C_J^{\mathrm{top}}$, and $C_T$.
Scalar and monopole bootstrap studies already constrain DQCP and QED$_3$ candidates
\cite{ChesterPufuQEDBootstrap,LiDQCPBootstrap,LiPolandGauge,AlbayrakQEDBootstrap,ChesterSuDQCP,LiShenDQCP}.
With currents as external operators, stress-tensor exchange in $J\times J$
also depends on $\gamma$
\cite{DymarskyCurrentBootstrap,ReehorstMixed,HeNonAbelianCurrents}.
Adding external stress tensors introduces
the $\TTT$ structures and $t_4$
\cite{DymarskyStressBootstrap,ChangEtAl,ErramilliMitchellStress,DommesEtAl}.
Mixed monopole--current systems would similarly use
$\gamma_{\mathrm{top}}$ in the
$j_{\mathrm{top}}\times j_{\mathrm{top}}$ channel, supplementing the
OPE coefficients fixed by charge Ward identities. Our results can guide large-flavor scans
and provide benchmarks for nonperturbative bounds. The good agreement
between our previous $O(N)$ predictions extrapolated to $O(2)$ and
the bootstrap region under weaker spectral assumptions is encouraging
\cite{GuoLiThreePoint,DommesEtAl}. Nevertheless, the large-$N$ result neither provides a controlled error estimate for gauge theories at small flavor number nor establishes the proposed DQCP dualities.

A natural extension is to include Chern--Simons couplings. For an
Abelian gauge field, the Euclidean action contains
$S_{\mathrm{CS}}=\frac{\ii k}{4\pi}\int a\wedge\dd a$,
where $k$ is the Chern--Simons level.
A useful large-flavor limit holds $k/N$ fixed, with $N=N_f$ or $N_b$,
so that the Chern--Simons term and the induced matter polarization
both enter the leading photon kernel \cite{ChesterCSMonopoles}.
The resulting propagator contains parity-even and parity-odd parts
and can be incorporated into the emission amplitudes and fluctuation
determinants. Although the Chern--Simons action is metric independent,
it changes the photon propagator and therefore the metric response
of the interacting theory.

The extension can also extract the additional parity-odd
separated-point structures allowed in $\TJJ$ and $\TTT$.
The energy-flux calculation should determine the full polarization
matrix, including interference between orthogonal polarizations.
These matrix elements probe the parity-odd coefficients, which obey
joint positivity bounds with the parity-even collider parameters
\cite{ChowdhuryParityCollider}.
An independent sphere extraction would require source profiles
sensitive to these additional structures and the parity-sensitive
part of the complex generating functional. The real free-energy
derivatives used here would not by themselves determine all the new
data. Local background Chern--Simons contact terms must also be
separated from the nonlocal three-point structures
\cite{ClossetContactTerms}.

Existing Chern--Simons--matter correlators and transport calculations
provide useful benchmarks
\cite{AharonyBosonization,GurAriYacobyCS,GurAriHartnollMahajan,GhoshMazumdarThermal},
although their limit of large gauge-group rank differs from our
Abelian large-flavor expansion. Extending the present calculation would
provide vacuum OPE data relevant to longitudinal and Hall response
near parity-breaking critical points, including gauge-theory
descriptions of chiral spin liquids
\cite{GuoThermalHall,MaityThermalHall}.
Quantitative finite-temperature transport would then require the
corresponding thermal calculation, including energy-magnetization
contributions for thermal Hall response.
\acknowledgments The research of Yongwei Guo and Zhijin Li was supported by the Startup Funding of Southeast University (No. 4007022314), the Southeast University Global Engagement of Excellence Fund (No. 3360682401E), and the National Natural Science Foundation of China (No. 12375061). The research of Ke Ren was supported by the National Natural Science Foundation of China (NSFC Grant No. 12405107). 

We acknowledge the assistance of ChatGPT Codex in the computations and the preparation of the manuscript draft.
\clearpage
\appendix
\section{Lorentzian cuts and detector normalization}
\label{app:phase-space}

\subsection*{Lorentzian detector and resolved cuts}\label{app:lorentzian-cuts}

We use the perturbative Schwinger--Keldysh detector prescription of
ref.~\cite[section~3.1.2]{CaronHuotDetectors}. Prepare the source
$\mathcal O(q)$ with a smooth wave packet near a future timelike
momentum $q$ and test the detector against a smooth angular weight $w$.
The two branches of the contour carry the time-ordered production
amplitude and its conjugate. Lines across the fold are positive-energy
Wightman functions. In the large-flavor expansion we resolve these
lines into cuts of elementary matter propagators.

To see the detector insertion in this organization, use light-cone
coordinates $x^\pm=t\pm\boldsymbol n\cdot\boldsymbol x$ and a null
axial gauge $a_-=0$. For this intermediate description one may take
the causal Mandelstam--Leibbrandt prescription for the axial pole
and impose the Ward identities before removing the regulator. The integrated
outgoing null component $T_{--}$ of the critical action then contains
the kinetic matter bilinear. Terms proportional to $g_{--}$, including
the scalar potential and Yukawa potential, vanish. The scalar
improvement is proportional to
$\partial_-^2(\sum_i\phi_i^*\phi_i)$ and integrates to a boundary term
that vanishes in the smeared matrix element. The bare Maxwell and
auxiliary kinetic operators are irrelevant at the fixed point.
If one works with the induced action, the metric variation at the
detector represents the corresponding matter insertion inside the
kernel. Metric variations at the production source still give the
mixed and induced vertices used in section~\ref{sec:stress}.

On a cut matter line of energy $\omega$, the Fourier phase at the
detector is $\exp[-\ii\omega u+\ii r\omega(\cos\chi-1)]$, where
$\chi$ is the angle from $\boldsymbol n$. At fixed $u$, the outgoing
stationary point is $\chi=0$. The antipodal term oscillates with
$2r\omega$ and drops out under energy smearing away from zero.
Integrating over $u$ pairs equal positive energies across the
bilinear. The terms with two creation or two annihilation modes
have no support at positive energies. With
$\int_{S^1}\dd\theta\,\mathcal E(\theta)=H$, this gives the one-line
weight $\omega w(\theta_k)$. Inserting the free cut basis at the
fold gives, in the sharp timelike momentum limit, the unnormalized
detector numerator in this perturbative prescription,
\begin{equation}
 \sum_X\int\dd\Phi_X(q)\,
 |\mathcal M_{\mathcal O\to X}|^2
 \sum_{a\in X} k_a^0\,w(\theta_a).
 \label{eq:resolved-detector}
\end{equation}
Here the sum includes flavors and spins and the phase space carries
the appropriate identical-particle factors. The sharp angular
detector uses $w(\theta)=\delta_{S^1}(\theta-\theta_{\boldsymbol n})$.

For a dressed line, let $K_R$ and $K_A$ denote the retarded and
advanced inverse propagators on the physical components. Then
\begin{equation}
 D_R-D_A=-D_R(K_R-K_A)D_A .
 \label{eq:induced-discontinuity}
\end{equation}
At leading order $K_R-K_A$ is the discontinuity of the matter bubble.
The induced propagators have a continuum at timelike momentum and
no isolated one-particle pole. At positive energy the Wightman
propagator is given by this discontinuity with its spectral
normalization. Inserting the two-body representation of the bubble
discontinuity therefore replaces a cut dressed line by a matter pair
with a dressed propagator on each side of the fold. Assigning the
detector weight after this replacement includes the energy carried by
each member of that pair. The Ward contractions in sections~\ref{sec:current}
and~\ref{sec:stress} eliminate longitudinal photon factors from the
production amplitudes.

For power counting, collapse each open matter chain and each closed
matter loop in a connected production graph to a vertex. A graph with
$m$ outgoing pairs and $\ell$ closed matter loops needs at least
$m+\ell-1$ dressed lines. Loop flavor sums contribute at most
$N^\ell$, and the dressed lines give at most $N^{-(m+\ell-1)}$.
The resulting amplitude at fixed outgoing flavors is of order
$N^{1-m}$ or smaller, with the source flavor factor understood.
The $N^m$ final flavor sum and the $O(N)$ leading singlet norm give
at most $N^{1-m}$ in the normalized flux. For the traceless flavor
current, the flavor sum and norm each lose one power of $N$.
The common BF and external photon factors cancel in the topological
ratio, leaving the singlet counting. For $m=2$, the leading graphs
are trees in this counting.
After absorbing matter bubbles into the propagators and tuning
tadpoles, they consist of emission and mixed source vertices on an
open chain, and induced source vertices connecting two dressed lines.
Linear source mixing with one dressed line vanishes by the flavor,
charge-conjugation or stress-tensor Ward identities, apart from the
external BF coupling.

The Feynman boundary values belong to the production branch.
For the hatted momenta of section~\ref{sec:flavor-projection}, use
$p_E=z\widehat p$ with $z=\exp[-\ii(\pi/2-0^+)]$ for $p=q,P,Q$.
In the interior of the phase space, the bubble and triangle
Feynman-parameter denominators are $z^2$ times positive combinations
of $q^2$, $P^2$ and $Q^2$. This fixes the square-root branch and
continues every homogeneous tensor integral with a common momentum
rotation. For example, the $Taa$ vertex has degree one and each
photon propagator has degree minus one. The induced term with two
propagators therefore has the same degree as the corresponding
emission term. For the $Taa$ and $T\sigma\sigma$ channels, and for
the topological amplitude in eq.~\eqref{eq:topcoherentamp}, the
open, contact and induced terms agree in degree within each coherent
sum. External spinors and polarizations are common to that sum.
Thus the rotation supplies only a common phase. The relative signs
follow from the source vertices and the variation
$\delta D=-D(\delta K)D$. The conjugate branch uses the conjugate
Feynman prescription.

At fixed regulator the tree two-matter form factor vanishes at
the selected angle, so its interference with the one-loop form
factor vanishes before any product with an infrared pole is taken.
The full projected four-matter squares are nonnegative. The finite
combined radial integrals in appendix~\ref{app:integrals} and
section~\ref{sec:topflat} then permit removal of the phase-space
excisions by monotone convergence applied to the full squares.
This establishes the finite selected coefficient within the
perturbative cut prescription. An identification independent of that
prescription would also require uniform control of the null limit at
unresolved endpoints of the regulated Wightman diagrams. The
squashed-sphere comparison gives a check on the selected coefficient
that uses a Euclidean calculation.

\subsection*{Four-particle phase space and flux}

We use the source momentum $q=(1,0,0)$ and the pair masses in
eq.~\eqref{eq:pairkinematics}. Let $E_P=(1+x^2-y^2)/2$,
$E_Q=1-E_P$, and $K=\sqrt{E_P^2-x^2}$.
The inequality $x+y\leq1$ makes $K$ real.
Choose the first pair momentum along the first spatial axis.
Boosting the pair decays from their rest frames then gives
\begin{align}
 k_1&=\tfrac12(E_P+K\cos\alpha,\ K+E_P\cos\alpha,\ x\sin\alpha),
 &k_2&=(E_P,K,0)-k_1,\nonumber\\
 k_3&=\tfrac12(E_Q-K\cos\beta,\ -K+E_Q\cos\beta,\ y\sin\beta),
 &k_4&=(E_Q,-K,0)-k_3.
 \label{eq:fourmomenta}
\end{align}
The angles range over $0\leq\alpha,\beta<2\pi$.
These momenta obey $k_a^2=0$, $\sum_a k_a=q$, and
$\sum_a k_a^0=1$.
For any function of the two decay angles we use
\begin{equation}
 \langle f\rangle_{\alpha,\beta}=
 \int_0^{2\pi}\frac{\dd\alpha}{2\pi}
 \int_0^{2\pi}\frac{\dd\beta}{2\pi}f(\alpha,\beta).
 \label{eq:angleaverage}
\end{equation}

For these kinematics we use the Lorentz-invariant measure
\begin{equation}
 \dd\Phi_m(q;k_1,\ldots,k_m)=
 (2\pi)^3\delta^{(3)}\!\left(q-\sum_{a=1}^mk_a\right)
 \prod_{a=1}^m\frac{\dd^2\boldsymbol k_a}{(2\pi)^2\,2k_a^0}.
 \label{eq:phasedefinition}
\end{equation}
For a two-body decay of a state of mass $M$, the radial delta function
gives $\dd\Phi_2=\dd\theta/(8\pi M)$ in three spacetime dimensions.
This expression also holds when the two daughter masses are nonzero,
within their allowed kinematic range.
Factorization into two pair decays therefore gives
\begin{align}
 \dd\Phi_4
 &=\frac{(2x\dd x)(2y\dd y)}{(2\pi)^2}
 \frac{\dd\theta_0}{8\pi}
 \frac{\dd\alpha}{8\pi x}\frac{\dd\beta}{8\pi y}\nonumber\\
 &=\frac{\dd x\,\dd y}{64\pi^2}
 \frac{\dd\theta_0}{2\pi}
 \frac{\dd\alpha}{2\pi}\frac{\dd\beta}{2\pi}.
 \label{eq:phasefactorization}
\end{align}
Here $\theta_0$ is the overall spatial rotation of the two pairs.
Rotational invariance lets us perform this integral by rotating the
source polarization relative to each detected particle.
The resulting measure is eq.~\eqref{eq:phasespace}.

\subsubsection*{Flavor counting and the selected flux}

For the normalized flavor current, the leading sum over distinct flavors
is $\sum_iq_i^2\sum_{j\ne i}1=N-1$.
The squared photon amplitude supplies $256/N^2$.
Their product is $256/N+O(N^{-2})$.
The free two-particle norm is $1/8$, so the selected current flux in
units of its angular average is
\begin{equation}
 \frac{256/N}{1/8}\frac{I_{J,a}}{64\pi^2}
 =\frac{32I_{J,a}}{\pi^2N},\qquad a=f,b.
 \label{eq:current-cut-normalization}
\end{equation}
It equals twice the opposite structure divided by the total current
normalization. Moving the flavor current to the other pair generates
an interference flavor sum $\sum_{i\ne j}q_iq_j=-1$.
That interference enters one order later than eq.~\eqref{eq:current-cut-normalization}.

The stress tensor is a flavor singlet and can create either pair.
Both pair attachments therefore occur in the amplitude, while the
unordered distinct-flavor sum is $N(N-1)/2$.
The squared prefactor gives $[N(N-1)/2](16/N)^2=128+O(N^{-1})$.
The leading norm of the state created by the stress tensor is $N/32$.
Consequently the selected energy flux in the stress-tensor state is
\begin{equation}
 \frac{128}{N/32}\frac1{64\pi^2}
 \int_0^1\dd x\int_0^{1-x}\dd y\,\mathcal S_a(x,y)
 =\frac{64}{\pi^2N}
 \int_0^1\dd x\int_0^{1-x}\dd y\,\mathcal S_a(x,y).
 \label{eq:stress-cut-normalization}
\end{equation}
The weight $\mathcal S_f$ is specified in eq.~\eqref{eq:stressweight}.
Its scalar counterpart uses
$-\sin(2\theta_a)\mathcal M_++\cos(2\theta_a)\mathcal M_\times$
inside the absolute square.
The selected flux equals twice the opposite stress-tensor structure divided
by $N$ at this order. Thus the opposite structure is
$32\int\mathcal S_a/\pi^2$, as used in section~\ref{sec:stress}.

\subsubsection*{Angular reduction}

For a detected particle, $\cos\theta_a=k_a^1/k_a^0$ and
$\sin\theta_a=k_a^2/k_a^0$.
Substitution of eq.~\eqref{eq:fourmomenta} expresses every detector
weight as a rational function of the decay-angle sines and cosines.
The denominators reduce to powers of $a-b\cos^2\alpha$ and their
$\beta$ counterparts. Partial fractions and the angular master in
eq.~\eqref{eq:angularmaster} then perform both angular integrations.
The remaining pair-mass domain is the triangle $x,y\geq0$, $x+y\leq1$.
The variables $u=x+y$, $v=x-y$ turn it into
$0\leq u\leq1$, $-u\leq v\leq u$.
Integrating the rational dependence on $v$ gives the one-dimensional
integrals evaluated in appendix~\ref{app:integrals}.

\section{Analytic loop and phase-space integrals}\label{app:integrals}

\subsection*{Loop integrals and tensor reduction}

The scalar convolution integral in three dimensions is
\begin{equation}
 \intk\frac1{(k^2)^a[(p-k)^2]^b}
 =\frac{(p^2)^{3/2-a-b}}{(4\pi)^{3/2}}
 \frac{\Gamma(a+b-3/2)\Gamma(3/2-a)\Gamma(3/2-b)}
 {\Gamma(a)\Gamma(b)\Gamma(3-a-b)}.
 \label{eq:bubblemaster}
\end{equation}
This equation first holds in its convergence domain.
Its meromorphic continuation specifies the analytic regulator.
The finite one-loop reductions use
\begin{equation}
 \intk\frac1{k^2(k-p)^2}=\frac1{8|p|},\qquad
 \intk\frac1{k^2(k-P)^2(k-q)^2}=
 \frac1{8|P||q-P||q|}.
 \label{eq:trianglemasters}
\end{equation}
Differences of denominators reduce contractions of the loop momentum
with external vectors. Rotational invariance then reduces the remaining
transverse moments. This procedure supplies the triangle and contact
bubble terms in eqs.~\eqref{eq:inducedfermion} and~\eqref{eq:inducedscalar}.

\subsection*{Phase-space integrals for photon exchange}

The angular integrals use
\begin{equation}
 \int_0^{2\pi}\frac{\dd\alpha}{2\pi}
 \frac1{a-b\cos^2\alpha}=\frac1{\sqrt{a(a-b)}},\qquad a>b\geq0.
 \label{eq:angularmaster}
\end{equation}
Differentiation in $a$ gives higher denominator powers.
Polynomial moments use
$\langle\cos^{2j}\alpha\rangle=\binom{2j}{j}/4^j$.
Partial fractions then reduce every decay-angle term to these integrals.

\subsubsection*{Current integrals}

For the pair-mass variables $u=x+y$ and $v=x-y$ introduced in
appendix~\ref{app:phase-space}, the domain is $0<u<1$, $-u<v<u$.
The Jacobian is
$\dd x\,\dd y=\dd u\,\dd v/2$.
The relative-mass integration gives
\begin{align}
 I_{J,f}&=\int_0^1\dd u\left[
 \frac{13u^7-20u^6-178u^5-324u^4-179u^3+12u^2+48u+12}
 {12u(u+1)^4}\right.\nonumber\\
 &\hspace{27mm}\left.-\frac72\log\frac{1-u}{1+u}
 +\frac{(u^2+1)^2}{2u^3}\log\frac{1-u^2}{1+u^2}\right],
 \label{eq:IJfradial}\\
 I_{J,b}&=\int_0^1\dd u\left[
 \frac{7u^7-20u^6-86u^5-52u^4+7u^3+60u^2+48u+12}
 {12u(u+1)^4}\right.\nonumber\\
 &\hspace{27mm}\left.-\frac52\log\frac{1-u}{1+u}
 +\frac{(u^2+1)^2}{2u^3}\log\frac{1-u^2}{1+u^2}\right].
 \label{eq:IJbradial}
\end{align}
\paragraph*{Primitives and endpoint terms}

The two integrals in eqs.~\eqref{eq:IJfradial} and~\eqref{eq:IJbradial}
have the same coefficient of the second logarithm.
A primitive of this coefficient is
\begin{equation}
 H_J(u)=\frac{u^2}{4}+\log u-\frac1{4u^2},\qquad
 H_J'(u)=\frac{(u^2+1)^2}{2u^3}.
 \label{eq:HJprimitive}
\end{equation}
The primitive vanishes at $u=1$.
Its products with the logarithm obey
\begin{align}
 \lim_{u\to1^-}H_J(u)\log\frac{1-u^2}{1+u^2}&=0,
 &\lim_{u\to0^+}H_J(u)\log\frac{1-u^2}{1+u^2}&=\frac12.
 \label{eq:Jendpoints}
\end{align}
These limits fix the boundary term in integration by parts.
At a lower cutoff $\delta>0$, integration by parts gives
\begin{align}
 &\int_\delta^1\dd u\,\frac{(u^2+1)^2}{2u^3}
 \log\frac{1-u^2}{1+u^2}
 =\left[H_J(u)\log\frac{1-u^2}{1+u^2}\right]_\delta^1
 +\int_\delta^1\dd u\,\frac{4uH_J(u)}{1-u^4}
 \nonumber\\
 &=\left[H_J(u)\log\frac{1-u^2}{1+u^2}\right]_\delta^1
 +\int_\delta^1\dd u\left[-\frac1u+
 \frac{4u\log u}{1-u^4}\right].
 \label{eq:Jparts}
\end{align}
The last equality substitutes eq.~\eqref{eq:HJprimitive}.
Its $-1/u$ term cancels the singular part of each rational integrand.
The primitives of the resulting regular rational functions are
\begin{align}
 F_{J,f}(u)={}&\frac{13u^3}{36}-3u^2+\frac{8u}{3}
 -\frac9{u+1}+\frac{10}{3(u+1)^2}-\frac2{3(u+1)^3}
 -7\log(u+1),\nonumber\\
 F_{J,b}(u)={}&\frac{7u^3}{36}-2u^2+\frac{16u}{3}
 +\frac5{u+1}-\frac{10}{3(u+1)^2}+\frac2{3(u+1)^3}
 -5\log(u+1).
 \label{eq:Jrationalprimitives}
\end{align}
Differentiating each line gives the rational term in its original
integrand minus $1/u$.
Their endpoint differences are therefore
\begin{align}
 \int_0^1 F_{J,f}'(u)\,\dd u
 &=F_{J,f}(1)-F_{J,f}(0)=\frac{47}{18}-7\log2,\nonumber\\
 \int_0^1 F_{J,b}'(u)\,\dd u
 &=F_{J,b}(1)-F_{J,b}(0)=\frac{53}{18}-5\log2.
 \label{eq:Jrationalendpoints}
\end{align}
The remaining logarithms can be integrated independently.
For the logarithm linear in $u$, the primitive gives
\begin{equation}
 \int_0^1\dd u\,\log\frac{1-u}{1+u}
 =\left[-(1-u)\log(1-u)-(1+u)\log(1+u)\right]_0^1
 =-2\log2.
 \label{eq:Jlinearlog}
\end{equation}
For the other term, $t=u^2$ gives
\begin{align}
 \int_0^1\dd u\,\frac{4u\log u}{1-u^4}
 &=\int_0^1\dd t\,\frac{\log t}{1-t^2}
 =\sum_{n=0}^{\infty}\int_0^1\dd t\,t^{2n}\log t
 \nonumber\\
 &=-\sum_{n=0}^{\infty}\frac1{(2n+1)^2}
 =-\left(1-\frac14\right)\zeta(2)=-\frac{\pi^2}{8}.
 \label{eq:Jquadraticlog}
\end{align}
The geometric series has nonnegative terms after multiplying by
$-\log t$, so monotone convergence justifies termwise integration.
Substitution of these integrals and the endpoint terms gives
\begin{align}
 I_{J,f}
 &=-\frac12+\frac{47}{18}-7\log2
 -\frac72(-2\log2)-\frac{\pi^2}{8}
 =\frac{19}{9}-\frac{\pi^2}{8},\nonumber\\
 I_{J,b}
 &=-\frac12+\frac{53}{18}-5\log2
 -\frac52(-2\log2)-\frac{\pi^2}{8}
 =\frac{22}{9}-\frac{\pi^2}{8}.
 \label{eq:Jcompleteevaluation}
\end{align}
These equalities evaluate the two starting integrals, including their
canceling lower-endpoint terms.

\subsubsection*{Stress-tensor integrals}

Use $I_{T,f}$ and $I_{T,b}$ for twice the physical mass integrals of the
photon weights for the stress-tensor source. The factor two accounts for the Jacobian
convention in the relative-mass integration. Their integral representations
are
\begin{equation}
 I_{T,a}=\int_0^1\dd u\left[\mathcal R_a(u)
 +P(u)\log\frac{1-u}{1+u}
 +Q(u)\log\frac{1-u^2}{1+u^2}\right],\qquad a=f,b,
 \label{eq:ITdefinition}
\end{equation}
where
\begin{align}
 \mathcal R_f(u)={}&\frac{1}{60u^2(u+1)^7}
 [15u^{12}+293u^{11}-2106u^{10}-2398u^9+35252u^8
 \nonumber\\
 &-57716u^7+17618u^6+78750u^5-32235u^4
 -44145u^3+9960u^2+2880u+360],\nonumber\\
 \mathcal R_b(u)={}&\frac{1}{60u^2(u+1)^7}
 [15u^{12}+233u^{11}-2646u^{10}-3218u^9+39592u^8
 \nonumber\\
 &-60176u^7+10878u^6+78370u^5-33135u^4
 -44265u^3+9960u^2+2880u+360],\nonumber\\
 P(u)={}&\frac1{8u^3(u+1)^4}
 [u^{11}+4u^{10}+12u^9+116u^8-1530u^7+4380u^6
 \nonumber\\
 &-8660u^5+8980u^4-4191u^3+128u^2+96u+24],\nonumber\\
 Q(u)={}&\frac{3u^4-2u^2+3}{u^3}.
 \label{eq:stressradial}
\end{align}
The logarithmic functions are identical in the two theories.
Their common primitives allow both integrals to be evaluated together.
\paragraph*{Primitives and endpoint terms}

The functions $P$ and $Q$ in eq.~\eqref{eq:stressradial} are common to
both matter theories.
Their primitives, chosen to vanish at $u=1$, are
\begin{align}
 H_P(u)={}&\frac{u^5}{40}+\frac{u^3}{4}+\frac{11u^2}{2}
 -\frac{1919u}{8}-\frac{17846}{15}
 +\frac{4000}{u+1}-\frac{2880}{(u+1)^2}
 +\frac{3488}{3(u+1)^3}\nonumber\\
 &-\frac3{2u^2}-2\log u+1440\log\frac{u+1}{2},\nonumber\\
 H_Q(u)={}&\frac{3u^2}{2}-2\log u-\frac3{2u^2}.
 \label{eq:stressprimitives}
\end{align}
Direct differentiation gives $H_P'=P$ and $H_Q'=Q$.
These derivatives put the scalar integral in the form
\begin{align}
 I_{T,b}
 =\lim_{\delta\to0^+}\bigg\{&
 \left[H_P(u)\log\frac{1-u}{1+u}
 +H_Q(u)\log\frac{1-u^2}{1+u^2}\right]_\delta^1
 \nonumber\\
 &+\int_\delta^1\dd u\left[
 \mathcal R_b(u)+\frac{2H_P(u)}{1-u^2}
 +\frac{4uH_Q(u)}{1-u^4}\right]\bigg\}.
 \label{eq:Tpartsregulated}
\end{align}
The boundary expression vanishes at the upper endpoint.
At the lower endpoint, its expansion is
\begin{equation}
 H_P(\delta)\log\frac{1-\delta}{1+\delta}
 +H_Q(\delta)\log\frac{1-\delta^2}{1+\delta^2}
 =\frac3\delta+3+O(\delta\log\delta).
 \label{eq:Tboundaryexpansion}
\end{equation}
This divergence cancels the $3/u^2$ term in the transformed integrand.
To show the cancellation explicitly, a primitive of its regular rational
part is
\begin{align}
 F_{T,b}(u)={}&\frac{u^4}{20}+\frac{32u^3}{45}
 -\frac{389u^2}{12}+\frac{1659u}{5}
 +\frac{5308}{15(u+1)}-\frac{16696}{3(u+1)^2}\nonumber\\
 &+\frac{34364}{5(u+1)^3}-\frac{12208}{3(u+1)^4}
 +\frac{1088}{(u+1)^5}-\frac{1280}{9(u+1)^6}.
 \label{eq:Trationalprimitive}
\end{align}
Substituting eq.~\eqref{eq:stressprimitives} and differentiating
this primitive gives the exact identity
\begin{align}
 \mathcal R_b(u)+\frac{2H_P(u)}{1-u^2}
 +\frac{4uH_Q(u)}{1-u^4}
 ={}&\frac3{u^2}+F_{T,b}'(u)
 +\left[-\frac{4u}{1+u^2}+\frac4{u-1}\right]\log u
 \nonumber\\
 &+1440\left[\frac1{u+1}-\frac1{u-1}\right]
 \log\frac{u+1}{2}.
 \label{eq:Tintegranddecomposition}
\end{align}
The integral of $3/u^2$ combines with eq.~\eqref{eq:Tboundaryexpansion}
as $-(3/\delta+3)+(3/\delta-3)=-6$ in the limit.
The remaining rational contribution is
\begin{equation}
 -6+\int_0^1F_{T,b}'(u)\,\dd u
 =-6+F_{T,b}(1)-F_{T,b}(0)
 =-6+\frac{53303}{45}=\frac{53033}{45}.
 \label{eq:Trationalendpoints}
\end{equation}
The first logarithmic contribution uses the same geometric-series
integrals as eq.~\eqref{eq:Jquadraticlog}:
\begin{align}
 \int_0^1\dd u\left[-\frac{4u}{1+u^2}+\frac4{u-1}\right]\log u
 &=-\int_0^1\dd t\,\frac{\log t}{1+t}
 -4\int_0^1\dd u\,\frac{\log u}{1-u}
 \nonumber\\
 &=\frac{\pi^2}{12}+\frac{2\pi^2}{3}=\frac{3\pi^2}{4}.
 \label{eq:Tfirstlog}
\end{align}
For the last contribution, the substitution $t=(1-u)/2$ in its second
term gives
\begin{align}
 &\int_0^1\dd u\left[\frac1{u+1}-\frac1{u-1}\right]
 \log\frac{u+1}{2}\nonumber\\
 &=\left[\frac12\log^2\frac{u+1}{2}\right]_0^1
 +\int_0^{1/2}\dd t\,\frac{\log(1-t)}t
 =-\frac{\log^22}{2}-\operatorname{Li}_2(1/2)
 =-\frac{\pi^2}{12}.
 \label{eq:Tsecondlog}
\end{align}
Here $\operatorname{Li}_2(z)=-\int_0^z\log(1-t)\,\dd t/t$.
Its reflection identity follows by differentiating
$\operatorname{Li}_2(z)+\operatorname{Li}_2(1-z)+\log z\log(1-z)$.
The derivative vanishes, and the limit at $z=0$ is $\zeta(2)$.
Evaluating the constant at $z=1/2$ gives the last equality in
eq.~\eqref{eq:Tsecondlog}.
Substituting all three contributions into eq.~\eqref{eq:Tpartsregulated}
therefore yields
\begin{align}
 I_{T,b}
 &=\frac{53033}{45}+\frac{3\pi^2}{4}
 +1440\left(-\frac{\pi^2}{12}\right)
 =\frac{53033}{45}-\frac{477\pi^2}{4},\nonumber\\
 I_{T,f}
 &=I_{T,b}+\int_0^1\dd u\,[\mathcal R_f(u)-\mathcal R_b(u)]
 =I_{T,b}+\frac23
 =\frac{53063}{45}-\frac{477\pi^2}{4}.
 \label{eq:Tcompleteevaluation}
\end{align}
The rational difference in the second line has primitive
\begin{align}
 &\frac{u^3}{3}+u^2-\frac{64u}{3}
 -\frac{384}{u+1}+\frac{576}{(u+1)^2}
 -\frac{1472}{3(u+1)^3}\nonumber\\
 &\qquad+\frac{224}{(u+1)^4}-\frac{128}{3(u+1)^5}.
 \label{eq:Tdifferenceprimitive}
\end{align}
Its endpoint difference is $2/3$.
Thus the original scalar and fermion stress-tensor integrals have both
been evaluated with their boundary terms included.

\subsection*{Auxiliary-scalar radial integrals}
\label{app:auxiliaryradial}

The auxiliary-scalar cuts have the same four-particle phase space as
the photon cuts. Their amplitudes use the auxiliary propagators in
eq.~\eqref{eq:sigmaprop}.
The different propagators change the radial integrands and their
normalization factors.

For the current cut, we remove the factor $1/2$ from each reduced
auxiliary emission amplitude. This multiplies its squared angular weight
by four. With this normalization, the relative-mass integral gives
\begin{align}
 I_{J,\sigma,f}={}&\int_0^1\dd u\left[
 \frac{13u^2}{6}-2u+\frac1u+
 \frac{(1-u^2)^2}{2u^3}\log\frac{1-u^2}{1+u^2}\right],\nonumber\\
 I_{J,\sigma,b}={}&\int_0^1\dd u\left[
 \frac{7u^2}{6}-2u+\frac1u+
 \frac{(1-u^2)^2}{2u^3}\log\frac{1-u^2}{1+u^2}\right].
 \label{eq:auxJradialdefinitions}
\end{align}
The common logarithmic coefficient has primitive
\begin{equation}
 H_{J,\sigma}(u)=\frac{u^2}{4}-\log u-\frac1{4u^2},\qquad
 H_{J,\sigma}'(u)=\frac{(1-u^2)^2}{2u^3}.
 \label{eq:auxJprimitive}
\end{equation}
Its logarithmic boundary product vanishes at $u=1$ and approaches
$1/2$ at $u=0$.
At a lower cutoff $\delta$, integration by parts therefore gives
\begin{align}
 I_{J,\sigma,f}
 &=\lim_{\delta\to0^+}\bigg\{
 \left[H_{J,\sigma}(u)\log\frac{1-u^2}{1+u^2}\right]_\delta^1
 +\int_\delta^1\dd u\left[
 \frac{13u^2}{6}-2u+\frac1u+
 \frac{4uH_{J,\sigma}(u)}{1-u^4}\right]\bigg\}
 \nonumber\\
 &=-\frac12+\int_0^1\dd u\left[
 \frac{13u^2}{6}-2u-\frac{4u\log u}{1-u^4}\right]
 =-\frac12+\frac{13}{18}-1+\frac{\pi^2}{8}
 =\frac{\pi^2}{8}-\frac79,\nonumber\\
 I_{J,\sigma,b}
 &=I_{J,\sigma,f}-\int_0^1u^2\,\dd u
 =I_{J,\sigma,f}-\frac13
 =\frac{\pi^2}{8}-\frac{10}{9}.
 \label{eq:auxJvalues}
\end{align}
The second line substitutes eq.~\eqref{eq:auxJprimitive} and cancels
$1/u$ before removing the cutoff.
Its logarithmic integral is the negative of eq.~\eqref{eq:Jquadraticlog}.
Restoring the factor from the emission amplitudes and the phase-space
normalization gives
\begin{align}
 (\nsJ)_\sigma
 &=\frac{4I_{J,\sigma,f}}{\pi^2N_f}
 =\frac1{N_f}\left(\frac12-\frac{28}{9\pi^2}\right),\nonumber\\
 (\nfJ)_\sigma
 &=\frac{4I_{J,\sigma,b}}{\pi^2N_b}
 =\frac1{N_b}\left(\frac12-\frac{40}{9\pi^2}\right).
 \label{eq:auxJstructureintegrals}
\end{align}
These expressions give the auxiliary current contributions in the
Dirac-flavor and complex-scalar conventions of this paper.

For the stress-tensor cut, write the auxiliary radial integral as
\begin{equation}
 I_{T,\sigma,a}=\int_0^1\dd u\left[
 \mathcal R_{\sigma,a}(u)
 +P_\sigma(u)\log\frac{1-u}{1+u}
 +Q_\sigma(u)\log\frac{1-u^2}{1+u^2}\right],\qquad a=f,b.
 \label{eq:auxTradialdefinition}
\end{equation}
The angular reduction gives
\begin{align}
 \mathcal R_{\sigma,f}(u)={}&\frac1{120u^3(u+1)^4}
 [30u^{10}+496u^9-2198u^8-2184u^7+7850u^6
 \nonumber\\
 &\qquad+1400u^5-6410u^4+1000u^3+600u^2+120u],\nonumber\\
 P_\sigma(u)={}&\frac{u^4}{8}-\frac{u^2}{4}+\frac{17u}{2}
 -\frac{479}{8}+\frac{176}{u+1}-\frac{144}{(u+1)^2}
 -\frac1u+\frac1{2u^3},\nonumber\\
 Q_\sigma(u)={}&\frac u2-\frac1u+\frac1{2u^3},\nonumber\\
 \mathcal R_{\sigma,b}(u)-\mathcal R_{\sigma,f}(u)
 ={}&-\frac{u(3u^5+15u^4+14u^3-34u^2+15u+3)}{3(u+1)^4}.
 \label{eq:auxTradialfunctions}
\end{align}
The difference in the last line is obtained by subtracting the scalar
and fermion angular integrands before their relative-mass integration.
This subtraction provides a separate check on the common logarithmic terms.

Primitives of those terms, vanishing at $u=1$, are
\begin{align}
 H_{P,\sigma}(u)={}&\frac{u^5}{40}-\frac{u^3}{12}
 +\frac{17u^2}{4}-\frac{479u}{8}+\frac{144}{u+1}\nonumber\\
 &-\frac1{4u^2}-\frac{241}{15}
 -\log u+176\log\frac{u+1}{2}.
 \label{eq:auxTprimitives}
\end{align}
The current primitive $H_{J,\sigma}$ in
eq.~\eqref{eq:auxJprimitive} has derivative $Q_\sigma$.
Integration by parts then gives
\begin{align}
 I_{T,\sigma,f}
 =\lim_{\delta\to0^+}\bigg\{&
 \left[H_{P,\sigma}(u)\log\frac{1-u}{1+u}
 +H_{J,\sigma}(u)\log\frac{1-u^2}{1+u^2}\right]_\delta^1
 \nonumber\\
 &+\int_\delta^1\dd u\left[
 \mathcal R_{\sigma,f}(u)+\frac{2H_{P,\sigma}(u)}{1-u^2}
 +\frac{4uH_{J,\sigma}(u)}{1-u^4}\right]\bigg\}.
 \label{eq:auxTparts}
\end{align}
The upper boundary is zero.
The lower boundary is $1/(2\delta)+1/2+O(\delta\log\delta)$.
Its pole cancels the integral of $1/(2u^2)$ in the transformed integrand.
The regular rational part has primitive
\begin{align}
 F_{T,\sigma,f}(u)={}&\frac{u^4}{20}+\frac{47u^3}{45}
 -\frac{967u^2}{60}+\frac{829u}{10}
 -\frac{1322}{15(u+1)}-\frac{112}{3(u+1)^2}
 -\frac8{3(u+1)^3}.
 \label{eq:auxTrationalprimitive}
\end{align}
Substitution into the transformed integrand gives
\begin{align}
 \mathcal R_{\sigma,f}(u)+\frac{2H_{P,\sigma}(u)}{1-u^2}
 +\frac{4uH_{J,\sigma}(u)}{1-u^4}
 ={}&\frac1{2u^2}+F_{T,\sigma,f}'(u)
 \nonumber\\
 &+\left[-\frac{2u}{1+u^2}+\frac2{u-1}\right]\log u
 \nonumber\\
 &+176\left[\frac1{u+1}-\frac1{u-1}\right]
 \log\frac{u+1}{2}.
 \label{eq:auxTdecomposition}
\end{align}
The boundary term and the $1/(2u^2)$ integral give $-1$ together.
The rational primitive gives
\begin{equation}
 -1+F_{T,\sigma,f}(1)-F_{T,\sigma,f}(0)=\frac{2543}{18}.
 \label{eq:auxTrationalendpoints}
\end{equation}
The two logarithmic integrals are respectively one half of
eq.~\eqref{eq:Tfirstlog} and 176 times eq.~\eqref{eq:Tsecondlog}.
For the scalar integral, the additional rational term gives
\begin{align}
 &\int_0^1\dd u\,[\mathcal R_{\sigma,b}(u)-\mathcal R_{\sigma,f}(u)]\nonumber\\
 &=\left[-\frac{u^3}{3}-\frac{u^2}{2}+\frac{16u}{3}
 +\frac{32}{u+1}-\frac{64}{3(u+1)^2}+\frac{16}{3(u+1)^3}\right]_0^1
 =-\frac16.
 \label{eq:auxTdifferenceprimitive}
\end{align}

Substituting these integrals into eq.~\eqref{eq:auxTparts} gives
\begin{align}
 I_{T,\sigma,f}
 &=\frac{2543}{18}+\frac{3\pi^2}{8}-\frac{44\pi^2}{3}
 =\frac{2543}{18}-\frac{343\pi^2}{24},\nonumber\\
 I_{T,\sigma,b}
 &=I_{T,\sigma,f}+\int_0^1\dd u\,
 [\mathcal R_{\sigma,b}(u)-\mathcal R_{\sigma,f}(u)]
 =I_{T,\sigma,f}-\frac16
 =\frac{1270}{9}-\frac{343\pi^2}{24}.
 \label{eq:auxTvalues}
\end{align}
The rational endpoint difference gives the second line of
eq.~\eqref{eq:auxTvalues}.

The norm of the state created by the stress tensor and the distinct-flavor pair sum give
\begin{align}
 (\nsT)_\sigma
 &=\frac8{\pi^2}I_{T,\sigma,f}
 =\frac{10172}{9\pi^2}-\frac{343}{3},\nonumber\\
 (\nfT)_\sigma
 &=\frac8{\pi^2}I_{T,\sigma,b}
 =\frac{10160}{9\pi^2}-\frac{343}{3}.
 \label{eq:auxTstructureintegrals}
\end{align}
These are the auxiliary contributions used in
eqs.~\eqref{eq:scalarstressresult} and~\eqref{eq:GNstressresult}.
Their normalization uses two-component Dirac fermions and complex scalars
throughout.

\section{Induced vertices and contact bubbles}\label{app:induced}

The induced vertices are the first functional derivatives of the matter
current kernel with respect to the metric. Their triangle integrals reduce to the bubble and
triangle masters in eq.~\eqref{eq:trianglemasters}.
For the cut kinematics, use $P=(E_P,-\ii K,0)$ and
$Q=(E_Q,\ii K,0)$, with $E_P=(1+x^2-y^2)/2$, $E_Q=1-E_P$,
and $K^2=E_P^2-x^2$.
The vectors $w_P=(\ii K,E_P,0)$, $w_Q=(-\ii K,E_Q,0)$,
and $z=(0,0,1)$ span the transverse current directions.
\subsubsection*{Photon-kernel triangles and contact terms}

For traceless $e$, the induced vertices in eqs.~\eqref{eq:inducedfermion}
and~\eqref{eq:inducedscalar} contain three nonzero contributions.
We separate these contributions by writing
\begin{equation}
 \Gamma_{a,e}=\mathcal T_{a,e}+\mathcal C_{a,e}^{Q}
 +\mathcal C_{a,e}^{P},\qquad a=f,b.
 \label{eq:inducedseparation}
\end{equation}
Here $\mathcal T_{a,e}$ denotes the integrated triangle term, including
both orientations and the prefactor in its defining vertex formula.
The other two terms are the integrated contact bubbles:
\begin{align}
 \mathcal C_{f,e}^{Q}=e\Pi(Q),&\qquad
 \mathcal C_{f,e}^{P}=\Pi(P)e,\nonumber\\
 \mathcal C_{b,e}^{Q}=(2e-\Tr(e)I)\Pi(Q),&\qquad
 \mathcal C_{b,e}^{P}=\Pi(P)(2e-\Tr(e)I).
 \label{eq:individualcontacts}
\end{align}
Each bubble uses the normalization in eq.~\eqref{eq:bubble}.
In the transverse basis specified above, its four nonzero
components are
\begin{equation}
\begin{array}{c|cc}
 \text{projection}&\mathcal C_{f,e}^{Q}&\mathcal C_{f,e}^{P}\\\hline
 w_P^T(\cdot)_+w_Q&yE_PE_Q/16&xE_PE_Q/16\\
 z^T(\cdot)_+z&-y/16&-x/16\\
 w_P^T(\cdot)_\times z&yE_P/16&xE_P/16\\
 z^T(\cdot)_\times w_Q&yE_Q/16&xE_Q/16
\end{array}
 \label{eq:contactprojectiontable}
\end{equation}
The scalar contact entries are twice these fermion entries because
$e_+$ and $e_\times$ are traceless.
The unlisted transverse entries vanish by reflection in the third axis.

The triangle integrals below use $u=x+y$ and $v=x-y$, as in the radial
calculation. Their dependence on $u,v$ follows from the exact tensor
reduction in eq.~\eqref{eq:trianglemasters}.
\begin{align}
w_P^T\mathcal T_{f,+}w_Q&=\frac{1}{128 \left(u + 1\right)^{2}}\bigl[u^{5} v^{2} - u^{5} + 2 u^{4} v^{2} - 2 u^{4}\nonumber\\
&\quad +u^{3} v^{4} + 4 u^{3} v^{2} - u^{3} + 2 u^{2} v^{4}\nonumber\\
&\quad +6 u^{2} v^{2} - 4 u^{2} - u v^{4} - u v^{2}\nonumber\\
&\quad - 2 u - 4 v^{4}\bigr].
\label{eq:triangle-f-ww}
\end{align}

\begin{align}
z^T\mathcal T_{f,+}z&=\frac{1}{32 \left(u + 1\right)^{2}}\bigl[- u^{3} v^{2} + u^{3} - 2 u^{2} v^{2} + 2 u^{2}\nonumber\\
&\quad +u v^{2} + 3 u + 4\bigr].
\label{eq:triangle-f-zz}
\end{align}

\begin{align}
w_P^T\mathcal T_{f,\times}z&=\frac{1}{32 \left(u + 1\right)^{2}}\bigl[u^{3} v^{2} - u^{3} + 2 u^{2} v^{2} - 2 u^{2} v\nonumber\\
&\quad - 2 u^{2} - 2 u v - 2 u + v^{3}\nonumber\\
&\quad - 2 v^{2} - v\bigr].
\label{eq:triangle-f-wz}
\end{align}

\begin{align}
z^T\mathcal T_{f,\times}w_Q&=\frac{1}{32 \left(u + 1\right)^{2}}\bigl[u^{3} v^{2} - u^{3} + 2 u^{2} v^{2} + 2 u^{2} v\nonumber\\
&\quad - 2 u^{2} + 2 u v - 2 u - v^{3}\nonumber\\
&\quad - 2 v^{2} + v\bigr].
\label{eq:triangle-f-zw}
\end{align}

\begin{align}
w_P^T\mathcal T_{b,+}w_Q&=\frac{1}{128 \left(u + 1\right)^{2}}\bigl[3 u^{5} v^{2} - u^{5} + 6 u^{4} v^{2} - 2 u^{4}\nonumber\\
&\quad +u^{3} v^{4} + 6 u^{3} v^{2} - 3 u^{3} + 2 u^{2} v^{4}\nonumber\\
&\quad +2 u^{2} v^{2} - 4 u^{2} - u v^{4} - u v^{2}\nonumber\\
&\quad - 4 u - 4 v^{2}\bigr].
\label{eq:triangle-b-ww}
\end{align}

\begin{align}
z^T\mathcal T_{b,+}z&=\frac{1}{32 \left(u + 1\right)^{2}}\bigl[- u^{3} v^{2} + 3 u^{3} - 2 u^{2} v^{2} + 6 u^{2}\nonumber\\
&\quad +u v^{2} + 5 u + 4 v^{2}\bigr].
\label{eq:triangle-b-zz}
\end{align}

\begin{align}
w_P^T\mathcal T_{b,\times}z&=\frac{1}{32 \left(u + 1\right)^{2}}\bigl[- u^{4} v + u^{3} v^{2} - 2 u^{3} v - 2 u^{3}\nonumber\\
&\quad +2 u^{2} v^{2} - 3 u^{2} v - 4 u^{2} - 2 u v^{2}\nonumber\\
&\quad - 2 u v - u - v^{3} - 2 v^{2}\nonumber\\
&\quad +v\bigr].
\label{eq:triangle-b-wz}
\end{align}

\begin{align}
z^T\mathcal T_{b,\times}w_Q&=\frac{1}{32 \left(u + 1\right)^{2}}\bigl[u^{4} v + u^{3} v^{2} + 2 u^{3} v - 2 u^{3}\nonumber\\
&\quad +2 u^{2} v^{2} + 3 u^{2} v - 4 u^{2} - 2 u v^{2}\nonumber\\
&\quad +2 u v - u + v^{3} - 2 v^{2}\nonumber\\
&\quad - v\bigr].
\label{eq:triangle-b-zw}
\end{align}

Adding the two bubble entries in eq.~\eqref{eq:contactprojectiontable}
to each fermion triangle gives the full vertex in eq.~\eqref{eq:inducedfermion}.
Adding twice those entries to each scalar triangle gives the scalar
full vertex.
The two scalar seagull terms are zero by
eq.~\eqref{eq:scalarseagullzero} and the scaleless tadpole integral.
Thus these formulas evaluate every diagram of the induced photon kernel
with one stress-tensor insertion in the transverse projections used by the cut.

In the scalar seagull bubble, both photons meet at one vertex, while the
improved stress tensor is inserted at the other.
For a symmetric polarization $h$ and arbitrary nonzero Euclidean
momentum $q$, its matter integral is
\begin{align}
 &\intk\frac{2k\cdot h\cdot(k-q)-\Tr(h)k\cdot(k-q)
 +(q\cdot h\cdot q-\Tr(h)q^2)/4}{k^2(k-q)^2}\nonumber\\
 &=2h_{ab}\frac{3q_aq_b-\delta_{ab}q^2}{64|q|}
 -2h_{ab}q_b\frac{q_a}{16|q|}
 +\Tr(h)\frac{q^2}{16|q|}
 +\frac{q\cdot h\cdot q-\Tr(h)q^2}{32|q|}\nonumber\\
 &=\frac{\Tr(h)q^2-q\cdot h\cdot q}{32|q|}
 +\frac{q\cdot h\cdot q-\Tr(h)q^2}{32|q|}=0.
 \label{eq:scalarseagullzero}
\end{align}
The first line uses the improved stress-tensor vertex.
The second substitutes the bubble moments
$\intk[k^2(k-q)^2]^{-1}=1/(8|q|)$,
$\intk k_a/[k^2(k-q)^2]=q_a/(16|q|)$, and
$\intk k_ak_b/[k^2(k-q)^2]=(3q_aq_b-\delta_{ab}q^2)/(64|q|)$.
The kinetic and curvature terms cancel in the last line.
Without the curvature improvement, this bubble would be nonzero.

The six labeled terms in this kernel therefore consist
of two triangle orientations, two mixed-vertex bubbles, the vanishing
seagull bubble in eq.~\eqref{eq:scalarseagullzero}, and one vanishing
tadpole. This accounts for the absence of the last two terms from the
nonzero induced vertex in eq.~\eqref{eq:inducedscalar}.

\subsubsection*{Stress-tensor vertices involving the auxiliary scalar}

The auxiliary-field contributions to the stress-tensor amplitudes contain emission from both pairs and
the induced $T\sigma\sigma$ vertex. In the same reduced normalization
as eq.~\eqref{eq:fullstressamp}, the fermion amplitude is
\begin{align}
 \mathcal M_{\sigma,f,e}={}&
 \frac{\bar u_1[(\gamma^a r_{1a})V_e(r_1,-k_2)/r_1^2
 -V_e(k_1,-r_2)(\gamma^a r_{2a})/r_2^2]v_2}{2y}
 (\bar u_3v_4)\nonumber\\
 &+(1,2\leftrightarrow3,4)
 +\frac4{xy}(\bar u_1v_2)(\bar u_3v_4)\Gamma_{\sigma,f,e},
 \label{eq:auxTfermionamp}\\
 \Gamma_{\sigma,f,e}={}&2\intk
 \frac{\operatorname{tr}[(\gamma^a e_{ab}(k-q/2)^b)
 (\gamma^c k_c)(\gamma^d(k-P)_d)(\gamma^r(k-q)_r)]}
 {k^2(k-P)^2(k-q)^2}.
 \label{eq:auxTfermionloop}
\end{align}
The pair exchange also exchanges $x$ and $y$ and the corresponding
internal fermion momenta. All momenta in the gamma matrices are hatted.
For the scalar amplitude and traceless $e$, the corresponding formulas are
\begin{align}
 \mathcal M_{\sigma,b,e}={}&\frac y2
 \left[\frac{V_{b,e}(r_1,k_2)}{r_1^2}
 +\frac{V_{b,e}(k_1,r_2)}{r_2^2}\right]
 +(1,2\leftrightarrow3,4)+4xy\Gamma_{\sigma,b,e},
 \label{eq:auxTscalaramp}\\
 \Gamma_{\sigma,b,e}={}&-2\intk
 \frac{2k\cdot e\cdot(k-q)+(q\cdot e\cdot q)/4}
 {k^2(k-P)^2(k-q)^2}.
 \label{eq:auxTscalarloop}
\end{align}
The physical amplitudes have the same prefactor $16/N$ as the photon
amplitudes. The relative factor $1/2$ in each emission comes from the
auxiliary propagator normalization.
The induced terms cancel the longitudinal stress-tensor source contractions
of the emissions. The same relative signs are used in the interference
test and in the squared amplitudes.

For the polarizations $e_+$ and $e_\times$, the scalar loop gives
\begin{equation}
 \Gamma_{\sigma,b,+}=
 \frac{(x-y-1)(x-y+1)(x+y-1)(x+y+2)}{16xy(x+y+1)},
 \qquad\Gamma_{\sigma,b,\times}=0.
 \label{eq:auxscalarprojections}
\end{equation}
The auxiliary-field radial integrals for the current and stress tensor in
appendix~\ref{app:auxiliaryradial} use these vertices with the same
phase-space and detector conventions as the photon calculation.

\section{Sphere kernels, saddle, and spectral continuation}
\label{app:sphere-details}

\subsection*{Resolvents and matter kernels}

The unit round-sphere scalar harmonics have Laplacian eigenvalues
$n^2-1$ and degeneracy $n^2$, with $n=1,2,\ldots$.
For the operator with eigenvalues $n^2-a^2$, the Green function is
\begin{equation}
 G_a(\chi)=\frac{\sin[a(\pi-\chi)]}{4\pi\sin\chi\sin(\pi a)},
 \qquad0<\chi<\pi.
 \label{eq:round-resolvent}
\end{equation}
Here $\chi$ is geodesic distance and $a$ is continued away from the
integer poles. The conformal scalar corresponds to $a=1/2$ and
$G_{1/2}=1/[8\pi\sin(\chi/2)]$.
Successive derivatives with respect to $a^2$ give the kernels of higher inverse powers:
\begin{equation}
 \frac1{r!}\left(\frac1{2a}\partial_a\right)^rG_a(\chi)
 =\bigl[(-\nabla^2+1-a^2)^{-r-1}\bigr](\chi).
 \label{eq:resolvent-powers}
\end{equation}
The equality follows by differentiating each spectral eigenvalue.
At $a=1/2$, writing $z=\tan(\chi/2)$ and $\eta=\pi-\chi$, the first
three differentiated kernels divided by $G_{1/2}$ are
\begin{align}
 r=1:&\quad\eta z,\nonumber\\
 r=2:&\quad\tfrac12(\pi^2-\eta^2-2\eta z),\nonumber\\
 r=3:&\quad\tfrac16[(-\eta^3+3\pi^2\eta+12\eta)z-6\pi^2+6\eta^2].
 \label{eq:resolvent-powers-explicit}
\end{align}
These kernels suffice for the cubic metric and mixed flavor responses.
The exact spectra in section~\ref{sec:sphere} specify their source
coefficients by expanding the inverse operator before harmonic projection.

Let $G_i$ and $P_i$ denote the scalar and spinor Green functions at the
source-dependent saddle. The scalar auxiliary kernel is proportional to
$\sum_iG_i(x,y)G_i(y,x)$, and the fermion auxiliary kernel is proportional
to $-\sum_i\operatorname{tr}[P_i(x,y)P_i(y,x)]$.
Their constants are fixed by the propagators in eq.~\eqref{eq:sigmaprop}.
The photon kernels are the second functional derivatives of the matter
effective action with respect to the gauge field. They include the scalar seagull before the gap equation
is imposed. Its local term is proportional to
$\sum_iG_i(x,x)$, which vanishes by that equation in the continued
prescription. Differentiating the gap equation also cancels the
source derivatives of this local term.

For a scalar fluctuation kernel $K$, its round-sphere projection can
be written as
\begin{equation}
 K_n=\frac1n\int_{S^3}\dd V_x\,
 \frac{\sin(n\chi)}{\sin\chi}K(x,1).
 \label{eq:character-projection}
\end{equation}
The character is the trace in spin $j=(n-1)/2$.
For a squashed metric, one projects the source derivatives on the
matrix elements of that representation. One-form kernels add a vector
index, so each fixed magnetic sector gives a matrix of dimension at
most three. The quadratic form and inner product are restricted to transverse
one-forms, which represent gauge-field configurations modulo gauge
transformations.
At zero source, the two nonzero photon branches have eigenvalues
$(n+1)/16$ and $(n-1)/16$ wherever those states are present.
The left-representation multiplicity is $n$.
These spectra provide a direct normalization test of the projected kernels.

\subsection*{Scalar saddle and determinant responses}\label{app:sphere-saddle}

\subsubsection*{Scalar saddle with metric and flavor sources}

To expand the gap equation, use
$A_n=n^2-1/4$ and
$\sigma_{rs}=N_b[\varepsilon^r\lambda^s]\sigma_*(\varepsilon,\lambda)$
for $(r,s)=(0,2),(1,2)$.
The magnetic sums needed at these orders are
\begin{equation}
 \sum_m1=n,\qquad
 \sum_m m^2=\frac{n(n^2-1)}{12},\qquad
 \sum_m m^4=\frac{n(n^2-1)(3n^2-7)}{240}.
 \label{eq:saddlemagneticmoments}
\end{equation}
The remaining sums can be generated from
\begin{align}
 \sum_{n=1}^{\infty}\frac1{n^2-a^2}
 &=\frac1{2a^2}-\frac\pi{2a}\cot(\pi a),\nonumber\\
 \sum_{n=1}^{\infty}\frac{n^2}{(n^2-a^2)^2}
 &=\frac1{2a}\frac{\dd}{\dd a}
 \left[a^2\sum_{n=1}^{\infty}\frac1{n^2-a^2}\right]
 =\frac{\pi^2}{4}\csc^2(\pi a)-\frac\pi{4a}\cot(\pi a).
 \label{eq:saddlegeneratingsum}
\end{align}
The first equality is the partial-fraction expansion of the cotangent.
Differentiating the second line generates the higher powers required here.
Writing
\begin{equation}
 S_k=\sum_{n=1}^{\infty}\frac{n^2}{A_n^k},\qquad k=2,3,4,
 \label{eq:saddlesumsdefinition}
\end{equation}
these derivatives give
\begin{align}
 S_2&=\frac{\pi^2}{4},\nonumber\\
 S_3&=\left.\frac1{4a}\frac{\dd}{\dd a}
 \left[\frac{\pi^2}{4}\csc^2(\pi a)-\frac\pi{4a}\cot(\pi a)\right]
 \right|_{a=1/2}=\frac{\pi^2}{4},\nonumber\\
 S_4&=\left.\frac16\left(\frac1{2a}\frac{\dd}{\dd a}\right)^2
 \left[\frac{\pi^2}{4}\csc^2(\pi a)-\frac\pi{4a}\cot(\pi a)\right]
 \right|_{a=1/2}=\frac{\pi^4}{12}-\frac{\pi^2}{2}.
 \label{eq:saddlesumsvalues}
\end{align}
All three sums are absolutely convergent.
The coefficient of $\lambda^2$ in the gap equation is consequently
\begin{align}
 0&=\sum_{n,m}n\left[
 \frac{16m^2}{A_n^3}-\frac{1+\sigma_{02}}{A_n^2}\right]
 =\frac43\left(S_2-\frac34S_3\right)-(1+\sigma_{02})S_2
 \nonumber\\
 &=\frac{\pi^2}{12}-(1+\sigma_{02})\frac{\pi^2}{4}
 \quad\Rightarrow\quad\sigma_{02}=-\frac23.
 \label{eq:saddlesourcecoefficient}
\end{align}
For the mixed coefficient, the linear metric perturbation of each scalar
eigenvalue is $-4m^2-1/4$.
Substituting this perturbation and $\sigma_{02}=-2/3$ into the inverse
operator expansion gives
\begin{align}
 0={}&\sum_{n,m}n\left[
 \frac{1-\sigma_{12}}{A_n^2}
 +\frac{2[(-4m^2-1/4)/3-16m^2]}{A_n^3}
 -\frac{48(-4m^2-1/4)m^2}{A_n^4}\right]
 \nonumber\\
 ={}&(1-\sigma_{12})S_2
 +\sum_{n=1}^{\infty}n^2\left[
 -\frac2{9A_n^2}-\frac{8(n^2-1)}{3A_n^3}
 +\frac{(n^2-1)(12n^2-23)}{5A_n^4}\right]
 \nonumber\\
 ={}&(1-\sigma_{12})S_2-\frac{22}{45}S_2-\frac{19}{5}S_3+3S_4
 \nonumber\\
 ={}&(1-\sigma_{12})\frac{\pi^2}{4}
 +\frac{\pi^4}{4}-\frac{463\pi^2}{180}
 \quad\Rightarrow\quad\sigma_{12}=\pi^2-\frac{418}{45}.
 \label{eq:saddlemixedcoefficient}
\end{align}
The second line uses eq.~\eqref{eq:saddlemagneticmoments}.
The third line substitutes $n^2=A_n+1/4$ and collects denominator powers.
The two solved coefficients establish eq.~\eqref{eq:mixedsaddle}
from the gap equation.

\subsubsection*{Metric derivatives of the determinant}

The source-dependent determinant is continued before differentiation.
To exhibit its metric response, write on each physical representation
\begin{equation}
 K_n(\varepsilon)=K_{0,n}+\varepsilon K_{1,n}
 +\varepsilon^2K_{2,n}+\varepsilon^3K_{3,n}+O(\varepsilon^4).
 \label{eq:metric-kernel-series}
\end{equation}
The subscripts label Taylor coefficients, including the factorials.
For the formal log-determinant, matrix differentiation gives
\begin{align}
 [\varepsilon^2]\tfrac12\Tr_n\log K_n
 &=\tfrac12\Tr_n\left[K_{0,n}^{-1}K_{2,n}
 -\tfrac12K_{0,n}^{-1}K_{1,n}K_{0,n}^{-1}K_{1,n}\right],\nonumber\\
 [\varepsilon^3]\tfrac12\Tr_n\log K_n
 &=\tfrac12\Tr_n\left[K_{0,n}^{-1}K_{3,n}
 -K_{0,n}^{-1}K_{1,n}K_{0,n}^{-1}K_{2,n}
 +\tfrac13(K_{0,n}^{-1}K_{1,n})^3\right].
 \label{eq:metric-matrix-traces}
\end{align}
These finite traces are not yet the continued infinite determinant.
For a positive-order kernel the continued determinant uses
\begin{equation}
 \log\det K=-\left.\partial_s\Tr K^{-s}\right|_{s=0}.
 \label{eq:zeta-determinant}
\end{equation}
The trace first converges for sufficiently large real part of $s$.
For the negative-order scalar auxiliary kernel one instead starts
from $\Tr K^s$ and takes $+\partial_s$ at zero.
Both prescriptions follow from the same eigenvalue logarithm.
Poles of the analytically continued sums give finite contributions
to the formal traces in
eq.~\eqref{eq:metric-matrix-traces}, just as in
eq.~\eqref{eq:mixedzetaexpansion}.

The photon determinant also contains the scalar Laplacian and the
global gauge-volume factor in eq.~\eqref{eq:quotientdet}.
Their Taylor coefficients follow from the nonconstant spectrum
\begin{equation}
 n^2-1-\frac{4m^2\varepsilon}{1+\varepsilon},
 \qquad n\geq2,\quad
 m=-\frac{n-1}{2},\ldots,\frac{n-1}{2},
 \label{eq:ghost-spectrum}
\end{equation}
with degeneracy $n$ for each $m$.
The first omitted mode is the constant gauge transformation.
The magnetic sums in eq.~\eqref{eq:saddlemagneticmoments} reduce its
metric derivatives to rational sums in $n$.
Their continuation, together with $\log V=\log(2\pi^2)+\frac12\log(1+\varepsilon)$,
gives eq.~\eqref{eq:ghostresponse}.

For numerical evaluation, the singular radial terms are subtracted
analytically before quadrature. The large-representation expansions
are then fitted and summed by zeta continuation.
We checked convergence by refining the numerical integration,
increasing the number of modes, and varying the extrapolation of the
spectral sums. The results are stable under these changes.
These checks were performed independently of the flat-space calculation.

\section{Metric response of photon--auxiliary mixing}
\label{app:topological}

The mixed trace in eq.~\eqref{eq:topmixtrace} is evaluated using the fixed
round-sphere measure. All kernels below include the density factors
needed for this measure. This convention makes the scalar and vector
bases independent of the squashing parameter.
Write their expansions to first order in the squashing parameter $\varepsilon$ as
\begin{equation}
 K(\varepsilon)=K_0+\varepsilon K_1+O(\varepsilon^2),\quad
 B(\varepsilon)=B_0+\varepsilon B_1+O(\varepsilon^2),\quad
 R(\varepsilon)=R_0+\varepsilon R_1+O(\varepsilon^2).
 \label{eq:topkernelexpansions}
\end{equation}
The matrix $R_0$ already contains one derivative with respect to the
background gauge-field amplitude $t$. The matrix $R_1$ contains an
additional derivative with respect to the squashing parameter $\varepsilon$.

For the scalar propagator, let $X$ be the right-invariant unit vector
field dual to $\omega$ at the round metric.
It has eigenvalue $2\ii m$ on a scalar harmonic with magnetic label $m$.
Let $L_0=-\nabla^2+3/4$ be the round conformal scalar operator.
Its inverse $G_{00}=L_0^{-1}$ has the kernel in
eq.~\eqref{eq:round-resolvent} at $a=1/2$.
The reduced inverse operator expands as
\begin{equation}
 G(\varepsilon,t)=G_{00}+\varepsilon G_{10}+tG_{01}
 +\varepsilon tG_{11}+O(\varepsilon^2,t^2).
 \label{eq:topscalarexpansion}
\end{equation}
The charged spectrum in eq.~\eqref{eq:chargedscalarspectrum} gives
the operator perturbations $L_{10}=X^2-1/4$,
$L_{01}=-2\ii X$, and $L_{11}=2\ii X$.
The scalar saddle has no linear term at these orders.
Differentiating the inverse operator therefore gives
\begin{align}
 G_{10}&=-G_{00}L_{10}G_{00}
 =-(X^2-1/4)L_0^{-2},\nonumber\\
 G_{01}&=-G_{00}L_{01}G_{00}
 =2\ii XL_0^{-2},\nonumber\\
 G_{11}&=G_{00}L_{10}G_{00}L_{01}G_{00}
 +G_{00}L_{01}G_{00}L_{10}G_{00}
 -G_{00}L_{11}G_{00}\nonumber\\
 &=-4\ii X(X^2-1/4)L_0^{-3}-2\ii XL_0^{-2}.
 \label{eq:topscalarinverse}
\end{align}
The last equalities use $[X,L_0]=0$.
The kernels of $L_0^{-2}$ and $L_0^{-3}$ follow by the derivatives with respect to $a^2$
in eq.~\eqref{eq:resolvent-powers}.

Let $V_a$ denote a left-invariant unit vector field, with $a=1,2,3$.
For compactness in the next equation, $G_{kl}^{\mathrm{rev}}(x,y)$
denotes $G_{kl}(y,x)$, and products are pointwise products of kernels.
The mixed second functional derivative of the matter effective action
with respect to the auxiliary and gauge fields inserts the scalar
density and the current. Its terms linear in $t$ give $R_0$ and $R_1$.
The auxiliary-field Hessian supplies $B_0$ and $B_1$. Expanding these
kernels gives, for $e=0,1$,
\begin{align}
 (R_e)_a={}&\ii\sum_{k=0}^e\sum_{l=0}^1
 \left[G_{kl}^{\mathrm{rev}}V_aG_{e-k,1-l}
 -G_{kl}V_aG_{e-k,1-l}^{\mathrm{rev}}\right]\nonumber\\
 &-2\delta_{a3}\sum_{k=0}^eG_{k0}^{\mathrm{rev}}G_{e-k,0}
 -e\delta_{a3}(R_0)_a,\nonumber\\
 B_e={}&\sum_{k=0}^eG_{k0}^{\mathrm{rev}}G_{e-k,0}.
 \label{eq:topscalarmixedderivative}
\end{align}
The second line includes the explicit background term in the scalar
current. Its last term differentiates the inverse metric in that vertex.
Both terms are required for the longitudinal Ward identity.

For fermions, write $P_{00}=D_0^{-1}$ for the round Dirac inverse.
The Dirac operator in eqs.~\eqref{eq:DiracSphere} and
\eqref{eq:chargedDirac} has perturbations
\begin{equation}
 D_{10}=-\sigma_3J_3-\frac14,\qquad
 D_{01}=\sigma_3,\qquad D_{11}=-\frac12\sigma_3.
 \label{eq:topDiracvariations}
\end{equation}
Here $J_3$ is the angular-momentum generator used in
eq.~\eqref{eq:DiracSphere}.
Writing the inverse coefficients as $P_{10},P_{01},P_{11}$ gives
\begin{align}
 P_{10}&=-P_{00}D_{10}P_{00},\qquad
 P_{01}=-P_{00}\sigma_3P_{00},\nonumber\\
 P_{11}&=P_{00}D_{10}P_{00}\sigma_3P_{00}
 +P_{00}\sigma_3P_{00}D_{10}P_{00}
 +\frac12P_{00}\sigma_3P_{00}.
 \label{eq:topDiracinverse}
\end{align}
These are the same inverse-operator differentiation identities used
for the scalar.
With $P_{kl}^{\mathrm{rev}}(x,y)=P_{kl}(y,x)$, the fermion kernels are
\begin{align}
 (R_e)_a&=\sum_{k=0}^e\sum_{l=0}^1
 \operatorname{tr}\left[\sigma_aP_{kl}^{\mathrm{rev}}P_{e-k,1-l}\right]
 -\frac e2\delta_{a3}(R_0)_a,\nonumber\\
 B_e&=-\sum_{k=0}^e
 \operatorname{tr}\left[P_{k0}^{\mathrm{rev}}P_{e-k,0}\right].
 \label{eq:topfermionmixedderivative}
\end{align}
The metric term in the first line differentiates the inverse vielbein.
The sign in the second follows from the Yukawa insertion $\ii\sigma$
in the convention for the Dirac operator explained after eq.~\eqref{eq:DiracSphere}.
The photon kernel is the second functional derivative of the matter
effective action with respect to the gauge field. Its expansion
coefficients $K_0,K_1$ use these same propagators and include the scalar
seagull when applicable.

\subsubsection*{Mixed trace and consistency checks}

Differentiating eq.~\eqref{eq:topmixtrace} and using
$\partial_\varepsilon K^{-1}=-K^{-1}(\partial_\varepsilon K)K^{-1}$
gives
\begin{align}
 [\varepsilon]u_{\mathrm{mix}}
 =\frac12\Tr\big[&
 B_0^{-1}R_1^\dagger K_0^{-1}R_0
 +B_0^{-1}R_0^\dagger K_0^{-1}R_1\nonumber\\
 &-B_0^{-1}B_1B_0^{-1}R_0^\dagger K_0^{-1}R_0
 -B_0^{-1}R_0^\dagger K_0^{-1}K_1K_0^{-1}R_0\big].
 \label{eq:topmixedmetrictrace}
\end{align}
The first line differentiates the two mixed vertices.
The second differentiates the internal auxiliary and photon propagators.
Thus every possible location of the metric derivative in this trace
is included.

The harmonic projection uses the same normalized scalar and one-form
representations as section~\ref{sec:sphere}.
At fixed $n$ and right magnetic label, $B_0$ is a scalar and $K_0$
is a matrix on at most three vector components.
Its longitudinal direction is removed before inversion.
The resulting summand includes the factor $n$ from the left magnetic
degeneracy, as in eq.~\eqref{eq:topmixroundsum}.

The numerical projections reproduce the known round-sphere results
and satisfy the longitudinal Ward identities within numerical precision.
The metric summands decay as $n^{-2}$, so their sum converges.
We checked numerical convergence by increasing the number of modes,
refining the integration, and varying the extrapolation of the remaining sum.
These changes affect the metric response by amounts well below the quoted precision.

An independent check differentiates the homogeneous scalar and Dirac
spectra before summing their eigenvalues.
For the constant scalar harmonic and photon direction $\omega$, this
calculation gives the metric derivatives of the mixed kernels
\begin{equation}
 (R_1)_{b,n=1}\simeq0.03406776693900632138,\qquad
 (R_1)_{f,n=1}\simeq0.01593223306099367862.
 \label{eq:tophomogeneouscheck}
\end{equation}
The independent calculation from the matter spectra agrees with the
position-space projections to the quoted precision.

The completed metric sums give eq.~\eqref{eq:topmixmetricresults}.
Together with the analytic round trace, they supply both coefficients
needed in eq.~\eqref{eq:topgammaextraction}.

\bibliographystyle{JHEP}
\bibliography{references}

\providecommand{\href}[2]{#2}\begingroup\raggedright\begin{thebibliography}{10}

\bibitem{GuoLiThreePoint}
Y.~Guo and Z.~Li, \emph{{$1/N$ corrections to $\langle TJJ\rangle$ and $\langle TTT\rangle$ in the critical $O(N)$ vector and Gross--Neveu--Yukawa models}},  \href{https://arxiv.org/abs/2611.07394}{{\ttfamily 2611.07394}}.

\bibitem{SenthilDQCP}
T.~Senthil, A.~Vishwanath, L.~Balents, S.~Sachdev and M.P.A.~Fisher, \emph{``deconfined'' quantum critical points}, \href{https://doi.org/10.1126/science.1091806}{\emph{Science} {\bfseries 303} (2004) 1490} [\href{https://arxiv.org/abs/cond-mat/0311326}{{\ttfamily cond-mat/0311326}}].

\bibitem{WangDQCP}
C.~Wang, A.~Nahum, M.A.~Metlitski, C.~Xu and T.~Senthil, \emph{{Deconfined quantum critical points: symmetries and dualities}}, {\emph{Phys. Rev. X} {\bfseries 7} (2017) 031051} [\href{https://arxiv.org/abs/1703.02426}{{\ttfamily 1703.02426}}].

\bibitem{BoyackQEDGN}
R.~Boyack, A.~Rayyan and J.~Maciejko, \emph{Deconfined criticality in the {QED$_3$ Gross-Neveu-Yukawa} model: The {$1/N$} expansion revisited}, \href{https://doi.org/10.1103/PhysRevB.99.195135}{\emph{Phys. Rev. B} {\bfseries 99} (2019) 195135} [\href{https://arxiv.org/abs/1812.02720}{{\ttfamily 1812.02720}}].

\bibitem{SeibergDualityWeb}
N.~Seiberg, T.~Senthil, C.~Wang and E.~Witten, \emph{{A duality web in $2+1$ dimensions and condensed matter physics}}, {\emph{Annals Phys.} {\bfseries 374} (2016) 395} [\href{https://arxiv.org/abs/1606.01989}{{\ttfamily 1606.01989}}].

\bibitem{KarchTongDuality}
A.~Karch and D.~Tong, \emph{{Particle-Vortex Duality from 3d Bosonization}}, \href{https://doi.org/10.1103/PhysRevX.6.031043}{\emph{Phys. Rev. X} {\bfseries 6} (2016) 031043} [\href{https://arxiv.org/abs/1606.01893}{{\ttfamily 1606.01893}}].

\bibitem{SenthilDualitiesReview}
T.~Senthil, D.T.~Son, C.~Wang and C.~Xu, \emph{Duality between {$(2+1)d$} quantum critical points}, \href{https://doi.org/10.1016/j.physrep.2019.09.001}{\emph{Phys. Rept.} {\bfseries 827} (2019) 1} [\href{https://arxiv.org/abs/1810.05174}{{\ttfamily 1810.05174}}].

\bibitem{HermeleSpinLiquid}
M.~Hermele, T.~Senthil, M.P.A.~Fisher, P.A.~Lee, N.~Nagaosa and X.-G.~Wen, \emph{{On the stability of $U(1)$ spin liquids in two dimensions}}, {\emph{Phys. Rev. B} {\bfseries 70} (2004) 214437} [\href{https://arxiv.org/abs/cond-mat/0404751}{{\ttfamily cond-mat/0404751}}].

\bibitem{SongSpinLiquids}
X.-Y.~Song, Y.-C.~He, A.~Vishwanath and C.~Wang, \emph{From spinon band topology to the symmetry quantum numbers of monopoles in {Dirac} spin liquids}, \href{https://doi.org/10.1103/PhysRevX.10.011033}{\emph{Phys. Rev. X} {\bfseries 10} (2020) 011033} [\href{https://arxiv.org/abs/1811.11182}{{\ttfamily 1811.11182}}].

\bibitem{HuDiracSpinLiquid}
S.~Hu, W.~Zhu, S.~Eggert and Y.-C.~He, \emph{{Dirac} spin liquid on the spin-1/2 triangular {Heisenberg} antiferromagnet}, \href{https://doi.org/10.1103/PhysRevLett.123.207203}{\emph{Phys. Rev. Lett.} {\bfseries 123} (2019) 207203} [\href{https://arxiv.org/abs/1905.09837}{{\ttfamily 1905.09837}}].

\bibitem{RantnerWenCuprates}
W.~Rantner and X.-G.~Wen, \emph{{Electron spectral function and algebraic spin liquid for the normal state of underdoped high $T_c$ superconductors}}, \href{https://doi.org/10.1103/PhysRevLett.86.3871}{\emph{Phys. Rev. Lett.} {\bfseries 86} (2001) 3871} [\href{https://arxiv.org/abs/cond-mat/0010378}{{\ttfamily cond-mat/0010378}}].

\bibitem{RantnerWenSpinCorrelations}
W.~Rantner and X.-G.~Wen, \emph{{Spin correlations in the algebraic spin liquid: Implications for high-$T_c$ superconductors}}, \href{https://doi.org/10.1103/PhysRevB.66.144501}{\emph{Phys. Rev. B} {\bfseries 66} (2002) 144501} [\href{https://arxiv.org/abs/cond-mat/0201521}{{\ttfamily cond-mat/0201521}}].

\bibitem{FranzTesanovicQED3}
M.~Franz and Z.~Te{\v{s}}anovi{\'{c}}, \emph{{Algebraic Fermi liquid from phase fluctuations: ``Topological'' fermions, vortex ``berryons,'' and QED$_3$ theory of cuprate superconductors}}, \href{https://doi.org/10.1103/PhysRevLett.87.257003}{\emph{Phys. Rev. Lett.} {\bfseries 87} (2001) 257003} [\href{https://arxiv.org/abs/cond-mat/0012445}{{\ttfamily cond-mat/0012445}}].

\bibitem{FranzTesanovicVafekQED3}
M.~Franz, Z.~Te{\v{s}}anovi{\'{c}} and O.~Vafek, \emph{{QED$_3$ theory of pairing pseudogap in cuprates: From $d$-wave superconductor to antiferromagnet via an algebraic Fermi liquid}}, \href{https://doi.org/10.1103/PhysRevB.66.054535}{\emph{Phys. Rev. B} {\bfseries 66} (2002) 054535} [\href{https://arxiv.org/abs/cond-mat/0203333}{{\ttfamily cond-mat/0203333}}].

\bibitem{ChowdhuryTransport}
D.~Chowdhury, S.~Raju, S.~Sachdev, A.~Singh and P.~Strack, \emph{{Multipoint correlators of conformal field theories: implications for quantum critical transport}}, {\emph{Phys. Rev. B} {\bfseries 87} (2013) 085138} [\href{https://arxiv.org/abs/1210.5247}{{\ttfamily 1210.5247}}].

\bibitem{KatzEtAl}
E.~Katz, S.~Sachdev, E.S.~S{\o}rensen and W.~Witczak-Krempa, \emph{{Conformal field theories at nonzero temperature: Operator product expansions, Monte Carlo, and holography}}, {\emph{Phys. Rev. B} {\bfseries 90} (2014) 245109} [\href{https://arxiv.org/abs/1409.3841}{{\ttfamily 1409.3841}}].

\bibitem{ChowdhuryDavidPrakash}
S.~Dutta~Chowdhury, J.R.~David and S.~Prakash, \emph{{Spectral sum rules for conformal field theories in arbitrary dimensions}}, {\emph{JHEP} {\bfseries 07} (2017) 119} [\href{https://arxiv.org/abs/1612.00609}{{\ttfamily 1612.00609}}].

\bibitem{ChesterPufuQEDBootstrap}
S.M.~Chester and S.S.~Pufu, \emph{{Towards bootstrapping QED$_3$}}, {\emph{JHEP} {\bfseries 08} (2016) 019} [\href{https://arxiv.org/abs/1601.03476}{{\ttfamily 1601.03476}}].

\bibitem{LiDQCPBootstrap}
Z.~Li, \emph{{Bootstrapping conformal QED$_3$ and deconfined quantum critical point}}, {\emph{JHEP} {\bfseries 11} (2022) 005} [\href{https://arxiv.org/abs/1812.09281}{{\ttfamily 1812.09281}}].

\bibitem{LiPolandGauge}
Z.~Li and D.~Poland, \emph{{Searching for gauge theories with the conformal bootstrap}}, {\emph{JHEP} {\bfseries 03} (2021) 172} [\href{https://arxiv.org/abs/2005.01721}{{\ttfamily 2005.01721}}].

\bibitem{AlbayrakQEDBootstrap}
S.~Albayrak, R.S.~Erramilli, Z.~Li, D.~Poland and Y.~Xin, \emph{{Bootstrapping $N_f=4$ conformal QED$_3$}}, {\emph{Phys. Rev. D} {\bfseries 105} (2022) 085008} [\href{https://arxiv.org/abs/2112.02106}{{\ttfamily 2112.02106}}].

\bibitem{ChesterSuDQCP}
S.M.~Chester and N.~Su, \emph{{Bootstrapping deconfined quantum tricriticality}}, {\emph{Phys. Rev. Lett.} {\bfseries 132} (2024) 111601} [\href{https://arxiv.org/abs/2310.08343}{{\ttfamily 2310.08343}}].

\bibitem{LiShenDQCP}
Z.~Li and T.~Shen, \emph{{Bootstrap Cone of the Multicritical Deconfined Quantum Critical Point}},  \href{https://arxiv.org/abs/2606.09985}{{\ttfamily 2606.09985}}.

\bibitem{DymarskyCurrentBootstrap}
A.~Dymarsky, J.~Penedones, E.~Trevisani and A.~Vichi, \emph{{Charting the space of 3D CFTs with a continuous global symmetry}}, \href{https://doi.org/10.1007/JHEP05(2019)098}{\emph{JHEP} {\bfseries 05} (2019) 098} [\href{https://arxiv.org/abs/1705.04278}{{\ttfamily 1705.04278}}].

\bibitem{ReehorstMixed}
M.~Reehorst, E.~Trevisani and A.~Vichi, \emph{{Mixed Scalar-Current bootstrap in three dimensions}}, {\emph{JHEP} {\bfseries 12} (2020) 156} [\href{https://arxiv.org/abs/1911.05747}{{\ttfamily 1911.05747}}].

\bibitem{HeNonAbelianCurrents}
Y.-C.~He, J.~Rong, N.~Su and A.~Vichi, \emph{{Non-Abelian currents bootstrap}}, {\emph{JHEP} {\bfseries 03} (2024) 175} [\href{https://arxiv.org/abs/2302.11585}{{\ttfamily 2302.11585}}].

\bibitem{DymarskyStressBootstrap}
A.~Dymarsky, F.~Kos, P.~Kravchuk, D.~Poland and D.~Simmons-Duffin, \emph{{The 3d Stress-Tensor Bootstrap}}, \href{https://doi.org/10.1007/JHEP02(2018)164}{\emph{JHEP} {\bfseries 02} (2018) 164} [\href{https://arxiv.org/abs/1708.05718}{{\ttfamily 1708.05718}}].

\bibitem{ChangEtAl}
C.-H.~Chang, V.~Dommes, R.S.~Erramilli, A.~Homrich, P.~Kravchuk, A.~Liu et~al., \emph{{Bootstrapping the 3d Ising stress tensor}}, {\emph{JHEP} {\bfseries 03} (2025) 136} [\href{https://arxiv.org/abs/2411.15300}{{\ttfamily 2411.15300}}].

\bibitem{ErramilliMitchellStress}
R.S.~Erramilli and M.S.~Mitchell, \emph{{Towards 3D CFT Cartography with the Stress Tensor Bootstrap}},  \href{https://arxiv.org/abs/2602.13383}{{\ttfamily 2602.13383}}.

\bibitem{DommesEtAl}
V.~Dommes, R.S.~Erramilli, C.~Flynn, A.~Homrich, P.~Kravchuk, D.~Poland et~al., \emph{{Bootstrapping currents and stress tensors in 3d CFTs}},  \href{https://arxiv.org/abs/2609.13370}{{\ttfamily 2609.13370}}.

\bibitem{HuhStrackSachdevCurrents}
Y.~Huh, P.~Strack and S.~Sachdev, \emph{{Conserved current correlators of conformal field theories in 2+1 dimensions}}, {\emph{Phys. Rev. B} {\bfseries 88} (2013) 155109} [\href{https://arxiv.org/abs/1307.6863}{{\ttfamily 1307.6863}}]. Erratum: \href{https://doi.org/10.1103/PhysRevB.90.199902}{\emph{Phys. Rev. B} \textbf{90} (2014) 199902}.

\bibitem{HuhStrackCentralCharges}
Y.~Huh and P.~Strack, \emph{{Stress tensor and current correlators of interacting conformal field theories in 2+1 dimensions: Fermionic Dirac matter coupled to U(1) gauge field}}, {\emph{JHEP} {\bfseries 01} (2015) 147} [\href{https://arxiv.org/abs/1410.1902}{{\ttfamily 1410.1902}}]. Erratum: \href{https://doi.org/10.1007/JHEP03(2016)054}{\emph{JHEP} \textbf{03} (2016) 054}.

\bibitem{GiombiTarnopolskyKlebanov}
S.~Giombi, G.~Tarnopolsky and I.R.~Klebanov, \emph{{On $C_J$ and $C_T$ in conformal QED}}, {\emph{JHEP} {\bfseries 08} (2016) 156} [\href{https://arxiv.org/abs/1602.01076}{{\ttfamily 1602.01076}}].

\bibitem{GuoLi}
Y.~Guo and Z.~Li, \emph{{Central charges $C_J$ and $C_T$ in QED$_d$-GNY model and scalar QED$_d$}}, \href{https://doi.org/10.1007/JHEP11(2026)136}{\emph{JHEP} {\bfseries 11} (2026) 136} [\href{https://arxiv.org/abs/2606.26623}{{\ttfamily 2606.26623}}].

\bibitem{HofmanMaldacena}
D.M.~Hofman and J.~Maldacena, \emph{{Conformal collider physics: Energy and charge correlations}}, {\emph{JHEP} {\bfseries 05} (2008) 012} [\href{https://arxiv.org/abs/0803.1467}{{\ttfamily 0803.1467}}].

\bibitem{HofmanEtAlProof}
D.M.~Hofman, D.~Li, D.~Meltzer, D.~Poland and F.~Rej\'{o}n-Barrera, \emph{{A proof of the conformal collider bounds}}, {\emph{JHEP} {\bfseries 06} (2016) 111} [\href{https://arxiv.org/abs/1603.03771}{{\ttfamily 1603.03771}}].

\bibitem{HartmanKunduTajdini}
T.~Hartman, S.~Kundu and A.~Tajdini, \emph{{Averaged null energy condition from causality}}, {\emph{JHEP} {\bfseries 07} (2017) 066} [\href{https://arxiv.org/abs/1610.05308}{{\ttfamily 1610.05308}}].

\bibitem{MyersSachdevSinghTransport}
R.C.~Myers, S.~Sachdev and A.~Singh, \emph{{Holographic quantum critical transport without self-duality}}, {\emph{Phys. Rev. D} {\bfseries 83} (2011) 066017} [\href{https://arxiv.org/abs/1010.0443}{{\ttfamily 1010.0443}}].

\bibitem{OsbornPetkou}
H.~Osborn and A.C.~Petkou, \emph{{Implications of conformal invariance in field theories for general dimensions}}, {\emph{Annals Phys.} {\bfseries 231} (1994) 311} [\href{https://arxiv.org/abs/hep-th/9307010}{{\ttfamily hep-th/9307010}}].

\bibitem{ErdmengerOsborn}
J.~Erdmenger and H.~Osborn, \emph{{Conserved currents and the energy momentum tensor in conformally invariant theories for general dimensions}}, {\emph{Nucl. Phys. B} {\bfseries 483} (1997) 431} [\href{https://arxiv.org/abs/hep-th/9605009}{{\ttfamily hep-th/9605009}}].

\bibitem{BMSMomentum}
A.~Bzowski, P.~McFadden and K.~Skenderis, \emph{{Implications of conformal invariance in momentum space}}, {\emph{JHEP} {\bfseries 03} (2014) 111} [\href{https://arxiv.org/abs/1304.7760}{{\ttfamily 1304.7760}}].

\bibitem{BMSRenormalised}
A.~Bzowski, P.~McFadden and K.~Skenderis, \emph{{Renormalised 3-point functions of stress tensors and conserved currents in CFT}}, {\emph{JHEP} {\bfseries 11} (2018) 153} [\href{https://arxiv.org/abs/1711.09105}{{\ttfamily 1711.09105}}].

\bibitem{IhrigQEDGN}
B.~Ihrig, L.~Janssen, L.N.~Mihaila and M.M.~Scherer, \emph{Deconfined criticality from the {QED$_3$-Gross-Neveu} model at three loops}, \href{https://doi.org/10.1103/PhysRevB.98.115163}{\emph{Phys. Rev. B} {\bfseries 98} (2018) 115163} [\href{https://arxiv.org/abs/1807.04958}{{\ttfamily 1807.04958}}].

\bibitem{ZerfQEDGNY}
N.~Zerf, P.~Marquard, R.~Boyack and J.~Maciejko, \emph{Critical behavior of the {QED$_3$-Gross-Neveu-Yukawa} model at four loops}, \href{https://doi.org/10.1103/PhysRevB.98.165125}{\emph{Phys. Rev. B} {\bfseries 98} (2018) 165125} [\href{https://arxiv.org/abs/1808.00549}{{\ttfamily 1808.00549}}].

\bibitem{BelitskyEventShapes}
A.V.~Belitsky, S.~Hohenegger, G.P.~Korchemsky, E.~Sokatchev and A.~Zhiboedov, \emph{{From correlation functions to event shapes}}, {\emph{Nucl. Phys. B} {\bfseries 884} (2014) 305} [\href{https://arxiv.org/abs/1309.0769}{{\ttfamily 1309.0769}}].

\bibitem{LiMeltzerPoland}
D.~Li, D.~Meltzer and D.~Poland, \emph{{Conformal collider physics from the lightcone bootstrap}}, {\emph{JHEP} {\bfseries 02} (2016) 143} [\href{https://arxiv.org/abs/1511.08025}{{\ttfamily 1511.08025}}].

\bibitem{LiSubgraphs}
Z.~Li, \emph{{Conformal 3-point correlators in momentum space, method of subgraphs and the $1/N$ expansion}}, \href{https://doi.org/10.1007/JHEP12(2025)066}{\emph{JHEP} {\bfseries 12} (2025) 066} [\href{https://arxiv.org/abs/2509.07106}{{\ttfamily 2509.07106}}].

\bibitem{LucasSierensWitczak}
A.~Lucas, T.~Sierens and W.~Witczak-Krempa, \emph{{Quantum critical response: from conformal perturbation theory to holography}}, {\emph{JHEP} {\bfseries 07} (2017) 149} [\href{https://arxiv.org/abs/1704.05461}{{\ttfamily 1704.05461}}].

\bibitem{BobevBuenoVreys}
N.~Bobev, P.~Bueno and Y.~Vreys, \emph{{Comments on squashed-sphere partition functions}}, {\emph{JHEP} {\bfseries 07} (2017) 093} [\href{https://arxiv.org/abs/1705.00292}{{\ttfamily 1705.00292}}].

\bibitem{BuenoEtAl}
P.~Bueno, P.A.~Cano, R.A.~Hennigar and R.B.~Mann, \emph{{Universality of squashed-sphere partition functions}}, {\emph{Phys. Rev. Lett.} {\bfseries 122} (2019) 071602} [\href{https://arxiv.org/abs/1808.02052}{{\ttfamily 1808.02052}}].

\bibitem{BuenoEtAl2020}
P.~Bueno, P.A.~Cano, R.A.~Hennigar, V.A.~Penas and A.~Ruip\'{e}rez, \emph{{Partition functions on slightly squashed spheres and flux parameters}}, {\emph{JHEP} {\bfseries 04} (2020) 123} [\href{https://arxiv.org/abs/2001.10020}{{\ttfamily 2001.10020}}].

\bibitem{ParmentierBobev}
K.~Parmentier and N.~Bobev, \emph{{CFTs on squashed spheres and the thermal effective action}}, {\emph{JHEP} {\bfseries 10} (2026) 079} [\href{https://arxiv.org/abs/2606.30716}{{\ttfamily 2606.30716}}].

\bibitem{LucasGazitPodolskyWitczak}
A.~Lucas, S.~Gazit, D.~Podolsky and W.~Witczak-Krempa, \emph{{Dynamical response near quantum critical points}}, {\emph{Phys. Rev. Lett.} {\bfseries 118} (2017) 056601} [\href{https://arxiv.org/abs/1608.02586}{{\ttfamily 1608.02586}}].

\bibitem{DamleSachdevTransport}
K.~Damle and S.~Sachdev, \emph{{Nonzero-temperature transport near quantum critical points}}, {\emph{Phys. Rev. B} {\bfseries 56} (1997) 8714} [\href{https://arxiv.org/abs/cond-mat/9705206}{{\ttfamily cond-mat/9705206}}].

\bibitem{ChesterCSMonopoles}
S.M.~Chester, L.V.~Iliesiu, M.~Mezei and S.S.~Pufu, \emph{{Monopole operators in $U(1)$ Chern--Simons-matter theories}}, {\emph{JHEP} {\bfseries 05} (2018) 157} [\href{https://arxiv.org/abs/1710.00654}{{\ttfamily 1710.00654}}].

\bibitem{ChowdhuryParityCollider}
S.~Dutta~Chowdhury, J.R.~David and S.~Prakash, \emph{{Constraints on parity violating conformal field theories in $d=3$}}, \href{https://doi.org/10.1007/JHEP11(2017)171}{\emph{JHEP} {\bfseries 11} (2017) 171} [\href{https://arxiv.org/abs/1707.03007}{{\ttfamily 1707.03007}}].

\bibitem{ClossetContactTerms}
C.~Closset, T.T.~Dumitrescu, G.~Festuccia, Z.~Komargodski and N.~Seiberg, \emph{{Comments on Chern--Simons contact terms in three dimensions}}, {\emph{JHEP} {\bfseries 09} (2012) 091} [\href{https://arxiv.org/abs/1206.5218}{{\ttfamily 1206.5218}}].

\bibitem{AharonyBosonization}
O.~Aharony, G.~Gur-Ari and R.~Yacoby, \emph{{Correlation functions of large $N$ Chern--Simons-matter theories and bosonization in three dimensions}}, {\emph{JHEP} {\bfseries 12} (2012) 028} [\href{https://arxiv.org/abs/1207.4593}{{\ttfamily 1207.4593}}].

\bibitem{GurAriYacobyCS}
G.~Gur-Ari and R.~Yacoby, \emph{{Correlators of large $N$ fermionic Chern--Simons vector models}}, {\emph{JHEP} {\bfseries 02} (2013) 150} [\href{https://arxiv.org/abs/1211.1866}{{\ttfamily 1211.1866}}].

\bibitem{GurAriHartnollMahajan}
G.~Gur-Ari, S.A.~Hartnoll and R.~Mahajan, \emph{{Transport in Chern--Simons-matter theories}}, {\emph{JHEP} {\bfseries 07} (2016) 090} [\href{https://arxiv.org/abs/1605.01122}{{\ttfamily 1605.01122}}].

\bibitem{GhoshMazumdarThermal}
S.~Ghosh and S.~Mazumdar, \emph{{Thermal correlators and bosonization dualities in large $N$ Chern--Simons matter theories}}, {\emph{JHEP} {\bfseries 02} (2023) 042} [\href{https://arxiv.org/abs/1912.06589}{{\ttfamily 1912.06589}}].

\bibitem{GuoThermalHall}
H.~Guo, R.~Samajdar, M.S.~Scheurer and S.~Sachdev, \emph{{Gauge theories for the thermal Hall effect}}, {\emph{Phys. Rev. B} {\bfseries 101} (2020) 195126} [\href{https://arxiv.org/abs/2002.01947}{{\ttfamily 2002.01947}}].

\bibitem{MaityThermalHall}
A.~Maity, H.~Guo, S.~Sachdev and V.~Tripathi, \emph{{Thermal Hall response of an abelian chiral spin liquid at finite temperatures}}, {\emph{Phys. Rev. B} {\bfseries 111} (2025) 205119} [\href{https://arxiv.org/abs/2411.15304}{{\ttfamily 2411.15304}}].

\bibitem{CaronHuotDetectors}
S.~Caron-Huot, M.~Kolo{\u{g}}lu, P.~Kravchuk, D.~Meltzer and D.~Simmons-Duffin, \emph{{Detectors in weakly-coupled field theories}}, {\emph{JHEP} {\bfseries 04} (2023) 014} [\href{https://arxiv.org/abs/2209.00008}{{\ttfamily 2209.00008}}].

\end{thebibliography}\endgroup
\end{document}